\documentclass[twocolumn]{aastex63}

\sfcode`A=1000 \sfcode`B=1000 \sfcode`C=1000 \sfcode`D=1000
\sfcode`E=1000 \sfcode`F=1000 \sfcode`G=1000 \sfcode`H=1000
\sfcode`I=1000 \sfcode`J=1000 \sfcode`K=1000 \sfcode`L=1000
\sfcode`M=1000 \sfcode`N=1000 \sfcode`O=1000 \sfcode`P=1000
\sfcode`Q=1000 \sfcode`R=1000 \sfcode`S=1000 \sfcode`T=1000
\sfcode`U=1000 \sfcode`V=1000 \sfcode`W=1000 \sfcode`X=1000
\sfcode`Y=1000 \sfcode`Z=1000

\received{November 11, 2020}
\accepted{June 3, 2021}
\submitjournal{ApJS}

\shorttitle{Rapid Variability of Sgr~A* across the Spectrum}
\shortauthors{Witzel,  et~al.}

\usepackage{graphicx}
\usepackage{mathrsfs}
\usepackage[lined]{algorithm2e}
\usepackage{amsmath}
\usepackage{placeins}

\usepackage{epsfig}
\usepackage{verbatim, amsmath, amsfonts, amssymb,amssymb}
\usepackage[utf8x]{inputenc}
\usepackage{float}
\usepackage{animate}
\usepackage{enumerate}
\usepackage{url}
\usepackage{bm}
\usepackage{color}
\usepackage[lined]{algorithm2e}
\usepackage{listings}
\usepackage{animate}
\usepackage{textcomp}
\usepackage{xfrac}
\usepackage{CJK}

\newcommand{\Sp}{{Spitzer\/}}
\newcommand{\SST}{{Spitzer Space Telescope\/}}

\newcommand{\Ch}{{Chandra}}
\newcommand{\Her}{{Herschel}}
\newcommand{\Sg}{Sgr~A*}

\newcommand{\Msun}{\mbox{M$_\sun$}}
\newcommand{\0}{\phantom{-}}

\begin{document}
\begin{CJK*}{UTF8}{gbsn}
\title{Rapid Variability of Sgr A* across the Electromagnetic Spectrum}

\author[0000-0003-2618-797X]{G.\ Witzel}
\affiliation{Max-Planck-Institut f\"ur Radioastronomie, Auf dem
  H\"ugel 69, 53121, Bonn, Germany}

\author[0000-0002-7476-2521]{G.\ Martinez}
\affiliation{Department of Physics and Astronomy, University of
  California, Los Angeles, Box 951547, Los Angeles, CA 90095-1547,
  USA} 

\author[0000-0002-9895-5758]{S.\ P.\ Willner}
\affiliation{Center for Astrophysics \textbar\ Harvard \& Smithsonian,
  60 Garden St., Cambridge, MA 02138 USA}



 \author{E.\ E.\ Becklin}
\affiliation{Department of Physics and Astronomy, University of
  California, Los Angeles, Box 951547, Los Angeles, CA 90095-1547,
  USA} 
 \affiliation{SOFIA Science Center, Moffett Field, CA USA}


\author{H.\ Boyce}
\affiliation{Department of Physics, McGill University, 3600
University St., Montreal, QC H3A 2T8, Canada}
\affiliation{McGill Space Institute, McGill University, Montreal, QC
H3A 2A7, Canada} 

\author[0000-0001-9554-6062]{T.\ Do}
\affiliation{Department of Physics and Astronomy, University of
  California, Los Angeles, Box 951547, Los Angeles, CA 90095-1547,
  USA} 

\author[0000-0001-6049-3132]{A.\ Eckart}
\affiliation{I.\ Physikalisches Institut, Universit\"at zu K\"oln
  Z\"ulpicher Str.\ 77 D-50937 Köln, Germany}
\affiliation{Max-Planck-Institut f\"ur Radioastronomie, Auf dem
  H\"ugel 69, 53121, Bonn, Germany}

\author[0000-0002-0670-0708]{G.\ G.\ Fazio}
\affiliation{Center for Astrophysics \textbar\ Harvard \& Smithsonian,
  60 Garden St., Cambridge, MA 02138 USA}

\author[0000-0003-3230-5055]{A.\ Ghez}
\affiliation{Department of Physics and Astronomy, University of
  California, Los Angeles, Box 951547, Los Angeles, CA 90095-1547,
  USA} 


\author[0000-0003-0685-3621]{M.\ A.\ Gurwell}
\affiliation{Center for Astrophysics \textbar\ Harvard \& Smithsonian,
  60 Garden St., Cambridge, MA 02138 USA}

\author[0000-0001-6803-2138]{D.\ Haggard}
\affiliation{Department of Physics, McGill University, 3600
University St., Montreal, QC H3A 2T8, Canada} 
\affiliation{McGill Space Institute, McGill University, Montreal, QC
  H3A 2A7, Canada} 

\author[0000-0002-7758-8717]{R.\ Herrero-Illana}
\affiliation{European Southern Observatory (ESO), Alonso de C\'ordova
  3107, Vitacura, Casilla 19001, Santiago de Chile, Chile}

\author[0000-0002-5599-4650]{J.\ L.\ Hora}
\affiliation{Center for Astrophysics \textbar\ Harvard \& Smithsonian,
  60 Garden St., Cambridge, MA 02138 USA}


\author[0000-0003-0355-6437]{Z.\ Li\ (李志远)}
\affiliation{School of Astronomy and Space Science, Nanjing University, 163 Xianlin Avenue, Nanjing 210023
  Jiangsu, China} 

\author[0000-0002-7615-7499]{J.\ Liu\ (刘俊)}
\affiliation{Max-Planck-Institut f\"ur Radioastronomie, Auf dem
  H\"ugel 69, 53121, Bonn, Germany}

\author[0000-0002-5523-7588]{N.\ Marchili}
\affiliation{Istituto di Radioastronomia---INAF, Via Piero Gobetti
  101, 40129, Bologna, Italy} 

\author[0000-0002-6753-2066]{Mark R.\ Morris}
\affiliation{Department of Physics and Astronomy, University of
  California, Los Angeles, Box 951547, Los Angeles, CA 90095-1547,
  USA} 

\author{Howard\ A.\ Smith}
\affiliation{Center for Astrophysics \textbar\ Harvard \& Smithsonian,
  60 Garden St., Cambridge, MA 02138 USA}

\author[0000-0001-6165-8525]{M.\ Subroweit}
\affiliation{I.\ Physikalisches Institut, Universit\"at zu K\"oln
  Z\"ulpicher Str.\ 77 D-50937 Köln, Germany}

\author[0000-0001-7470-3321]{J.\ A.\ Zensus}
\affiliation{Max-Planck-Institut f\"ur Radioastronomie, Auf dem
  H\"ugel 69, 53121, Bonn, Germany}
  \affiliation{I.\ Physikalisches Institut, Universit\"at zu K\"oln
  Z\"ulpicher Str.\ 77 D-50937 Köln, Germany}

\begin{abstract}

Sagittarius A* (Sgr~A*) is the variable radio, near-infrared
(NIR), and X-ray source associated with accretion onto the
Galactic center black hole. We have analyzed a comprehensive
submillimeter (including new observations simultaneous with NIR
monitoring), NIR, and 2--8~keV dataset.  Submillimeter variations
tend to lag those in the NIR by $\sim$30~minutes. An approximate
Bayesian computation (ABC) fit to the X-ray first-order structure
function shows significantly less power at short timescales in the
X-rays than in the NIR. 
Less X-ray variability at short timescales combined with the observed NIR--X-ray
correlations means the variability can be described as the result of two strictly correlated 
stochastic processes, the X-ray process being the low-pass-filtered version of the NIR process.
The NIR--X-ray
linkage suggests a simple radiative model: a compact,
self-absorbed synchrotron sphere with high-frequency cutoff close
to NIR frequencies plus a synchrotron self-Compton scattering
component at higher frequencies. This model, with parameters fit
to the submillimeter, NIR, and X-ray structure functions, reproduces the
observed flux densities at all wavelengths, the statistical
properties of all light curves, and the time lags between bands.
The fit also gives reasonable values for physical parameters such
as magnetic flux density $B\approx13$~G, source size
$L \approx2.2R_{S}$, and high-energy electron density
$n_{e}\approx4\times10^{7}$~{cm}$^{-3}$.  An animation illustrates
typical light curves, and we make public the parameter chain of
our Bayesian analysis, the model implementation, and the
visualization code.

\end{abstract}


\section{Introduction}
Since the discovery of rapid flaring of Sagittarius~A* (Sgr~A*) in
the X-rays and near infrared (NIR) in the early 2000s
\citep{2001Natur.413...45B,2003Natur.425..934G,
2004ApJ...601L.159G}, many monitoring programs have been
executed to understand the origin and properties of the variable
emission. As a result, more than 70 papers have been published
describing observations and modeling light
curves in the submillimeter (submm), NIR, and X-rays. Most of these publications
have focused on statistical analyses of flux densities and timing
properties, multi-wavelength observations and modeling of the
spectral energy distribution (SED), and NIR and submm polarization.
(See reviews by \citealt{2010RvMP...82.3121G} and \citealt{2012RAA....12..995M},
the comprehensive \citealt{2018ApJ...863...15W} introduction, and
references in their paper.) This fascination with 
\Sg's variability has a reason: light-crossing-time arguments
link rapid changes in flux density to spatial scales that---until
recently---were not accessible otherwise. For example,
\cite{2009ApJ...698..676D} found sudden flux density
changes of a factor ${\sim}2$ in less than 47~seconds, and
\cite{2019ApJ...882L..27D} saw changes of a factor ${\sim}9$ in less than
2~minutes.  These times correspond to spatial scales of $1.2R_{S}$ and 
$3R_{S}$, respectively.\footnote{Here $R_S$ means Schwarzschild radius,
  $R_S=1.23\times10^{10}$~m for mass $M=4.15\times10^6$~\Msun\ \citep{2019A&A...625L..10G}.
  The scales mentioned are not to be
  taken as distances to the black hole. They are rather upper
  limits on the characteristic size of the volume where the
  radiation originates, e.g., of a region in the accretion disc or in
  a jet likely well away from the event horizon.}

Event-horizon spatial scales can now be studied by two types of large
interferometers.  One is  VLTI/GRAVITY, which
operates in the NIR and has observed
the variable emission source moving in the plane of the sky
(\citealt{2018A&A...615L..15G,2018A&A...618L..10G,2019A&A...625L..10G,2020A&A...635A.143G,2020A&A...636L...5G}).  
The other is very long baseline interferometry (VLBI) at millimeter
wavelengths (reviewed by \citealt{Boccardi2017}).  The Event Horizon Telescope (EHT) aims to image
Sgr~A* with a resolution close to $R_{S}$ at 1.3--3.5~mm wavelengths
(\citealt{Dexter2014,2014arXiv1406.4650T, 2018ApJ...859...60L}).
The detection of circular trajectories of the center of light during
flares of Sgr~A* by VLTI/GRAVITY and corresponding loops in the Stokes
$(Q,U)$ plane (\citealt{2018A&A...618L..10G}) suggests compact source
structure of ${<}5R_{S}$ on an orbit around an average position
at ${\sim}9R_{S}$ (\citealt{2020A&A...635A.143G}). At 1.3~mm
wavelength, VLBI studies also found compact structure with upper
limits on the intrinsic source size of ${\la}4R_{S}$
(\citealt{2008Natur.455...78D}), likewise with some evidence that the
center of light is not centered on the black hole itself.

In light of the interferometric results, the
variability data obtained over the last two decades are valuable
as a complementary source of information about the physical processes
at event-horizon scales. Many multi-wavelength campaigns in the submm,
NIR, and the X-rays have been organized in the hope of
determining---or at least constraining---the underlying radiative processes
(\citealt{2001Natur.413...45B,2004A&A...427....1E,2006A&A...450..535E,2006ApJ...640L.163G,2006ApJ...650..189Y,2006ApJ...644..198Y,2008A&A...492..337E,2008A&A...479..625E,2008ApJ...682..373M,2008ApJ...682..361Y,2009ApJ...698..676D,2009ApJ...706..348Y,2011AA...528A.140T,2012A&A...537A..52E,2012AJ....144....1Y,2012A&A...540A..41H,2016A&A...589A.116M,2016A&A...587A..37R,2017MNRAS.468.2447P}).

Several models have been proposed to explain the variability at
different wavelengths. All models
assume that the NIR is dominated by optically
thin synchrotron radiation, but they differ in the mechanism
for the X-ray emission. Some make the case for optically thin
synchrotron radiation with a cooling break to explain the X-rays
(\citealt{2009ApJ...698..676D,2017MNRAS.468.2447P}), while others suggest
synchrotron self-Compton scattering (synchrotron--SSC)
(\citealt{2009ApJ...698..676D,2012A&A...537A..52E,2016A&A...589A.116M})
or inverse Compton scattering of photons by a second population of
electrons (\citealt{2009ApJ...698..676D}). 

While some authors have claimed evidence for a close relation between submm and NIR variability, the phenomenology and degree of correlation remain inconclusive. \cite{Dexter2014}, for example, found a submm variability timescale of $\sim$8~hours, significantly longer than in the NIR, and concluded that different mechanisms might be creating the variability in the two wavelength regimes.  In contrast, \citealt{2012A&A...537A..52E} linked the submm and NIR through the evolution of the optical depth caused by adiabatic expansion.
\added{Evidence for variability peaks propagating from submm to radio frequencies has been reported (e.g., \citealt{2006ApJ...650..189Y, 2008ApJ...682..373M}) and convincingly modeled in the framework of adiabatic expansion (\citealt{2009ApJ...706..348Y}, their Fig.~27).}

\cite{2006A&A...450..535E}, \cite{2011AA...532A..26B}, and
\cite{2018ApJ...863...15W} found evidence for an (exponential) cooling
cutoff of the SED at NIR frequencies.  If the cooling-cutoff energy varies with
source luminosity, that could explain spectral
index changes as a function of flux
density.  
\cite{2010ApJ...725..450D} developed the first
time-dependent models for Sgr~A* variability, integrating the
differential equations for the electron energy distribution under
injection and escape, resulting in sequences of SEDs.

While past studies are very informative about many basic properties
of the covariability of the flux densities at different wavelengths,
multi-wavelength campaigns are difficult to organize and have to rely
on favorable weather and operational conditions at all observing
sites and on the luck of \Sg\ varying above the detection
limit during the campaign. As a result, only a small number of simultaneous
multi-wavelength observations are available today. Many of these have
been obtained during  24- and 16-hour observations with
\Sp/IRAC combined with some of Keck, \Ch, SMA, and ALMA
(\citealt{2018ApJ...864...58F,2018ApJ...863...15W,2019ApJ...871..161B}). Open
questions  still include the exact nature of the NIR--submm and
NIR--X-ray correlations and the power spectral density (PSD) of the
X-ray variability.

Despite the critical role played by the variable component, it does
not account for all of \Sg's emission. The radio emission at
wavelengths $\ga$3~mm \replaced{has little or no}{shows much lower fractional} 
variability
\citep{2010RvMP...82.3121G}, and there is a non-varying, spatially
extended \added{($\sim 1''$)} X-ray component
\citep{2003ApJ...591..891B,2006ApJ...640..319X}.  The angular size of
the constant radio component is unknown because of interstellar
scattering.  This paper addresses only the \added{(intra-day)} variable component of
emission.  The other components must come from a separate process or
processes, and the spatial extent shows that the constant X-ray
component at least must originate in a different volume of the source
than the compact volume giving rise to the intra-day variable emission.

The goal of the present work is to put the statistical properties of
\Sg's rapid variability at all wavelengths into the context of a
single radiative model by using available long-duration observations to best
advantage and without requiring  simultaneity of the individual
observations. This approach has already been developed by
\cite{2018ApJ...863...15W} in  modeling the covariability of
2.2 and 4.5~\micron. Given the clear
evidence of  structure with size ${<}5R_{S}$, we consider a one-zone
model.  The model combines a 
self-absorbed synchrotron spectrum with cooling cutoff and
synchrotron--SSC
scattering. This model choice is partly motivated by the
timing properties of the correlated NIR and X-ray light curves, which
cannot easily be
reproduced by the other models mentioned above. Additionally,
\citet{2012A&A...537A..52E}, \cite{2016A&A...589A.116M}, and
\cite{2020ApJ...898..138S} showed convincingly that peak fluxes of 
simultaneously observed NIR and X-ray flares as well
as the flux density distributions are well described by a
synchrotron--SSC model. \cite{2012A&A...537A..52E} pointed out
that bright, compact synchrotron sources show the precise
conditions to exhibit self-absorption and self-scattering. In
their model, the observed pairs of flux density peaks
suggest source sizes of a few $R_{S}$ and magnetic flux
densities of a several tens of gauss.
A source of this size and magnetic flux density shows a flux density
spectrum that peaks in the submm (naturally contributing to
variability at submm wavelengths as well), and X-ray photons
are mainly up-scattered submm photons. In contrast to
earlier studies, our analysis  takes into
account the auto-correlation and cross-correlation properties of the
light curves and the effects of the cooling cutoff in the NIR by fitting
a time-dependent, analytic model to the body of submm, NIR,
and X-ray data.

This paper has two parts. The first  analyzes---for the first
time---the PSD of the X-ray variability.
The PSD model is based on a generic statistical (i.e., not
physically motivated) flux density model. The paper's second part,
motivated by the PSD analysis, proposes a single-zone radiative model and
analyzes the variability at submm, NIR, and X-ray wavelengths.
Both
parts discuss prior distributions, a model, and the resulting
posteriors of the model parameters.
Section~\ref{data} describes the datasets used in this
analysis.  Section~\ref{XPSD} shows evidence that at the shortest
variability timescales, power
in the X-ray PSD  appears
suppressed  compared to the NIR PSD. Further,
the correlation between the NIR and the X-rays (characterized by
strict correlation in arrival time and lack of correlation in
flux density levels) can be understood as the effect of the
difference between the two PSDs. Section~\ref{radmodel} shows that the
separation of timescales is a natural result of the equation for SSC
flux densities \citep{1983ApJ...264..296M} if the source of the
fast NIR variability is the fast-varying cooling cutoff close to
NIR frequencies. We additionally discuss modes of co-development of the
synchrotron self-absorption with NIR and X-ray flux densities and fit
a simple, semi-analytic version of the model to the observed
first-order structure functions in all bands. Section~\ref{res}
presents the model-fit results
including  animations. Section~\ref{discussion} discusses the merits
and shortcomings of the 
model, and Section~\ref{summary} summarizes our findings.

As a convention, $H$-, $K$-, $L$-, and $M$-band refer to the
respective NIR bands. Radio bands will be denoted by their central
wavelengths. We use the term spectral energy distribution (SED) even
when the quantity is expressed as flux density ($S_\nu$) rather than
energy ($\nu S_\nu$).

%
%


\begin{table*}
\begin{center}
\caption{New SMA and ALMA epochs} \label{submmobs}
\begin{tabular}{ccccccll}
\hline
\hline
Date & Start Time & Stop Time & Baselines\tablenotemark{a} & \# Ant. & Tuning &
BW\tablenotemark{b}  & Calibrators \\
UT & UT & UT & k$\lambda$ & & GHz & GHz\0\\
\hline
\multicolumn{1}{c}{ SMA}\\
2014 June 18 & 07:23:48 & 13:28:47 & 6.2--84.8 & 8 &  343.0 & 8 (2 GHz/sb x 2 sb x 2 pol)\0\0  &
Neptune,  NRAO~530 \\

2015 May 14 & 09:52:41  & 15:52:24   & 6.1--133.2 & 6 & 226.9  & 8 (4 GHz/sb x 2 sb x 1 pol)\0\0   & Titan,
Callisto, \\ 
&&&&&&& NRAO~530 \\
2016 July 13 & 05:34:48  & 12:04:48  & 46.7--412 & 6 & 236.1  & 16 (4 GHz/sb x 2 sb x 2 pol)\0\0   &
NRAO~530, J1924$-$292\tablenotemark{c}\\ 
2017 July 16 & 04:46:05  & 12:27:14  & 4.9--54.4 & 8 & 228.0  & 32 (8 GHz/sb x 2 sb x 2 pol)\0\0   & NRAO~530,
J1924$-$292\\ 
2017 July 26 & 05:19:43  & 11:22:07  & 6.3--53.8 & 7 & 228.0  & 32 (8 GHz/sb x 2 sb x 2 pol)\0\0   & NRAO~530,
J1924$-$292\\ 
\hline
\multicolumn{1}{c}{ ALMA}\\
2016 July 12 & 22:58:12  & 03:38:14  & 9.4--674.4 & 40\0 & 232  & 7.45\0   &
Titan, PKS 1741$-$312\\ 
2016 July 18 & 23:14:51  & 24:10:34  & 9.4--674.4 & 40\0 & 232  & 7.45\0   &
Titan, PKS 1741$-$312 \\
2016 July 19 & 02:05:37  & 05:43:36  & 9.4--674.4 & 40\0 & 232  & 7.45\0   &
Titan, PKS 1741$-$312\\ 
\hline
\end{tabular}
\end{center}
\tablenotetext{a}{Baselines $<$35-40~k$\lambda$ (\added{exact limit different for different SMA integrations}) were not used for the
  final flux densities in order to avoid contamination by extended
  structure.} 
\tablenotetext{b}{Effective spectral bandwidth
  including all polarizations, not necessarily continuous.}
\tablenotetext{c}{Flux density calibration was based on secondary
  calibrations of the gain calibrators listed.}
\end{table*}

\section{The Data}\label{data}
Our model is based on a coanalysis of a comprehensive
multi-wavelength variability dataset of Sgr~A*. With
the exception of two of the ALMA and four of the SMA observations,
all data were published before.
While comprehensive, this dataset is not complete, but it
uses the large NIR and X-ray samples analyzed in previous
publications because they are statistically well characterized already.

\begin{figure*}[h!]
\begin{center}
\includegraphics[width=\textwidth, angle=0]{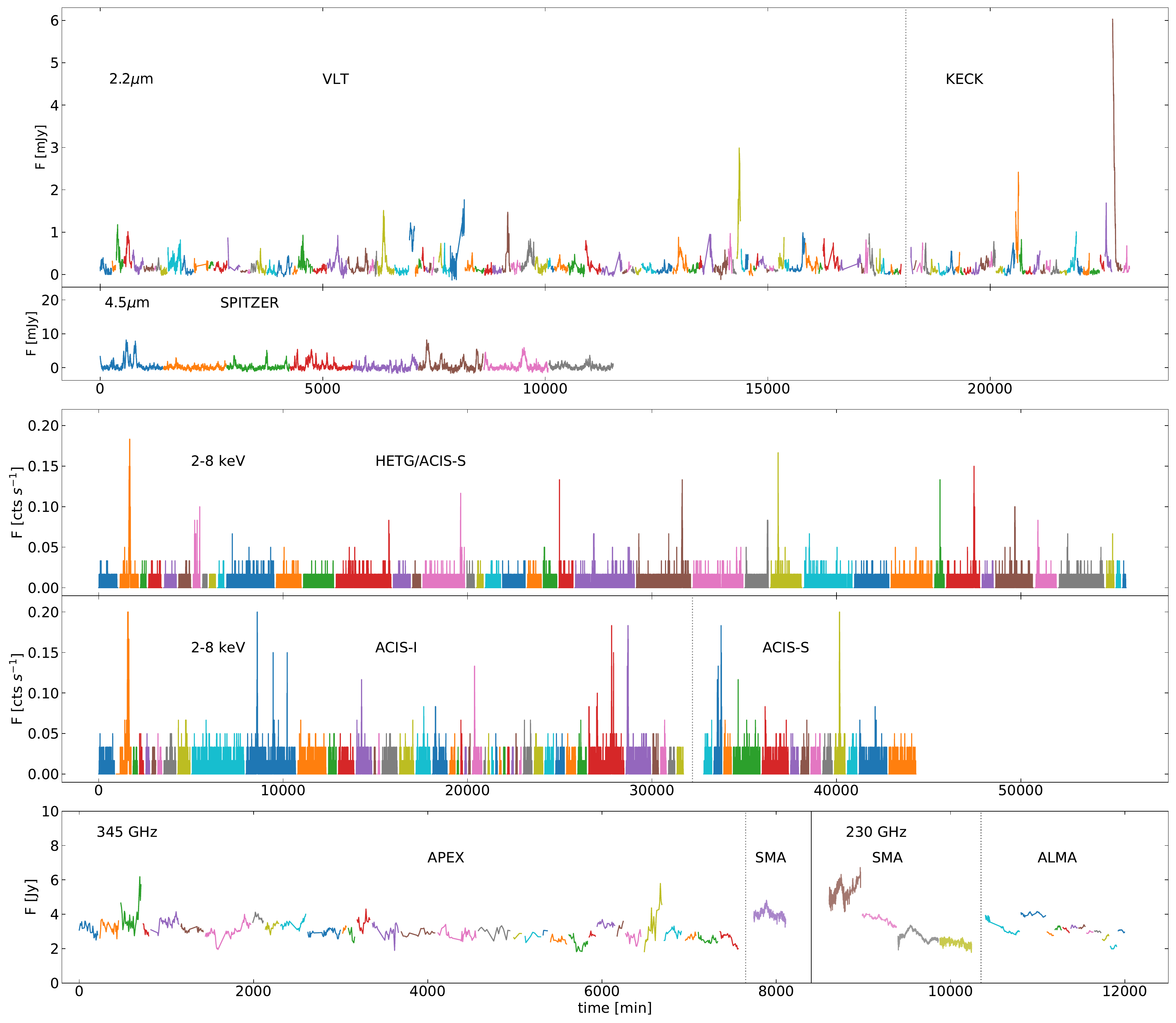}


\end{center}
\setlength{\abovecaptionskip}{-12pt}
\caption{Observed light curves for \Sg.  The data are presented
  without observational gaps between the epochs, but each epoch is in
  a different color. \added{(colors do not indicate common epochs across different panels)}.   Gap durations are hours to years, but they
  should not affect the present analysis.  Time scales differ among
  panels.  To the accuracy of the calibrations, all flux densities
  are for a point source, excluding extended structures, and all data
  are presented as observed with no correction for interstellar
  extinction.  Top panel shows 2.2~\micron\ with the vertical dashed
  line separating VLT data on the left from Keck data on the
  right. Second panel shows 4.5~\micron\ \Sp/IRAC data in one-minute
  bins. Third panel shows \Ch/ACIS-S/HETG data, and fourth panel shows
  \Ch/ACIS-I data on the left and \Ch/ACIS-S data on the right.
  X-ray data are shown with one-minute binning and no pileup
  correction. The different \Ch\ instruments cover nearly the same
  energy range but with different sensitivities, which are accounted
  for in the analysis.  Fifth panel shows 345~GHz data from
  APEX/LABOCA and SMA on the left and 230~GHz data from SMA and ALMA on the right. }
\label{nir_data}
\label{X_data}
\label{submm_data}
\end{figure*}

\subsection{Near Infrared}

The NIR dataset used in this analysis is a combination of the
extensive 2.12, 2.18, and 4.5~\micron\ data of
\cite{2018ApJ...863...15W} and the 2.12~\micron\ dataset
of \cite{2019ApJ...882L..27D}. The original publications explain the
reduction and  statistics. (For the historic $K$-band dataset see also
\citealt{2012ApJS..203...18W} and \citealt{2014ApJ...791...24M}, and
for the fundamentals of observing Sgr~A* with \Sp/IRAC see
\citealt{2014ApJ...793..120H}.) Our dataset contains eight 24-hour
epochs of \Sg\ at 4.5~\micron\ with IRAC on the \SST, 93 epochs of
2.18~\micron\ data from Naos Conica at the Very Large Telescope, 34
epochs of 2.12~\micron\ data from the NIRC2 camera at the Keck
Observatory, in total 95\,307 NIR measurements. Figure~\ref{nir_data}
shows the light curves of these three observatories. The average cadence of the \Sp\ data is
8.4~s, and typical cadences of the VLT/Keck light curves are about one
image per 1.1--1.2 minutes with integration times of 30--40~s and
28~s, respectively. The typical uncertainties of the individual data
points are 0.66~mJy for \Sp, 0.033~mJy for the VLT, and about
0.017~mJy for the Keck data. The light curves of all three
instruments are prone to contamination with an additive background
flux density level from the resolved and unresolved stellar
population at the Galactic center. The \Sp\ data are---by nature of the applied
data reduction algorithm---a differential measurement with an
arbitrary zero point. The VLT and Keck data show a typical
photometric offset of about 0.06~mJy and 0.03~mJy, respectively. These
offsets play an important role in  spectral index
measurements but do not affect the first-order structure
function, which quantifies differences in
flux densities rather than absolute levels.

\begin{figure*}
\begin{center}
\includegraphics[scale=0.5, angle=0]{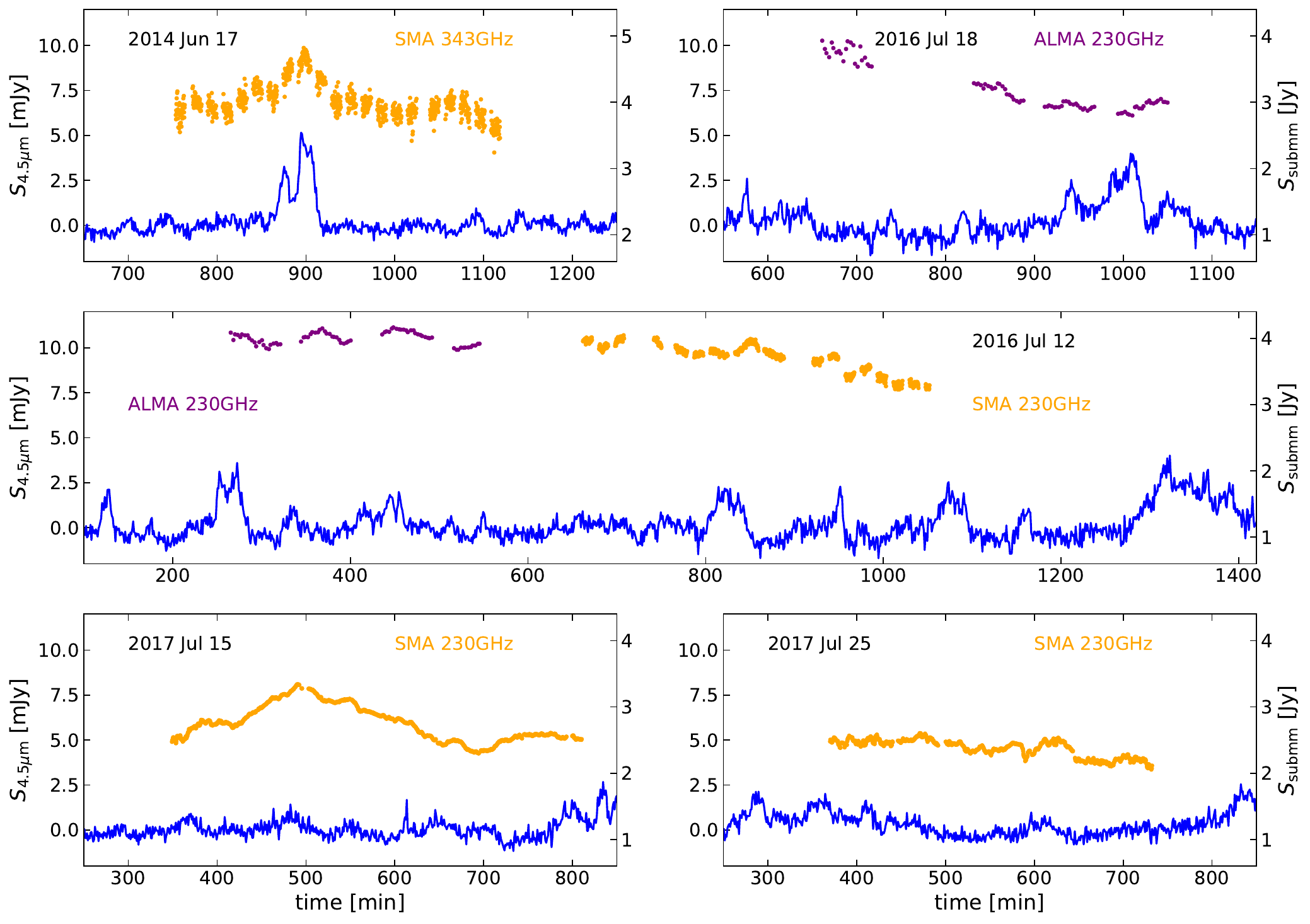}
\end{center}
\setlength{\abovecaptionskip}{-5pt}
\caption{Simultaneous submm and $4.5\micron$ datasets obtained during
  the \Sp\ campaigns. Blue lines show the 4.5~\micron\ \Sp\
  light curves with one minute binning.  Flux densities are as
  observed, not corrected for reddening.  Purple shows 230~GHz ALMA
  light curves, and orange shows 
  230~GHz and 345~GHz SMA light curves as labeled. Times are arbitrary
  but matched to 
  heliocentric for all curves. The 345~GHz SMA light curve of 2014
  Jun~17 is the same one
  published by \cite{2018ApJ...864...58F}. Other submm
  data are presented here for the first time. For this figure as well as for the correlation analysis, the 2017 SMA data has been slightly smoothed with a 3-minute smoothing kernel.}\label{synchdata}
\end{figure*}

\subsection{\Ch\ Data}

Sgr A* has been  observed often by \Ch\ starting in 1999.
We include all data available from the \Ch\ archive
through 2017 having aimpoint  within 1~arcmin
of \Sg. This ensured an optimal point-spread function and best
photometric performance for a compact source like Sgr~A*. The \Ch\
instruments and observing modes have changed over time:
\begin{itemize}
\item In 1999--2011, all data were taken with ACIS-I, and there were
  two additional ACIS-I observations taken in 2013. These 49
  ACIS-I exposures sum  to ~1.5~Ms.  Pileup  affects these observations.
\item In 2012, in the framework of the X-ray Visionary Program (XVP),
  the data were taken with the ACIS-S/HETG. There are three
  additional observations taken with ACIS-S/HETG in 2013. These
  41 ACIS-S/HETG exposures sum to ${\sim}$3~Ms. Only the
  zeroth-order image is used here.  Pileup is small but not always zero.
\item From 2013 to 2017, almost all observations were done with ACIS-S
  (no grating) in sub-array mode, making them basically unaffected by
  pileup. The 39 ACIS-S exposures sum to 
  $\sim$1.4~Ms. However, during 2013 and 2014 (the first 25  epochs), the
 magnetar PSR~J1745$–$2900 \citep{Coti2017} contributed significantly to the flux.
  Because this varying source contributed additional
photon noise, we ignored all ACIS-S epochs prior to 2015-10-21.  This
leaves about a two-year gap between the last ACIS-I
observation (2013) and the first ACIS-S observation (2015-10), when
the magnetar became  dim enough  not to 
affect our photometry.  The 14 good epochs give $\sim$0.6~Ms of data.
\end{itemize}
All told, we have 103 epochs and 5.3~Ms of 2--8~keV \Ch\ data. 

We reduced the  \Ch\ data ourselves to
guarantee consistency across all epochs.
All archival data were downloaded and reprocessed using the \Ch\
Interactive Analysis of Observations software package (CIAO v4.10)
and the \Ch\ Calibration Database (CALDB v4.7.8) following the
standard procedures as outlined by \cite{2019ApJ...875...44Z}, who
used the same \Ch\ dataset to study a candidate parsec-scale jet
from \Sg. Photons were extracted from within a radius of 1\farcs25
from the best-guessed centroid of \Sg\ and within the 2000--8000~eV
range to be consistent with  most previous work.  This gave a
total of 34\,000 photons in the useful epochs. We used
unbinned data of the arrival times of individual photons
(counts) and corrected the times to the barycenter of
the solar system.

Figure~\ref{X_data} shows the light curves of all three modes. 
The effective area of each mode is a weighted mean over
the 2--8~keV band, and it assumes an incident source
spectrum. If we changed the assumed spectral model, the
absolute values of the effective area would change, but the relative
values among observations using the same detector should be
insensitive to the model. The relative values between different
detectors (I versus S) could be  more sensitive but are
assumed to be constant as well.

\subsection{APEX Data}

We used 32 epochs (6641 minutes) of 345~GHz data from
the LABOCA bolometer at the APEX telescope
\citep{2017A&A...601A..80S}. This dataset was generated using
on-the-fly mapping, resulting in fully sampled maps of
0\fdg5$\times$0\fdg17 with 280~s integration time. The data
were taken over the course of seven years and have a typical cadence
of about 8 minutes. An average map was created for each epoch
and, after subtracting a Gaussian point source at the position of 
\Sg, subtracted from each individual image. The flux density of Sgr~A* was
derived by modeling a Gaussian source at the position of Sgr~A* in
each residual image and using two secondary calibrators (G10.62$-$0.38, 
IRAS~16293$-$2422). The relative uncertainty of the flux density calibration is
about $4\%$ or 0.1~Jy, while the absolute uncertainty is expected to
be of order $15\%$.
\cite{2008A&A...492..337E}, \cite{2011ApJ...738..158G}, and
\cite{2017A&A...601A..80S} gave detailed descriptions of the data
reduction, calibration, and atmospheric opacity monitoring.  The lower panel of figure~\ref{submm_data} shows the resulting 345 GHz light curves.

\subsection{SMA Data}

We include one published (365~minutes---\citealt{2018ApJ...864...58F})
epoch of 343~GHz data and four new epochs (1573~minutes) of 230~GHz data
from the Submillimeter Array (SMA). Observation details are in
Table~\ref{submmobs}. The SMA was operated in a dual-receiver
polarization track with double-sideband observations using sideband
separation implemented in the correlator. The continuum visibility
was calculated by averaging the two same-sense polarization
signals. Final flux-density measurements were determined by
vector-averaging the measured visibility data for instantaneous
baselines greater than 35--40~k$\lambda$ to filter out
large-scale emission structure around Sgr~A*.  Water vapor was
$\sim$1.5~mm for most epochs but nearer 0.9~mm in 2016 and higher and
unstable on 2016 Jul~19.
The lower panel of Figure~\ref{submm_data} shows the resulting
343~GHz and 230~GHz SMA light curves.

\subsection{ALMA Data}

This analysis used 12 epochs (1374~minutes) of data from the ALMA
observatory.  The first two epochs (project code 2015.A.00021.S,
PI G.~Witzel) were taken in 2016
while \Sp\ was observing at 4.5~\micron.
Details are in Table~\ref{submmobs}. 
The data were calibrated with the standard
ALMA pipeline using the Common Astronomy Software Application package
\citep[CASA:][]{McMullin2007}.  A few
spectral windows showed suspicious absorption features and were
excluded from the analysis.
Because ALMA has relatively fewer short baselines than SMA, all baselines were included, and the source was imaged with uniform weighting using task {\sc  clean}.
\Sg\ is a strong source at mm wavelengths, and its visibility was
mostly flat as a function of baseline length. That means the source
is compact, and with 
ALMA's good U--V coverage and uniform weighting, light curves
could be extracted by simply measuring
the peak flux densities of the image. The off-source RMS
gave the uncertainty. The robustness of the results
was investigated using 1--30-minute averaging times. The
final choice of 3~minutes is a good balance between
signal-to-noise and sampling  the light curve. We obtained two
long light curves, the first 389~minutes with a gap and the second
280~minutes as shown in Table~\ref{submmobs}.

The remaining 10 ALMA epochs are 70-minute light curves at 234~GHz
\citep{2020ApJ...892L..30I}. The epochs were observed in
2017 October over ten days, i.e., they are separated by roughly one
day and  have  average cadence of
${\sim}1.6$~minutes. \cite{2020ApJ...892L..30I} gave details
of the data reduction.
Figure~\ref{submm_data} shows all 12 epochs of
230~GHz ALMA data.

\subsection{Simultaneous NIR and Submm Observations}

A majority of the submm data were taken during the
\Sp\ observations, resulting in the largest set of synchronous
submm (230 or 345~GHz) and NIR (4.5~\micron) light curves.
(\Ch\ observed simultaneously as well, but there was no X-ray event during times of SMA or ALMA coverage.)
In total, we obtained  $>$1800~minutes on-source within a total
duration of 2247~minutes of simultaneous data. The combined ALMA
and SMA data on 2016 Jun~12 cover 787~minutes with a gap of 116~minutes
between the datasets, making this the longest simultaneous light curve so
far. Figure~\ref{synchdata} shows all simultaneous data taken during the \Sp\
campaigns.

\begin{figure}
\begin{center}
\includegraphics[scale=0.38,  angle=0]{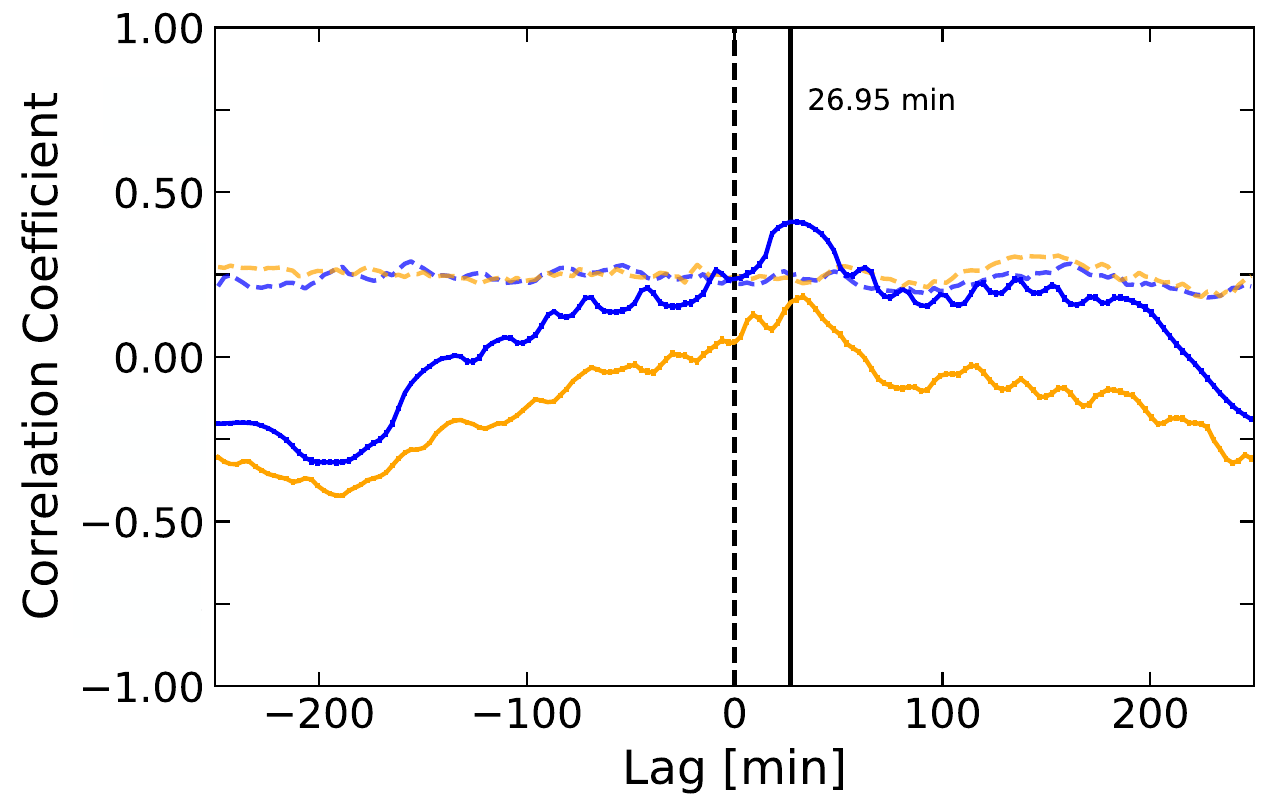}
\end{center}
\setlength{\abovecaptionskip}{-5pt}
\caption{Discrete cross-correlation (DCC) functions of the 230~GHz
  and 4.5~\micron\ light curves shown in Figure~\ref{synchdata}. 
  Orange shows the DCC of the unaltered data, and blue with
    shaded error bars shows the DCC
  of 230~GHz with the logarithm of the low-pass-filtered 4.5~\micron\
  data. The vertical dashed line indicates a time
  lag of zero. The DCC reaches its 
  maximum at a time lag of about 33~minutes for the unaltered data and
  at $27\pm8$~minutes (solid vertical line) with the transformed data.  
  Positive lags indicate the 4.5~\micron\ data lead the submm. 
  \replaced{The quoted uncertainty is that of the peak position of the
  part of the DCC that lies above the 95\% false alarm probability
  (dashed horizontal curves), which was derived from our Bayesian posterior
  (Section~\ref{results}).}{The quoted uncertainty of the time lag is 2.345 times the full width at half maximum (FWHM) of the part of the transformed DCC that lies above the 95\% false-alarm probability (dashed horizontal curve). The false alarm probability was derived from uncorrelated pairs of light curves generated from our Bayesian posterior
  (Section~\ref{res}).}
  }\label{cross_corr}
\end{figure}

Figure~\ref{cross_corr} shows the
discrete cross-correlation (DCC) of the entire 230~GHz dataset with
the 4.5~\micron\ light curves. Because the
data were not regularly sampled and show large gaps, we used the
algorithm of \cite{1988ApJ...333..646E} as implemented by
\cite{2015MNRAS.453.3455R}. We calculated the DCC for both the data
as presented in Figure~\ref{synchdata} and for a transformed version
of the 4.5~\micron\ data. The transform was to convolve the data with a
Hanning smoothing kernel with size of 61~minutes and then take
the logarithm.\footnote{\added{Smoothing the NIR data suppresses fast NIR variability,  which is uncorrelated with the submm and the X-rays. The logarithm accounts for the non-linear relation between the submm and the NIR,  a consequence of the radiative mechanism discussed in Section~\ref{discussion}.}} \deleted{ (Section~\ref{discussion}  explains the reasons for
doing this.)} Both DCCs have their
maxima at time lags  of 20--30~minutes, positive lag meaning
that the 4.5~\micron\ data  led the submm.

\section{The X-ray Power Spectrum and the NIR--X-ray Correlation}\label{XPSD}

The first step of our analysis was to determine the PSD
of the X-ray variability of Sgr~A* under the
preliminary assumption of a log-normal PDF for the X-ray
flux densities. 
We did this by forward-modeling
light curves for a range of PSDs, comparing the model light curves
with observed ones through a suitable summary statistic (a ``distance
function''), and exploring the parameter space by approximate
Bayesian computation (ABC). ABC is appropriate for problems
with no analytic likelihood function. All three
parts---the forward modeling, distance function, and implementation of
the ABC---were described in detail by
\citet[][their Appendix~B]{2018ApJ...863...15W}. 
The present analysis closely follows their procedure 
but was adapted for X-rays by
deriving photon statistics from the modeled variable flux densities
plus a constant source contribution.  Details are given below.

A log-normal PSD was chosen because a) it can---as the
results show---describe the observed data successfully, and b) a
log-normal was the preferred model for NIR light
curves  \citep{2018ApJ...863...15W}.  That analysis showed that the
NIR PSD is a single broken power law with a slope of 2 (red
noise) and a precisely measured break timescale $\tau_b =
243^{+82}_{-57}$~minutes. In all such analyses, the inferred PSD parameters  depend on the
model for the flux-density PDF, and using a log-normal PSD lets us
directly compare 
X-ray PSD parameters with the NIR. The  analysis in
Section~\ref{radmodel} replaces the PSD 
model assumption with a physical model.


\subsection{Forward Modeling of Light Curves}
We used the FFT-based method  \citep{1995A&A...300..707T} to generate
random light curves from a given PSD and a set of independently drawn
random numbers. This method results in realizations of a Gaussian
process $g(t)$  exhibiting periodograms (i.e., PSD estimators)
consistent with the input PSD. We parameterized the PSD as a
broken power law of the form:
\begin{equation}\label{psddef}
 \mathrm{PSD}(f) \propto \left\{
 \begin{array}{lcl}
  f^{-\gamma_0} & \mathrm{for} & f < f_{b}\\
  f^{-\gamma} & \mathrm{for} & f \geq f_{b}\quad,
 \end{array}\right.
\end{equation}
where $f$ is temporal frequency, and we assume $\gamma_0=0$ (\citealt{2009ApJ...694L..87M}).
In order to generate realistic light curves, we transformed the values
$g(t)$ to make the resulting distribution of $S(t) = T[g(t)]$ 
consistent with the distribution of observed flux densities. 
With a log-normal distribution as our target,
\begin{equation}\label{transdef}
T(g(t)) = \exp{(\sigma_{\rm{logn}} \cdot g(t) + \mu_{\rm{logn}})} \;\; ,
\end{equation}
$\mu_{\rm{logn}} $ and $\sigma_{\rm{logn}} $ being the log-normal parameters.

For \Ch\ data, we modeled the light curves in count rate
$\Lambda(t) = T[g(t)]$ (counts per second, cps) instead of
flux density $S(t)$. The conversion factor between flux
density (as observed at Earth, i.e., after suffering interstellar extinction)
and count rate is absorbed by the log-normal mean $\mu$ and
otherwise does not affect log-normality. We assumed an effective area
$e_{I} = 1$ for the ACIS-S/HETG data and Gaussian priors for
$e_{G}/e_{I}$
and $e_{S}/e_{I}$, the relative effective areas of the other two modes.\footnote{Our priors on $e_{G}/e_{I}$ and
  $e_{S}/e_{I}$ are based on the spectral model used by
  \cite{2018ApJS..235...26Z}, i.e., an absorbed bremsstrahlung with a
  plasma temperature of 10~keV and a foreground absorption column
  density of $10^{23}$~{cm}$^{-2}$.} For each mode, we included an
additive term ($\chi_{G}$,$\chi_{S}$,$\chi_{\rm{I}}$) to
account for the contribution from the extended, non-varying X-ray
source at the position of Sgr~A*. The resulting
total count rate (before measurement errors) from the 
position of Sgr~A* is then:
\begin{equation}\label{cpseq}
\Lambda_{\rm{tot}}(t) = e_{i}/e_{I} \cdot \left[\Lambda(t) + \chi_i\right] \; \; ,
\end{equation}
where $i$ is $S$, $I$, or $G$ for the three \Ch\ instruments used,
and $\chi_i$ is the count rate of the constant source as seen by the
respective instrument.
Pileup was included in the forward modeling for  ACIS-I and
ACIS-S/HETG data for all count rates $>$0.02~cps.  (It is negligible
below that.)
We used Equations~3 and~4 of \cite{Yuan2015}:
\begin{align}
\Lambda_{\rm{out}, I}(t) = &\Big[ 4.180 \Lambda_{\rm{tot}}(t)^{-0.07387} \nonumber \\
 & + 0.5381 \Lambda_{\rm{tot}}(t)^{-1.160} \Big]^{-1}  \;\; ,
\end{align}
and
\begin{align}
\Lambda_{\rm{out}, G}(t) = &\Big[ 3.933 \Lambda_{\rm{tot}}(t)^{-0.03541} \nonumber \\
 & + 0.6564 \Lambda_{\rm{tot}}(t)^{-1.107} \Big]^{-1}  \;\; .
\end{align}
For ACIS-I count rates between 0.1 and 0.2~cps, the corrections
are 27 to 64\%.
We finally modeled the measured count rate $\Lambda_{\rm{meas}}$,
sampled once per minute, by a Poisson process:
\begin{equation}\label{cpsmeaseq}
\Lambda_{\rm{meas}}(t) = \rm{Pois}\left[\Lambda_{\rm{out}}(t) \cdot 60  \right] / 60  \;\;.
\end{equation}
With this 1-minute binning, the effect of the \Ch\ frame
time (3.2~s for ACIS-I and HETG, 0.4~s for ACIS-S subarray) is negligible.

\subsection{The Distance Function}
\begin{figure}[t]
\begin{center}
\includegraphics[scale=0.55, angle=0]{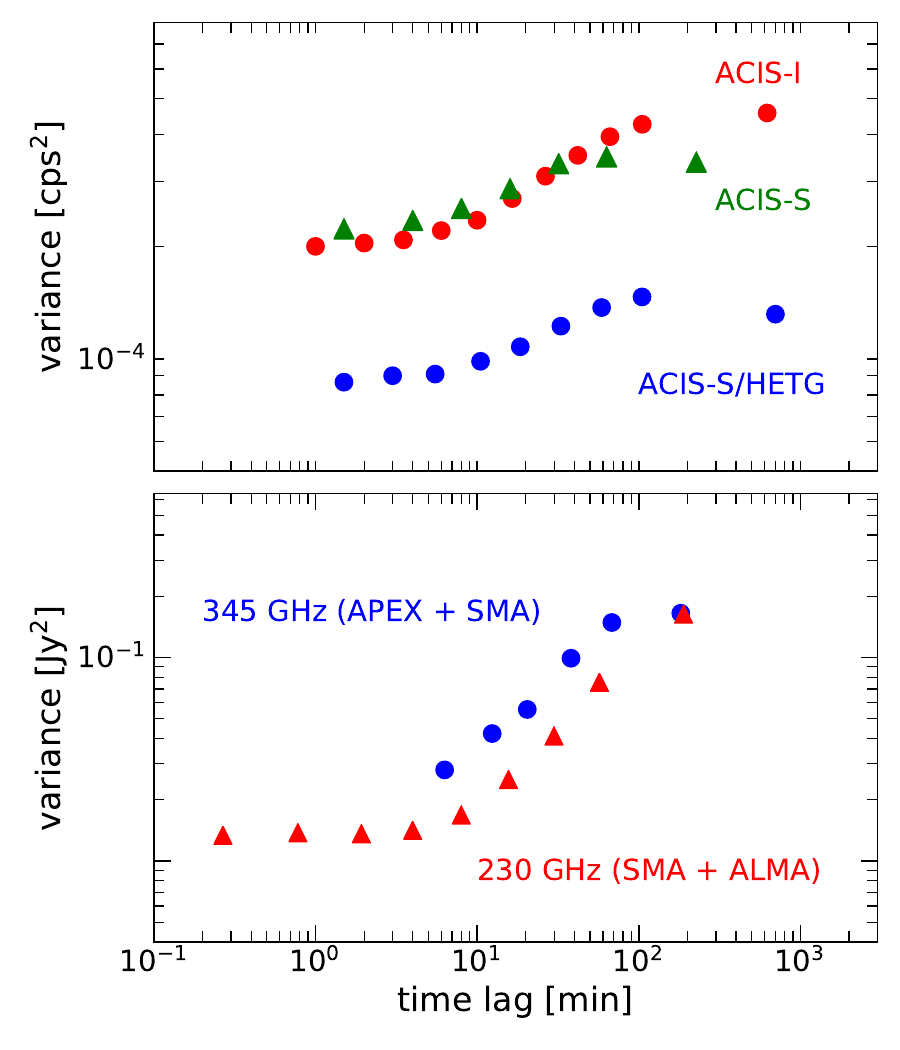}
\end{center}
\setlength{\abovecaptionskip}{-5pt}
\caption{Logarithmically binned structure functions
  (Eq.~\ref{eq:strfundefa}). Upper panel shows the X-ray SFs, red for
  ACIS-I, green for ACIS-S, and blue fpr ACIS-S/HETG. Lower panel
  shows the 230~GHz  and 345~GHz SFs, red the combined SF from SMA and ALMA data, and blue the APEX
  data combined with one epoch of SMA data. }\label{Xsf}
\end{figure}

Following \citet[][their Appendix B.2]{2018ApJ...863...15W}, we used the
first order structure function as the distance function. The
structure function quantifies the variance of the flux density at any
given time separation and contains information on the
PSD as well as the flux-density distribution of the variability
process. The structure function is defined as:
\begin{align}
\label{eq:strfundefa}
V(\tau_i) &= 
  \frac{1}{n_i}\sum_{t_j, t_k}\left[F(t_j) - F(t_k)\right]^{2}\\ 
  &\quad\mathrm{for}\; \tau_i \leq  (t_j - t_k) < \tau_{i+1}\quad,\nonumber
\end{align}
that is, the sum of $\left[F(t_j) - F(t_k)\right]$ over all
measurement pairs whose time lags $(t_j - t_k)$ fall within the bin
$[\tau_i, \tau_{i+1}]$, there being $n_i$ such pairs.  The structure functions for the three modes
of \Ch\ data are shown in Figure~\ref{Xsf}.

\added{A detailed discussion of the choice of the $\tau_i$ can be found in Section~3 and Appendix~B.2 of \cite{2018ApJ...863...15W}. In short, with increasing time lag, a decreasing
number of point pairs contribute to the structure function bins.  For time lags longer than half the observing
window, not all flux-density measurements contribute to every structure function bin,
and the variance of the structure function increases dramatically without carrying much information about the intrinsic variability. Therefore we chose
a logarithmic binning scheme, roughly equally spaced in
logarithmic time lags, with a spacing large enough to
allow for a similar number of points in the long-time-lag
bins. The maximum lag bin was defined by the point where the variance of the linearly binned structure function starts to increase significantly. This bin is larger by a factore  3 (for the X-ray data) than the other bins in order to mitigate the increase of the variance.}

We defined the distance between two light curves as the weighted L2
norm of the difference between the logarithms of the respective
structure function's binned values:
\begin{equation}\label{distdef}
\phi(V_1, V_2) = \sum_{i} w_{i} 
  (\log\left[V_1(\tau_i)/V_2(\tau_i) \right])^{2}\quad
\end{equation}
with $w_{i}$ the weights for the chosen binning. The
weights adopted here were unity for each structure-function bin
except  the single wide bin at large time lags, which had 
$w_{i}=3$.
These values gave uniform and reasonably quick convergence of the fit.

There is a general problem of consistency
when comparing variability timescales
quoted in the literature. The definition
of ``timescale'' can depend on the method used for its
estimation, and ignoring that can lead to 
errors in the interpretation of  time series analysis results, especially
in their conversion into physical quantities. The most substantial
difference in the definition of timescales concerns the domain in
which they were estimated. In the spectral domain, a timescale
naturally refers to the period of a sinusoidal signal. In the time
domain, in particular when the estimation is through a structure
function, ``timescale'' refers to a characteristic  interval of coherent
variation. (Roughly speaking, this is the average time of monotonic
increase or decrease of the analyzed quantity.) The 
\replaced{Ornstein–Uhlenbeck-process}{Ornstein-Uhlenbeck process} (OU-process) timescale
given by \cite{Dexter2014} is in the time domain and is
consistent with the definition provided by the structure function. For a
sinusoidal function of frequency $f$, the time-domain timescale
equals $0.5/f$. The equation $\tau_{\mathrm{time \;domain}} = 0.5 /
f_{\mathrm{spectral\; domain}}$ is in fact valid for any function because,
through a Fourier analysis, we can
describe any function as a superposition of
sinusoids.
As long as the consistency issue is taken into account,
different definitions of timescales are equivalent. 
The relevant question is which timescale is more informative about
physical properties. The ``memory'' or coherence time of a physical
process can be better described by a structure-function
timescale rather than by a spectral-domain one. Structure-function
timescales are also the appropriate choice for the calculation of
source sizes through causality arguments. Periodicities, on the other
hand, are naturally quantified in the spectral domain.
This paper uses the structure function to define a distance between mock data and observed data. However, our timescale parameter $f_b$ is defined in Equation~\ref{psddef} and refers to the spectral domain.

Caution should also be exercised in comparing
structure-function slopes. While originally defined as in
Equation~\ref{eq:strfundefa}, the structure function is sometimes
expressed as
\begin{align}
\label{eq:strfundefb}
V(\tau_i) &= \sqrt{\frac{1}{n_i}\sum_{t_j, t_k}\left[F(t_j) - F(t_k) \right]^{2}} \\
&\quad\mathrm{for}\quad \tau_i \leq (t_j - t_k) < \tau_{i+1}\quad,\nonumber
\end{align}
(e.g., \citealt{Dexter2014}). Conversion of the
structure function slope into a PSD power-law index
depends on the structure function definition used.

\subsection{Approximate Bayesian Computation}
ABC is an iterative method based on prior distributions of model
parameters, two Monte Carlo sampling steps---one for picking a random
parameter set and one for drawing a light curve realization for this
parameter set---and an acceptance step. The acceptance is based on
the comparison of the distance between simulated and observed data
with a threshold decreasing with each iteration. For
each iteration, the two sampling steps are executed many times until
$n$ (e.g., $n = 500$)  realizations are accepted. The
parameters  (called ``particles'') of the accepted realizations represent the current best
estimate of the posterior. This  estimate is used to inform the
parameter sampling of the next iteration. If the distance function is
informative for the model at hand, the prior distribution is
transformed into a close approximation of the posterior after a
sufficient number of iterations. A detailed description and
explanation of the algorithm and its implementation is given by
\citet[][their Appendix~B.4]{2018ApJ...863...15W}.

\begin{table*}[h!]
\begin{center}
\caption{Priors and Posteriors of Bayesian Analysis} \label{results}
\begin{tabular}{llcl}
\hline
\hline
& & Median of &\\
Parameter & Prior & Posterior & Description\\
\hline
\null\\[-1.5ex]
\multicolumn{4}{l}{\quad Analysis of X-ray PSD, log-normal model}\\

$\gamma$ & flat on [0.2, 6.0] & $> 2.0$ & PSD slope \\
$f_{b}$ [$10^{-3} \cdot \rm{minutes}^{-1}$] & flat on [1.0, 600.0] &
$\04.0^{+1.0}_{-1.0}$ & correlation frequency\\ 

$\mu_{\rm{logn}}$\tablenotemark{a} & flat on [$-$13.5,0.0] & $-11.4^{+1.6}_{-1.3}$ &  log-normal mean\\
$\sigma_{\rm{logn}}$\tablenotemark{a} & flat on [0.0001, 	6.0]  & $\03.1^{+0.5}_{-0.6}$ &
log-normal standard deviation\\ 

$\chi_\mathrm{I}$ [$10^{-3} \cdot$ cps]\tablenotemark{a} & Gaussian ($\mu = 4.87$, $\sigma=0.5$)&
$\04.9\pm{0.5}$ &  contribution of steady X-ray source,
ACIS-S/HETG\\ 
$e_{\rm{G}}/e_{I}$ & Gaussian ($\mu = 0.38$, $\sigma=0.05$)
& $\00.41\pm{0.04}$ &effective area of ACIS-S/HETG rel.\ to
ACIS-I \\ 
\hline

\null\\[-1.5ex]
\multicolumn{4}{l}{\quad Analysis of X-ray, NIR, and submm data,
  synchrotron--SSC model}\\ 
$\gamma_{e}$ & flat on [1.5, 4.0] & $\02.95^{+0.21}_{-0.19}$ &
power-law index of electron energy distribution\\ 
$B_0$ [G]& log flat on [$10^{-2}$, $10^3$] & $\08.5^{+1.8}_{-1.4}$ &
minimum magnetic flux density\\ 
$L_0$ [$R_S$] & log flat on [$10^{-2}$, $10^3$] &
$\02.7\pm{0.3}$ & maximum physical source diameter \\ 
$\gamma_{\rm{slow}}$ & flat on [1.2, 6.5] & $\04.9\pm{0.8}$ & PSD
slope of slow Gaussian process\\ 
$\gamma_{\rm{fast}}$ & flat on [1.2, 9.5] & $\02.1^{+0.4}_{-0.3}$ & PSD
slope of fast Gaussian process\\ 
$f_{b,\rm{slow}}$ [$10^{-2} \cdot \rm{minutes}^{-1}$] & flat on [0.001,
2.0] & $\00.74^{+0.13}_{-0.12}$ & correlation frequency of slow
Gaussian process\\ 
$f_{b,\rm{fast}}$ [$10^{-2} \cdot \rm{minutes}^{-1}$] & flat on [0.001,
2.0] & $\01.21^{+0.29}_{-0.27}$ & correlation frequency of fast
Gaussian process\\

$\mu_{\rm slow}$\tablenotemark{b} & flat on [0,10] & $-5.77^{+0.19}_{-0.26}$ &  log-normal
mean of slow process\\ 
$\sigma_{\rm slow}$\tablenotemark{b} & flat on [0.001, 	1.5]  &
$\00.616^{+0.075}_{-0.059}$ &  log-normal standard deviation of slow
process\\ 
$\mu_{\rm fast}$\tablenotemark{b} & flat on [$-$7.5, 4.0] & $-1.65^{+0.66}_{-0.57}$ &
log-normal mean of fast process\\ 
$\sigma_{\rm fast}$\tablenotemark{b} & flat on [0.001, 3.0] & $\01.55^{+0.46}_{-0.40}$ &
log-normal standard deviation of fast process\\ 

$\chi_I$ [$10^{-3} \cdot$ cps]\tablenotemark{a} & flat on [0.0,5.0]  &
$\01.1^{+0.8}_{-0.5}$ & contribution of steady, extended X-ray source\\ 

$\sigma_\mathrm{Keck}$ [mJy] &Gaussian ($\mu = 0.015$,
$\sigma=0.008$)  & $\00.018\pm{0.006}$ & measurement noise of
the Keck observations\\ 
$\sigma_\mathrm{VLT}$ [mJy] & Gaussian ($\mu = 0.031$,
$\sigma=0.008$) & $\00.034\pm{0.005}$ & measurement noise of
the VLT observations\\ 
$\sigma_\mathrm{IRAC}$ [mJy] & Gaussian ($\mu = 0.65$, $\sigma=0.4$)
& $\00.654^{+0.041}_{-0.047}$ & measurement noise of the IRAC
observations\\ 

$\sigma_{345\,\mathrm{GHz}}$ [Jy] &Gaussian ($\mu = 0.1$,
$\sigma=0.1$)  & $\00.10\pm{0.01}$ & measurement noise of the
345~GHz (APEX and SMA)\\ 
$\sigma_{230\,\mathrm{GHz}}$ [Jy] & Gaussian ($\mu = 0.15$,
$\sigma=0.1$) & $\00.08\pm{0.01}$ & measurement noise of the
230~GHz (SMA and ALMA)\\ 
$e_{S}/e_I$& Gaussian ($\mu = 1.12$, $\sigma=0.04$) &
$\01.12\pm{0.03}$ & effective area of ACIS-S relative to ACIS-I\\ 
$e_{G}/e_I$& Gaussian ($\mu = 0.38$, $\sigma=0.02$) &
$\00.38\pm{0.01}$ & effective area of ACIS-S/HETG relative to ACIS-I\\ 

$\Delta A_{{K}}$& Gaussian ($\mu = 0.0$, $\sigma=0.1$) &
$-0.01\pm{0.07}$ & modification of $K$-band extinction\\ 
$\Delta A_{{M}}$& Gaussian ($\mu = 0.0$, $\sigma=0.14$) &
$\00.09^{+0.08}_{-0.07}$ & modification of $M$-band extinction\\ 

\hline

\null\\[-1.5ex]
\multicolumn{4}{l}{\quad Posteriors (medians
  and 16\% and 84\% quantiles) for the synchrotron-SSC model}\\
\multicolumn{4}{l}{\quad derived
  from 1000 light curves of 700 minutes duration each}\\
$B(t)$ [G]& & $\012.7^{+4.1}_{-2.9}$ & time-dependent magnetic flux density\\
$L(t)$ [$R_S$]& & $\02.20^{+0.32}_{-0.30}$ & time-dependent source size\\
$n_e(t)$ [$10^7$~{cm}$^{-3}$]& & $\04.3^{+6.9}_{-2.7} $ &
time-dependent electron density\\ 
$\alpha_{\rm{NIR}}(t)$& & $-1.65^{+0.29}_{-0.30}$ & time-dependent NIR spectral index (2.2 to 4.5 $\micron$)\\
$\nu_m(t)$ [GHz]& & $\0308^{+96}_{-75}$ & time-dependent self-absorption turnover\\
$\nu_2(t)$ [THz]& & $\054^{+111}_{-29}$ & time-dependent cooling cutoff frequency\\

\hline
\end{tabular}
\end{center}
\tablenotetext{a}{$\mu_{\rm{logn}}$, $\sigma_{\rm{logn}}$, and $\chi$ are data descriptive quantities representing the observed count rate in cps at the \Ch\ detector, i.e., after interstellar extinction (Equations~\ref{transdef} and \ref{cpseq}). \added{They are linearly correlated with $\sigma_{\rm{logn}} \approx -0.40 \cdot \mu_{\rm{logn}} -1.45$.}}
\tablenotetext{b}{$\mu$ and $\sigma$ are the log-normal parameters in Equations~\ref{slowprodef} and~\ref{fastprodef}. They describe the time series of optically thin synchrotron flux density $S_{\rm{thin}}(t)$ in Jy at $\nu_{\rm{NIR}} = 136269$~GHz, and the exponential cutoff frequency $\nu_2(t)$ in units of $\nu_{\rm{NIR}}$.}

\end{table*}

\begin{figure*}[t!]
\begin{center}
\includegraphics[scale=0.32, angle=0]{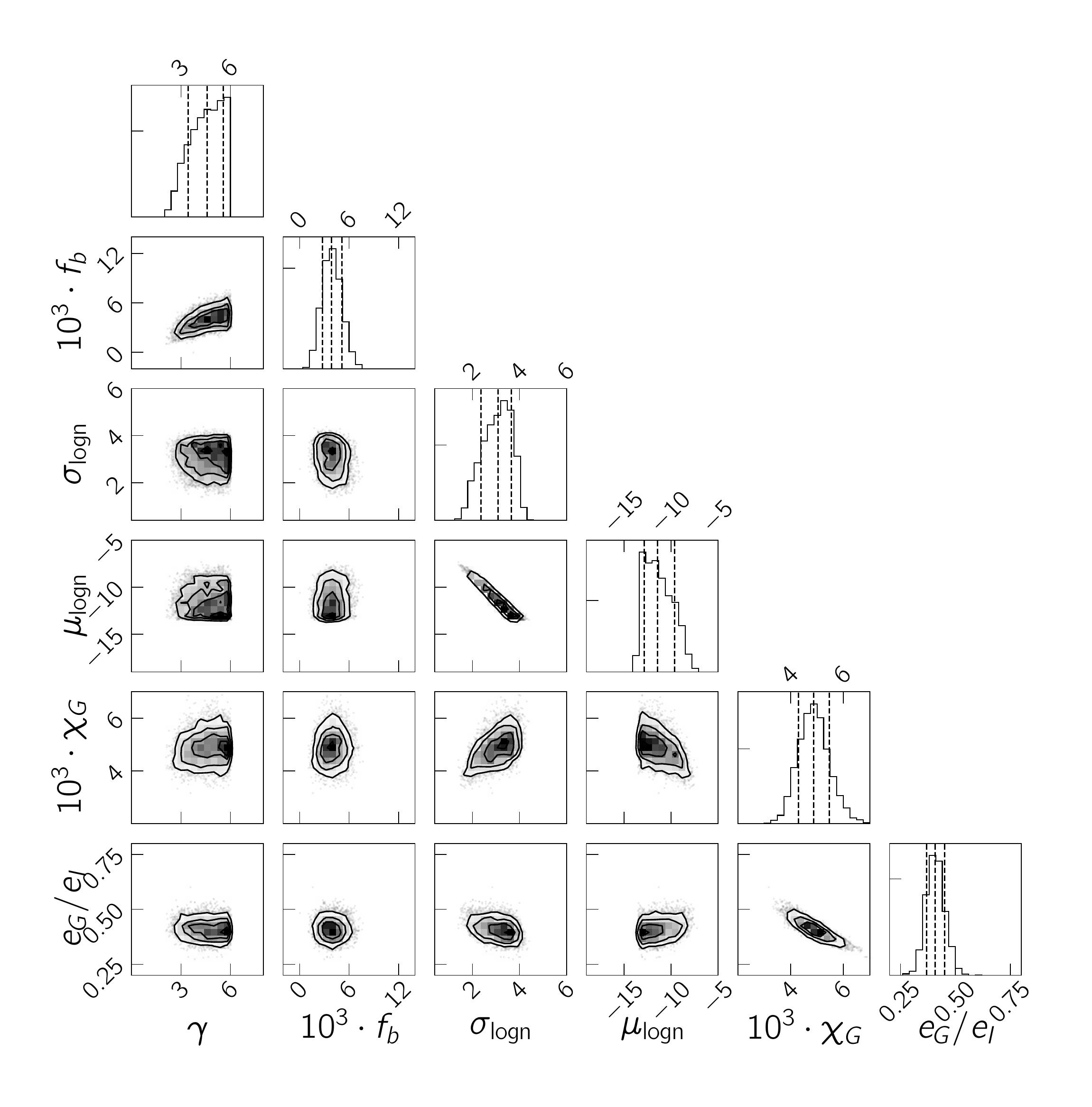}
\end{center}
\setlength{\abovecaptionskip}{-5pt}
\caption{Results for the Bayesian structure function fit for the
  X-ray assuming a log-normal flux-density probability density function. Contours show the
  joint (posterior) probability density for each parameter pair, and
  panels along the right upper edge show histograms of the
  marginalized posterior of each parameter defined in
  Table~\ref{results}. For each histogram, the dashed lines mark the
  16\%, 50\%, and 84\% quantiles.}\label{corner_plot_X}
\end{figure*}

Our custom C++
implementation of the forward modeling and ABC
algorithm includes a fast algorithm for
repetitive calculations of structure functions
(\citealt{2018ApJ...863...15W}, Appendix C) and is
MPI-integrated\footnote{MPI means ``message-passing interface,'' a
  standard for parallel computing.} to be
run on computing clusters.  For this analysis, we  modified the model to
include photon statistics according to Equations~\ref{cpseq} 
to~\ref{cpsmeaseq}.
We ran the ABC on 680 cores of the VLBI correlator computer
cluster of the Max Planck Institute for Radio Astronomy (Bonn) with
100\,000 initial light curve drawings.  From these we selected the
particles having the smallest distance values.
The final ABC run is the result of 16 iterations with $n=800$ particles and two iterations with
$n=5\,000$ particles. 

\subsection{PSD Parameters of the X-ray Variability}\label{XPSDsub}

\added{While testing the algorithm it became clear that the ACIS-S dataset did not add much information to this part of the analysis and could be safely ignored, which reduced the number of parameters by two ($\chi_S$, $e_{S}/e_{I}$).
Additionally, we set $\chi_G = \chi_I$.}

Model parameters and their priors and
posteriors are listed in 
Table~\ref{results} and shown in Figure~\ref{corner_plot_X}. 
While the posterior of $e_{G}/e_{I}$ is a
mere minor alteration of its prior, the posterior of the break frequency $f_b$
is a constrained, peaked distribution. \added{The posterior of the PSD slope $\gamma$ is constrained at the lower end, but prior-dominated at higher values, and in Table~\ref{results} we report its lower limit.}
The log-normal parameters
$\mu_{\rm{logn}}$ and $\sigma_{\rm{logn}}$ are highly correlated ($\sigma_{\rm{logn}} \approx -0.40 \cdot \mu_{\rm{logn}} -1.45$). A more precise
determination of these parameters is not possible because
of photon noise and pileup, which make the log-normal distributions for a range of
$\mu_{\rm{logn}}$ and $\sigma_{\rm{logn}}$ combinations indistinguishable\footnote{\added{Log-normal distributions sensitively depend on the exact position of the peak of the distribution, which in the presence of background photon noise is uncertain. In the absence of a precise location of the peak, the correlation of both parameters is governed by the variance.}}. The background levels $\chi_i \cdot e_i/e_I$ are reasonable when compared 
to X-ray spectra accumulated over all periods of time without flares.

\begin{figure}[h!]
\begin{center}
\includegraphics[scale=0.55, angle=0]{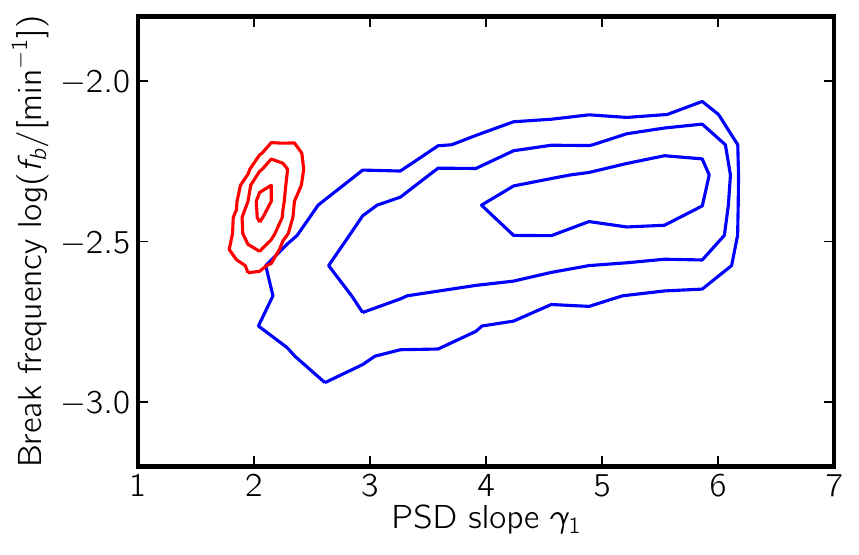}
\end{center}
\setlength{\abovecaptionskip}{-5pt}
\caption{Credible contours (68\%, 95\%, 99\%) for the parameters
  $\gamma$ and $f_b$. Blue shows the results of the present analysis for
  X-ray data (Table~\ref{results}), and red shows results of the NIR
  analysis of \cite[][their Case~3]{2018ApJ...863...15W}. In both
  analyses, the posteriors have been marginalized over all other
  parameters.}\label{X_NIR_contours}
\end{figure}

The parameters of main interest are $\gamma$ and $f_{b}$, 
shown in Figure~\ref{X_NIR_contours} in more detail. 
Determinations of both parameters are significantly less precise for the \Ch\ data than for the NIR. This is not
surprising considering  that in 18 years only $\sim$34\,000 X-ray photons
were detected. However, despite the lack of precision, the comparison
with the NIR is fruitful: the X-ray contours are displaced towards
higher slopes and slightly smaller break frequencies. While the two sets
of contours can be interpreted as marginally consistent,
Figure~\ref{X_NIR_contours} suggests that the X-ray variability does
not show as much power at high frequencies as the
NIR. This interpretation of suppressed power of the fastest
variability turns out to be a clue to understanding the NIR--X-ray
correlations and allows us to identify a radiative
model that can explain many aspects of the rapid (i.e., minutes to
hours) variability of Sgr~A* from the submm to the X-rays.

\subsection{Simulations of NIR--X-ray Correlations}\label{xncorr}

\begin{figure*}[h!]
\begin{center}
\includegraphics[scale=0.29, angle=0]{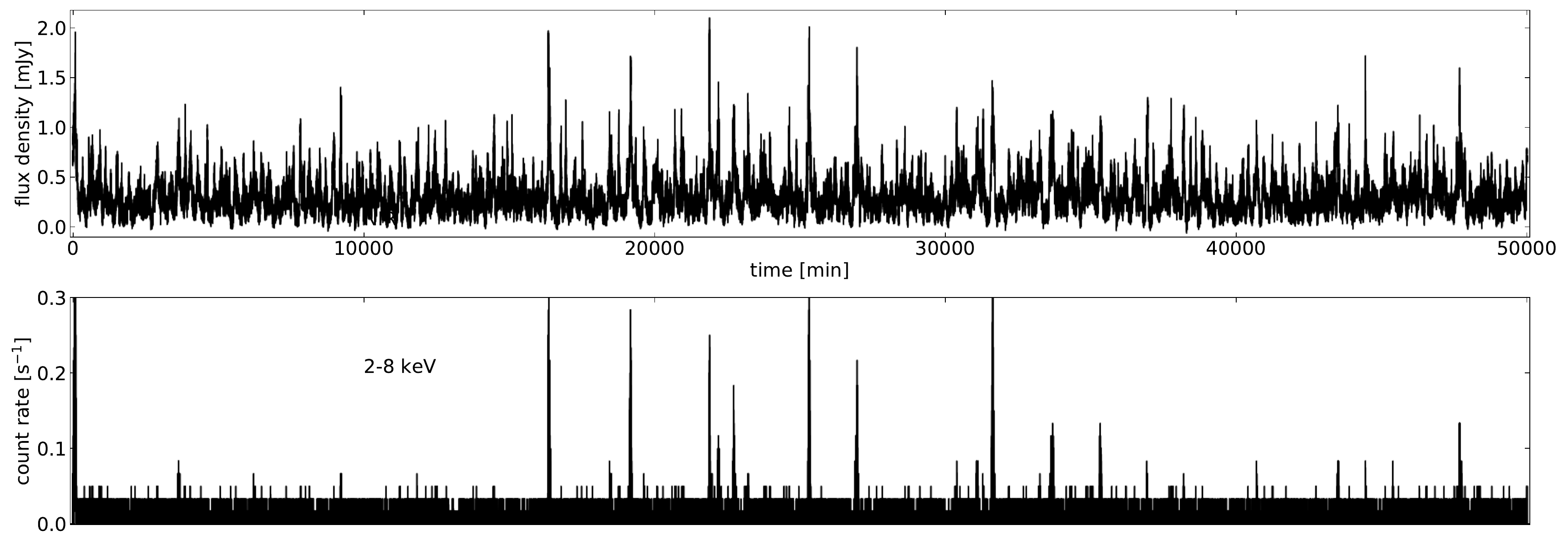}
\end{center}
\setlength{\abovecaptionskip}{-5pt}
\caption{Multiwavelength  mock light curves
  generated from the X-ray data model. The \replaced{lower}{upper}
  panel shows a 50\,000-minute realization of log-normal NIR flux
  densities, and the \replaced{upper}{lower} panel presents X-ray count rates during the
  same time interval.  \replaced{Count rates are
  as observed }{Simulated count rates are modeled to match observations} and include
  the steady X-ray
  background, pileup, and Poisson noise.  (Flux-density normalization
  and the extinction correction are implicit in the model parameter $\mu$.)
  Figure~\ref{cartoon34}a illustrates the procedure for
  generating the mock light curves.}\label{lc_gen} 
\end{figure*}

\begin{figure*}[h!]
\begin{center}
\includegraphics[scale=0.4, angle=0]{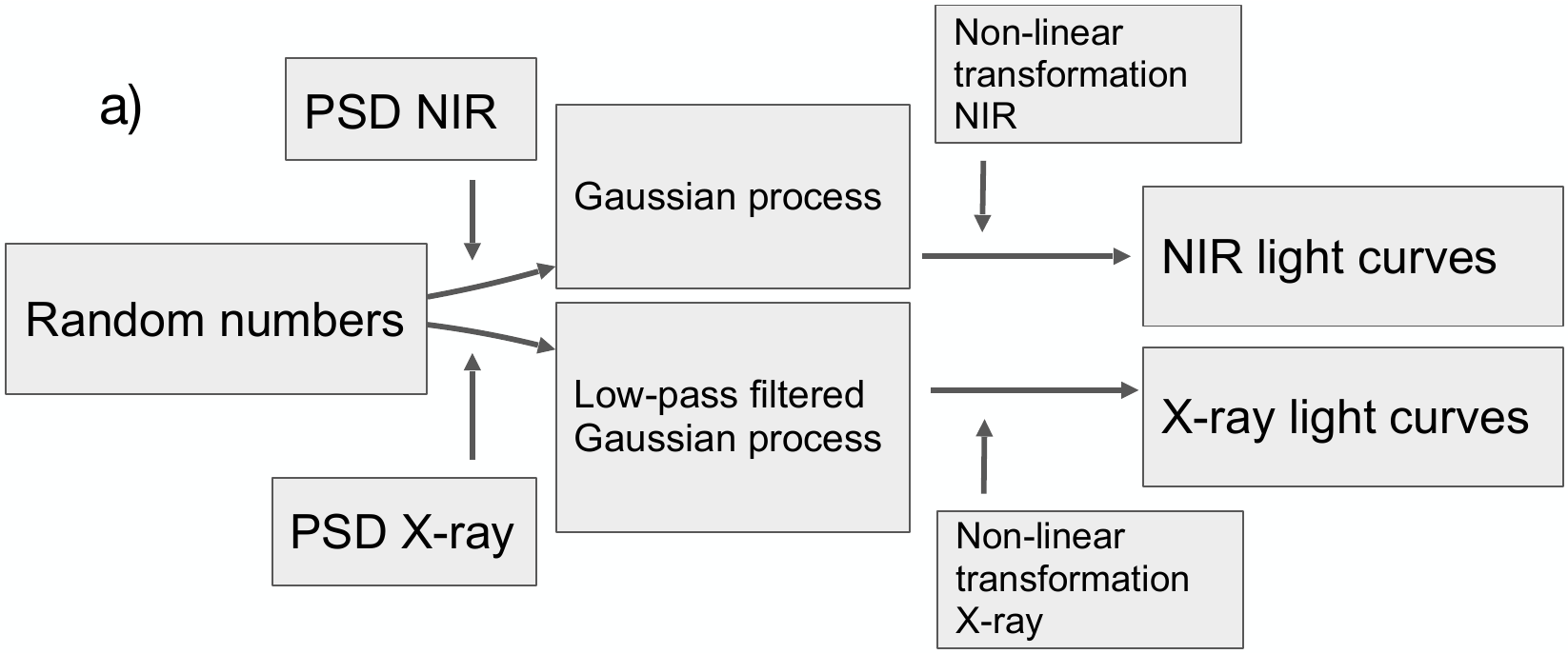}
\par
\vspace{0.7cm}
\includegraphics[scale=0.4, angle=0]{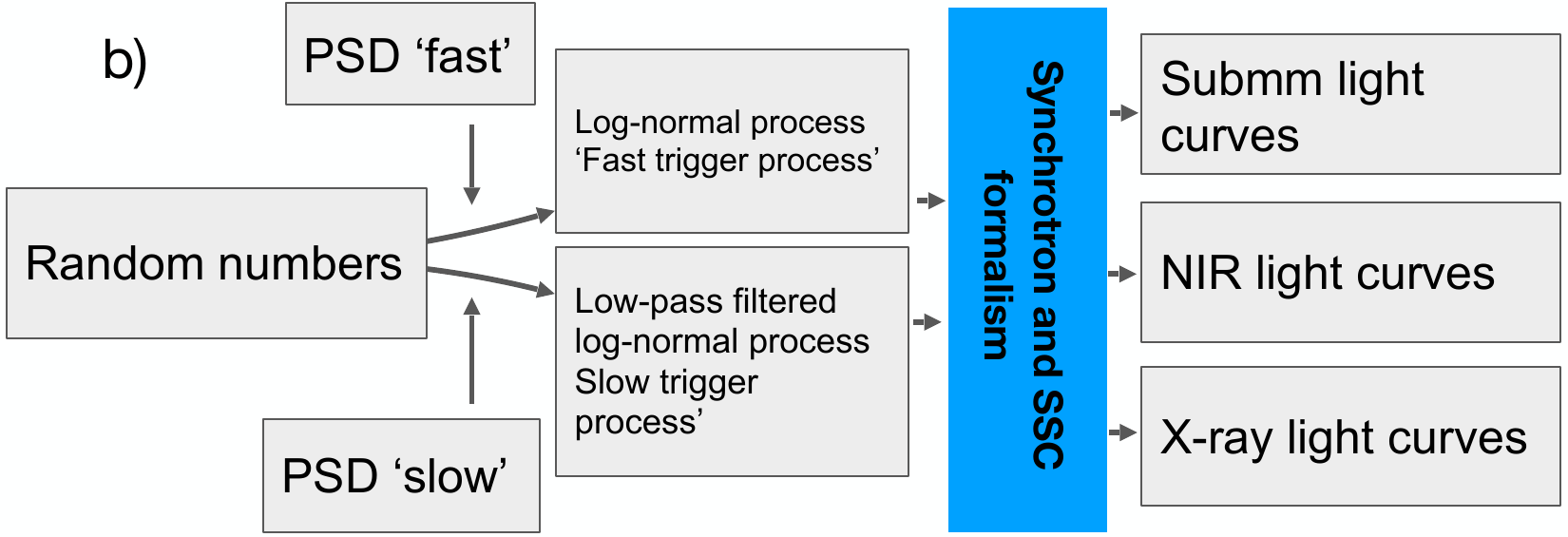}
\end{center}
\setlength{\abovecaptionskip}{-5pt}
\caption{Flow chart of simulation algorithms for modeling
  light curves. Upper panel a) shows an example of low-pass correlated
  Gaussian processes under transformation with log-normal
  distributions. The generation of each Gaussian process from random
  numbers follows the procedure described by
  \cite{1995A&A...300..707T}. The two PSDs applied are both broken
  power laws, one of which has---by means of a steeper slope and/or a
  smaller break frequency---significantly lower power at short
  variability timescales. Lower panel b) shows a similar
  low-pass correlated set of log-normal trigger processes that are
  then used to inform our synchrotron--SSC model. The radiative
  model combines the fast and slow processes according to the
  equations developed in Section~\ref{impl} and lets us derive time
  series of flux densities at any given wavelength.}\label{cartoon34}
\end{figure*}

Figure~\ref{lc_gen} demonstrates how PSD parameters affect NIR--X-ray
correlations  The figure shows two light
curves from the same set of Gaussian random numbers with two
different PSDs. In the previously mentioned method of
\cite{1995A&A...300..707T}, a random number---one for each relevant
frequency in the Fourier representation of the light curve---is
multiplied by the square root of the PSD at the corresponding
frequency. By, let's say, using a  broken power-law PSD model for one of
the light curves and a second PSD with the same break frequency but
a steeper slope for the other, we can generate pairs of light curves.
Figure~\ref{cartoon34}a) shows a flow chart for the
procedure. Because the curves use
the same random numbers, they are strongly correlated. However, one of
them is the low-pass filtered version of the other.
When we apply two, e.g., log-normal
transformations (with the one applied to the slower process showing a
heavier tail), the slower, more non-linear process (representative of
the X-ray variability) will always have a counterpart in the faster
process (representative of the NIR), while the faster process shows a
lot of peaks having a wide range of levels with (almost) no
counterpart. When peaks are seen in both curves, their arrival times are
strictly correlated, but there is little correlation in peak flux. This phenomenology mimics the NIR--X-ray
observations \citep{2010RvMP...82.3121G,2018ApJ...864...58F} and becomes even more realistic
when we include photon statistics, pileup effect, and a constant
Poissonian background for the slower
process.

\section{Synchrotron Self-Compton Scattering and the Evolution of the Optical Depth}
\label{radmodel}

The theory of astronomical synchrotron sources was developed in the
1950s and 1960s
(\citealt{1951DoSSR..76..377G,1952AZh....29..418S,1953DoSSR..90..983S,1965ARA&A...3..297G}).
This paper uses the later, systematic development of
synchrotron and synchrotron-SSC theory by
\cite{1962AZh....39..393K}, \cite{1966Natur.211.1131V},
\cite{1979A&A....76..306G}, \cite{1983ApJ...264..296M}, and
\cite{1985ApJ...298..128B}. In particular, we assume familiarity with
the excellent overview article by \cite{1975gaun.book..211M}.

Based on  flux-density levels, timing properties, and correlations,
SSC scattering is a good candidate to explain
Sgr~A*'s X-ray variability. SSC
becomes relevant under the same circumstances---high luminosity in a
source of small size \citep{1975gaun.book..211M}---as synchrotron
self-absorption,  and therefore we discuss the co-evolution of SSC
scattering and synchrotron opacity. A simple model
of injection, compression, and expansion
can account for many aspects of the time-variable SED of
Sgr~A*. A semi-analytic approach
allows us to co-fit the structure functions at all wavelengths
considered in this study.

\subsection{The NIR--X-ray Correlation}

\begin{figure}[h!]
\begin{center}
\includegraphics[width=\columnwidth,clip=true,trim=6 6 18 24]{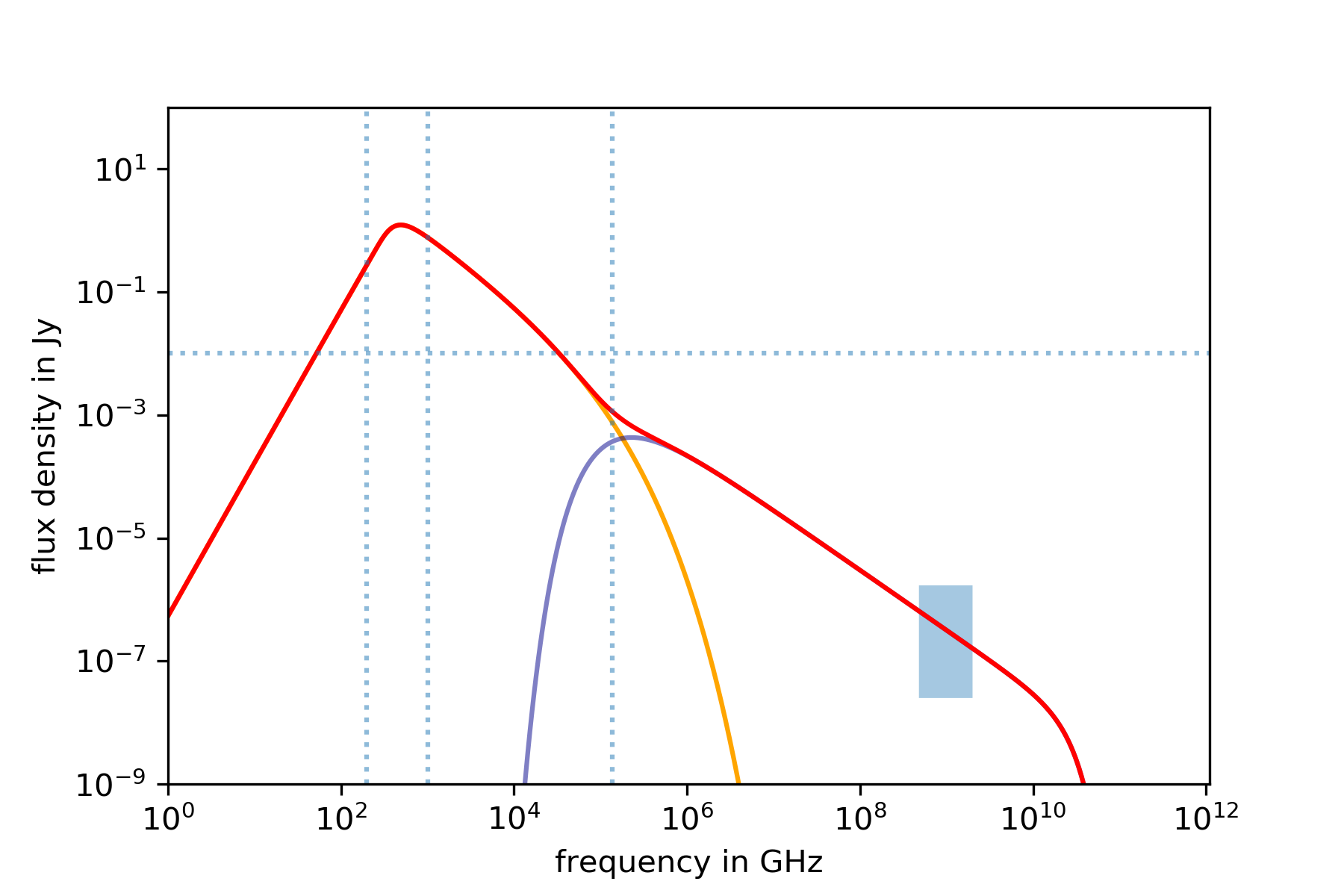}
\end{center}
\setlength{\abovecaptionskip}{-5pt}
\caption{Model radio-to-X-ray SED for Sgr A*. Orange
  shows the synchrotron component, blue the SSC component, and red
  their sum. The SED is shown as if interstellar extinction were zero.  Model
  parameters (see Table~\ref{results} for 
  definitions) are
  $n_{e} = 1.066\times10^8$~cm$^{-3}$, $B = 10$~G,
  $L_0=2$~$R_{S}$, $\nu_2 = 20$~THz, $E_{\rm{min}}=0.02$~GeV,
  and $\gamma = 2.4$. The
  vertical dashed lines mark the frequencies 230~GHz, 345~GHz, and
  $\nu_{\rm{NIR}}$. The blue rectangle marks the 2--8~keV X-ray band and
  shows typical peak flux densities during X-ray flares. The
  increasing and decreasing flanks of the SSC spectrum (where the
  spectrum is not described by a simple power law according to
  Equation~\ref{SSCEq}) are illustrated by arbitrary exponential
  cutoffs. The precise shape of the SSC spectrum in these flanks
  depends on many assumptions including the density profile of the
  source and is beyond the scope of this study. The spectrum shown is not a fit to
  any  actual data. In particular it combines low-level NIR flux
  densities with detectable X-ray emission, a condition that has
  not been observed.}\label{SSC_SED}
\end{figure}

As mentioned before, several studies have found the flux-density peak
levels of synchronous NIR and X-ray flares to be consistent with the
prediction of simple SSC models. Furthermore,
\cite{2018ApJ...863...15W} showed evidence that the NIR variability is
at least partially caused by a variable, exponential synchrotron
cooling cutoff close to the NIR; i.e., the synchrotron spectrum of
flux densities is of the form:
\begin{equation}
  S_{\nu}(t) = I_{\nu} \exp\left[-\left(\frac{\nu}{\nu_2(t)}\right)^{\frac{1}{2}}\right]  \Delta\Omega \;\; ,
\label{fluxdensity}
\end{equation}
where the source subtends a solid angle of $\Delta \Omega = \pi/4\cdot (L
/ D_s)^2$ with $L$ the source diameter and $D_s$ the distance to the
source. $I_{\nu}$ is the synchrotron intensity at frequency
$\nu$, and $\nu_2(t)$ is the variable cutoff frequency due to
synchrotron cooling.  Building on these ideas, 
the analytical equation for SSC radiation \citep{1983ApJ...264..296M} also depends
on $\nu_2(t)$ but very weakly because it is the argument of a
logarithmic term:
\begin{equation}\label{SSCEq}
S(E_{\rm{keV}}) = d(-\alpha) \ln{(\nu_2/\nu_m)} \theta^{(4\alpha-6)} \nu_m^{(3\alpha-5)}  S_m^{(4-2\alpha)} E_{\rm{keV}}^{\alpha} \;\; ,
\end{equation}
where $\alpha$ is the optically thin spectral index,\footnote{We define
  $\alpha$ such that $S_\nu\propto\nu^{\alpha}$.
  \citet{1983ApJ...264..296M} used the opposite sign convention.} $d(-\alpha)$ are coefficients given by \cite{1983ApJ...264..296M},
$\theta = L/D_s$ is the
angular diameter of the source, $\nu_m$ is the self-absorption turnover
frequency, $S_m$ is the self-absorption flux density, and $E = h\nu$ is the
photon energy in keV with $h \approx 4.136 \times 10^{-9}$~keV/GHz  the
Planck constant. Equation~\ref{SSCEq} is valid for
\begin{equation}\label{SSC_limits}
5.5 \times 10^{-9} \gamma_1^2 \nu_m \lesssim E_{\rm{keV}} \lesssim 
0.2 [b(-\alpha)]^{-1} \theta^{-4} \nu_m^{-5} S_m^2
\end{equation}
with $b(-\alpha)$ coefficients given by \cite{1983ApJ...264..296M}
and $\gamma_1$ the minimum Lorentz factor of the electrons. For
energies in this range, the SSC flux density
is a power law with the same spectral index as the
NIR. Figure~\ref{SSC_SED} shows a
synchrotron--SSC SED for realistic \Sg\
parameters. (Section~\ref{opac} explains how to calculate 
$I_{\nu}$ from these parameters.)

The logarithmic dependence of $S(E_{\rm{keV}})$ on the 
cutoff $\nu_2(t)$ implies that if $\nu_2$ is the origin
of the fast variability in the NIR, this variability power is
suppressed in the X-ray light curves. However, it is not obvious that
Equation~\ref{SSCEq} indeed suggests a clear separation of
variability power: the X-ray flux density is highly variable, and
this variability must be related to changes of the synchrotron source
and spectrum itself, i.e., to changes of $\nu_m$, $S_m$, and
$\theta$. This in turn means that  $I_{\nu}$ and
$\Delta\Omega$ in Equation~\ref{fluxdensity}
are time-dependent with similarly slow
variability as the SSC flux density. Thus, following the idea
presented in Section~\ref{xncorr} and if SSC is responsible for the
X-ray emission, the NIR flux density $S_{\nu_{\rm{NIR}}}(t)$ is the
product of two correlated processes,
$I_{\nu_{\rm{NIR}}}(t)\Delta\Omega(t)$ and
$\exp{\left[-\left(\frac{\nu_{\rm{NIR}}}{\nu_2(t)}\right)^{{1}/{2}}\right]}$,
the former being the low-pass-filtered counterpart of the latter.
This situation---the slow process (postulated from the X-rays)
feeding back into the fast process (the NIR variability, described by
the product)---is slightly more complex than the simple case
considered in Section~\ref{xncorr}. However,  for the
right model for $I_{\nu}(t)$ and proper model parameters, the fast
process will dominate this product, and the result will indeed be
similar to the light curves of Figure~\ref{lc_gen}.

According to Equation~\ref{SSCEq}, $S(E_{\rm{keV}})$ is not a direct
function of the optically thin flux density at the $K$-band
frequency $\nu_{\rm{NIR}} = 136269$~GHz,
\begin{equation}
S_{\rm{thin}}  = I_{\nu_{\rm{NIR}}}(t)\Delta\Omega(t)  \;\; ;
\end{equation}
the self-absorption turnover flux density and turnover
frequency also matter. In order to generate light curves in the
submm that show the observed correlations and delays relative
to the NIR, we need to discuss possible scenarios of co-evolution of
$S_{\rm{thin}}$ and the synchrotron opacity. In particular,
it is not correct to assume $S_m \propto S_{\rm{thin}}$ in
the context of Equation~\ref{SSCEq} and to treat $\nu_m$ and $\theta$
as independent parameters (as often done in the literature) and then
to argue that $S(E_{\rm{keV}})$ depends on the optically thin flux
density in a highly non-linear way due to the term $S_m^{(4-2\alpha)}$.

\subsection{Evolution of the Synchrotron Source and Synchrotron Self-absorption}\label{opac}

Following the formalism and notation of
\cite{1975gaun.book..211M}, the solution of the radiative transfer
equation $dI_{\nu} = (\epsilon_{\nu}-\kappa_{\nu}I_{\nu})\;dz$
through a homogeneous slab of material with constant and isotropic
emissivity $\epsilon_{\nu}$, absorption $\kappa_{\nu}$, and thickness
$z_t$ is
\begin{equation}
  I_{\nu}(z_t) =  \frac{\epsilon_{\nu}}{\kappa_{\nu}}[1 - \exp{(-\kappa_{\nu} z_t)}]~~.\label{radtrans}
\end{equation}
For a synchrotron source of homogeneous electron density in a
tangled magnetic field, assuming a power-law electron energy
distribution and an isotropic pitch angle distribution,
\begin{align}
  \epsilon_{\nu} & =  \frac{1}{2} C_2 n_0 B^{(\gamma_{e}+1)/2} (C_1/\nu)^{(\gamma_{e}-1)/2}  \nonumber \\
  &~~\cdot G'(\nu/\nu_1,\nu/\nu_2, \gamma_{e})~~,
\label{emis}
\end{align}
and
\begin{align}
  \kappa_{\nu} & = \frac{1}{2} c^2 C_2 C_1^{\gamma_{e}/2} n_0 B^{(\gamma_{e}+2)/2} \nu^{-(\gamma_{e}+4)/2} (\gamma_{e} +2) \nonumber \\
  & ~~ \cdot G'(\nu/\nu_1,\nu/\nu_2, \gamma_{e}+1) ~~.
\label{absorp}
\end{align}
Here $c$ is the speed of light, $C_1$ and $C_2$ are constants given in
Table~\ref{const}, $\gamma_{e} = 1-2\alpha$ is the power-law index of
the electron energy distribution, $n_0$ is the scaling factor of the
electron energy distribution, and $B$ is the magnetic flux density in
gauss. $G'(\nu/\nu_1,\nu/\nu_2, \gamma_{e})$ is defined as
\begin{align}
  G'(\nu/\nu_1,\nu/\nu_2, \gamma_{e}) & \equiv \frac{1}{2}\pi^{1/2}\frac{\Gamma[(\gamma_{e}+5)/4]}{\Gamma[(\gamma_{e}+7)/4]} \nonumber  \\
  & ~~ \cdot \int_0^{\infty}x^{(\gamma_{e} -3)/2}F(x)dx \;\; ,
\end{align}
with
\begin{equation}\label{besselint}
F(x) = x \int_x^{\infty} K_{5/3}(\eta)d\eta \;\; ,
\end{equation}
$K_{5/3}(\eta)$ the modified Bessel function of the second kind, and
$\Gamma(x)$ the Gamma function. Here we set $\nu/\nu_1 = \infty$ and
$\nu/\nu_2 = 0$. However, this does not imply that the maximal
Lorentz factor $\gamma_{\rm{max}}$ of the electrons is
infinite. Equation~\ref{fluxdensity} includes an exponential term to
approximate the effect of a finite $\gamma_{\rm{max}}$. This approach
allows us later to use a lookup table for $G'(\infty,0,
\gamma_{e})$ without having to re-evaluate the integral for every
frequency.

\begin{figure}[h!]
\begin{center}
\includegraphics[scale=0.60, angle=0]{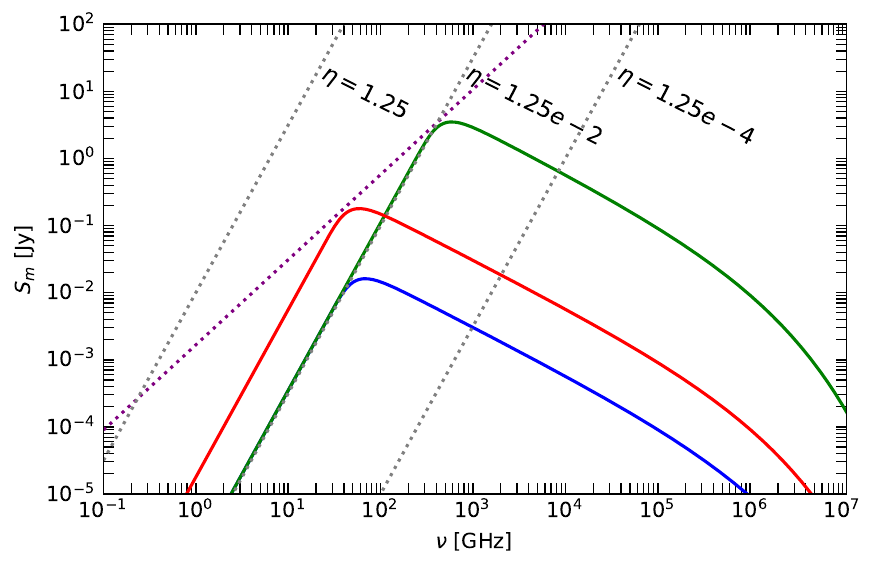}
\end{center}
\setlength{\abovecaptionskip}{-5pt}
\caption{Example SEDs derived from Equations~\ref{fluxdensity}
  and~\ref{radtrans} to~\ref{besselint}. SEDs shown in green and blue
  were derived assuming the same ratio $\eta = \theta/B^{1/4}$
  but different electron densities $n_{e}$. For a given $\eta$, the
  self-absorption turnover with the coordinates ($\nu_m$, $S_m$) lies
  on one of the dashed grey lines having a power-law index of 2.5
  (the slope of the optically thick part of the
  spectrum), independent of $n_{e}$ or the optically thin
  flux-density
  levels. The purple dashed line marks the trajectory of the
  turnover under adiabatic expansion, i.e., the red SED could be the
  result of adiabatically expanding a synchrotron sphere that starts
  off with the green SED. The slope of the turnover trajectory
  related to adiabatic expansion can be significantly flatter than
  the iso-$\eta$ trajectories (the exact slope being given by
  Equation~\ref{toadj}).}\label{SEDs}
\end{figure}

A main result of synchrotron theory is that the SED is
peaked and asymptotically approaches two different power laws: the
optically thick spectrum $I_{\nu} = \epsilon_{\nu}/\kappa_{\nu}$
with slope $+2.5$ on the low-frequency side of the peak and the
optically thin spectrum $I_{\nu} = \epsilon_{\nu} \cdot z_t$ with spectral
index $\alpha = (1- \gamma_{e})/2$ on the high-frequency
side. Typical 
values observed for radio galaxies are  $\alpha \sim -0.7$. The
peak is described by  ($\nu_m$,$S_m$), where $S_m$ is the
intersection of the two power laws, i.e., an extrapolation that is not
actually reached by the SED.  Figure~\ref{SEDs} shows some example
synchrotron spectra.

Equation~\ref{SSCEq} for the SSC flux density makes the assumption of
spherical symmetry. As mentioned, Equation~\ref{radtrans} is formally
derived for a slab of material, i.e., if $\Delta \Omega = \pi/4 \cdot
(L / d)^2$, $S_{\nu}$ of Equation~\ref{fluxdensity} describes a
cylinder \added{of diameter and height $L$} homogeneously filled with electrons. However,
\cite{1985ApJ...298..128B} demonstrated that the difference between
spectra derived with a slab approximation and with
spherical symmetry is very small. (Of course, $n_0$ needs
to be multiplied by $1.5$ in order to account for the ratio between
the volume of a cylinder and the volume of a sphere.) Therefore
we will proceed with the slab approximation and use the
correction factor of $1.5$ when calculating electron densities from
flux densities.

\cite{1975gaun.book..211M} derived a central equation to
link the self-absorption turnover of the synchrotron spectrum
described by the observable quantities $\nu_m$ and $S_m$ with the
physical parameters $\theta$ and $B$:
\begin{equation}
\nu_m^{5/2}  = c^2C_1^{1/2}F'S_m  B^{1/2}/(\pi/4 \cdot \theta^2)
\end{equation}
with
\begin{equation}\label{Fprime}
  F' = \frac{(\gamma_{e} + 2) \cdot G'(\nu/\nu_1,\nu/\nu_1, \gamma_{e} + 1)}{G'(\nu/\nu_1,\nu/\nu_1, \gamma_{e})} \;\;.
\end{equation}
This equation is similar to an equation of state and allows us to
describe how the observable quantities change under evolution of the
physical quantities. Because $\nu_m$ and $S_m$ are not
observable in our case---the self absorption turnover of the compact
component responsible for the X-ray and NIR variability is veiled
by emission from the constant radio component---we express
$S_m$ in terms of the optically 
thin synchrotron flux density $S_{\rm{thin}}$:
\begin{align}
  S_{\rm{thin}} & = \pi/4 \cdot  \epsilon_{\nu_{\rm{NIR}}} \cdot L
  \cdot \theta^3  \;\; ,~{\rm and}\\
  \nu_m^{5/2 - \alpha} & = c^2C_1^{1/2}F'S_{\rm{thin}} \cdot
  (\nu_{\rm{NIR}})^{-\alpha} B^{1/2}/(\pi/4 \cdot \theta^2) \;\;
  . \label{eqstatethin}
\end{align}
Similarly, the SSC flux density defined in Equation~\ref{SSCEq}
\begin{align}
  S(E_{\rm{keV}}) & \propto \nu_m^{(3\alpha-5)} S_m^{(4-2\alpha)} \theta^{(4\alpha-6)} \\
  & \propto \nu_m^{(3\alpha-5)-2\alpha(\alpha+2)}
  S_{\rm{thin}}^{(4-2\alpha)} \theta^{(4\alpha-3)} \;\;.
\end{align}

We are now in position to discuss illustrative scenarios of source
evolution and their consequences for the dependence of the X-ray and
submm flux densities on $S_{\rm{thin}}$.  While none of these is
directly applicable to the final model, they are useful to show the
dependence of observables on physical quantities. In the following, we define
\begin{equation}
\eta \equiv \theta/B^{1/4} \;\;.
\end{equation}

\subsubsection{Case 1}
Let us assume $\Delta \eta = 0$, $\Delta S_{\rm{thin}} \neq 0$. In this
case, $S_m \propto (\nu_m)^{2.5}$, i.e., the turnover moves along the
iso-$\eta$ lines shown in Figure~\ref{SEDs}. These iso-$\eta$ lines
have the same slope as the optically thick part of the
spectrum, and the flux densities at frequencies below the peak don't
change. From
Equation~\ref{eqstatethin},
\begin{equation}
\nu_m \propto (S_{\rm{thin}})^{\frac{2}{5-2\alpha}} \;\;,
\end{equation}
and with $\Delta\theta = 0$ (i.e., no changes other than in $n_0$)
and assuming $\alpha = -1$,
\begin{equation}
S(E_{\rm{keV}}) \propto (S_{\rm{thin}})^2\;\; .
\end{equation}
This occurs because the increase in $\nu_m$ partially counters that in
$S_m$. This gives the dependence of
$S(E_{\rm{keV}})$ on $S_{\rm{thin}}$  a significantly lower
exponent than the value 6 apparent from
Equation~\ref{SSCEq}.

\subsubsection{Case 2}
Next consider the case $\Delta \nu_m = 0$, $\Delta S_{\rm{thin}} \neq 0$,
i.e., a constant turnover frequency under changing
$S_{\rm{thin}}$. In this case, $S_m \propto S_{\rm{thin}}$,
$S_{\rm{thick}} \propto S_{\rm{thin}}$, and
\begin{equation}
\eta \propto (S_{\rm{thin}})^{1/2} \;\;.
\end{equation}
If the change in $\eta$ is mainly a change in
$\theta$---a corresponding change in $B$ would have to be higher by 
the fourth power---$\Delta B = 0$ and $\alpha = -1$ imply
\begin{equation}
S(E_{\rm{keV}}) \propto S_{\rm{thin}} \;\;.
\end{equation}

\subsubsection{Case 3}
As mentioned in the Introduction, a case of particular interest is
source evolution through adiabatic expansion. The fundamental
assumptions have been stated by \cite{1960AZh....37..256S} in the
context of supernova-remnant evolution:
\begin{equation}
B \propto \theta^{-2} \; \; ,
\end{equation}
and 
\begin{equation}\label{adjcooling}
E \propto \theta^{-1} 
\end{equation}
as $\theta$ changes with time.
In contrast to synchrotron cooling, cooling by adiabatic
expansion applies to electrons of all energies at the same rate set
by the expansion speed. \cite{1966Natur.211.1131V} showed
that in this case,
\begin{equation}
S_m(\nu_{m}) \propto 
\left(\nu_{m}\right)^{-\frac{7\gamma_{e}+3}{4\gamma_{e}+6}}
\;\;. 
\label{toadj}
\end{equation}
For reasonable values of $\gamma_{e}$, $S_m$ has a flatter index
than the iso-$\eta$ lines (Figure~\ref{SEDs}). Other relations are:
\begin{align}
\nu_m & \propto (S_{\rm{thin}})^{\frac{2}{5 - 2\alpha}} \cdot
\eta^{\frac{-4}{5-2\alpha}}\;\; ,\\ 
B &\propto (S_{\rm{thin}})^{\frac{2}{3-4\alpha}}\;\; , \;\; {\rm and}\\
\theta & \propto (S_{\rm{thin}})^{-\frac{1}{3-4\alpha}} \;\; .
\end{align}
For the X-ray flux density:
\begin{align}
S(E_{\rm{keV}}) &\propto 
  \eta^{-4(5-3\alpha+2\alpha(\alpha-2))/(5-2\alpha)} \nonumber \\
& \cdot (S_{\rm{thin}})^{(4-2\alpha) + 2(3\alpha-5-2\alpha(\alpha+2))/(5-2\alpha)} \nonumber \\
& \cdot \theta^{(4\alpha-6)}
\end{align}
For $\alpha=-1$,
\begin{equation}
  S(E_{\rm{keV}})\propto (S_{\rm{thin}})^{1.71} \;\;.
\end{equation}
While $S_m$ decreases with $S_{\rm{thin}}$,  $\nu_m$ decreases as well.
Therefore the optically thick flux densities show a temporary
increase some time after the time of maximum flux density at optically thin
frequencies.

In all three cases discussed here, under the physical
constraints expressed in Equation~\ref{eqstatethin}, the actual dependence
of the SSC flux density on $S_{\rm{thin}}$ is only
weakly non-linear with an exponent of 1--2 and does not come close to
the apparent exponent of $(4-2\alpha)$ in Equation~\ref{SSCEq}.

None of the discussed cases of source evolution can by itself
reproduce the submm to NIR phenomenology. Case~1 does
not predict NIR-correlated submm variability at all. Case~2, a
scenario of non-adiabatic compression and expansion, predicts a
direct proportionality of submm flux-density changes to $S_{\rm{thin}}$,
which is not observed. Case~3 describes only the cooling of the
synchrotron source by expansion, i.e., decaying $S_{\rm{thin}}$, and
the resulting propagation of delayed peaks towards longer observing
wavelengths. The opposite, increasing $S_{\rm{thin}}$ under adiabatic
compression, would be difficult to understand
physically. Furthermore, adiabatic compression would result in
leading maxima at submm wavelengths, also not observed.
 
\subsection{A Simple Source Model}\label{simple}
To generate submm light curves that correspond qualitatively and
quantitatively to the observed data, we here propose a simple
three-step process of electron injection, compression of magnetic field lines,
and expansion. For simplicity, we model this three-step process as a
cyclic process in a single zone.
Analysis of this process makes the assumption that if $S_{\rm{thin}}$ is rising
fast, the variability is injection dominated, and the opacity is
developing according to Case~1. If $S_{\rm{thin}}$ is falling
quickly, the source is adiabatically expanding according to
Case~3. Just before the time of peak flux density, the source is subjected
to an episode of (non-adiabatic) compression and particle escape,
i.e., a compression of the magnetic field lines mainly without
electron heating (similar to Case~2 but with changing $B$). Here,
particle escape is necessary because close to the peak, $S_{\rm{thin}}$
changes little, but in our model $B$
increases. Because $\epsilon_{\nu}$ depends on $B$ and on
the product $n_0  \theta^3$ (which is constant under particle
conservation), particles must escape\footnote{\added{A decreasing source size with increasing magnetic flux density $B$ can be created naturally when magnetic flux lines are compressed with the bulk of electrons not following that compression. As a consequence, a smaller volume filled with electrons is interacting with the stronger magnetic field. The electrons outside this active zone then have ``escaped'' the synchrotron region. The details of this compression phase depend on the characteristics of the processes that govern the electron budget (injection, escape, cooling).
}}.

\added{The reason for using a cyclic model is simplicity. However, the cycles we are modelling are not “sequential” in the sense that an individual cycle has to return to its starting point before a new cycle can start. A change of $\dot{S}_{\rm{thin}}(t)$ from negative to positive will start a `new cycle'. To translate this to a picture of multiple regions, a new region will start to dominate where the old left off. While a new source region might start with a different set of source parameters, statistically the presented approach is equivalent, at least with respect to the posteriors of the source parameters marginalized over time.}

\begin{figure}
\begin{center}
\includegraphics[scale=0.35, angle=0]{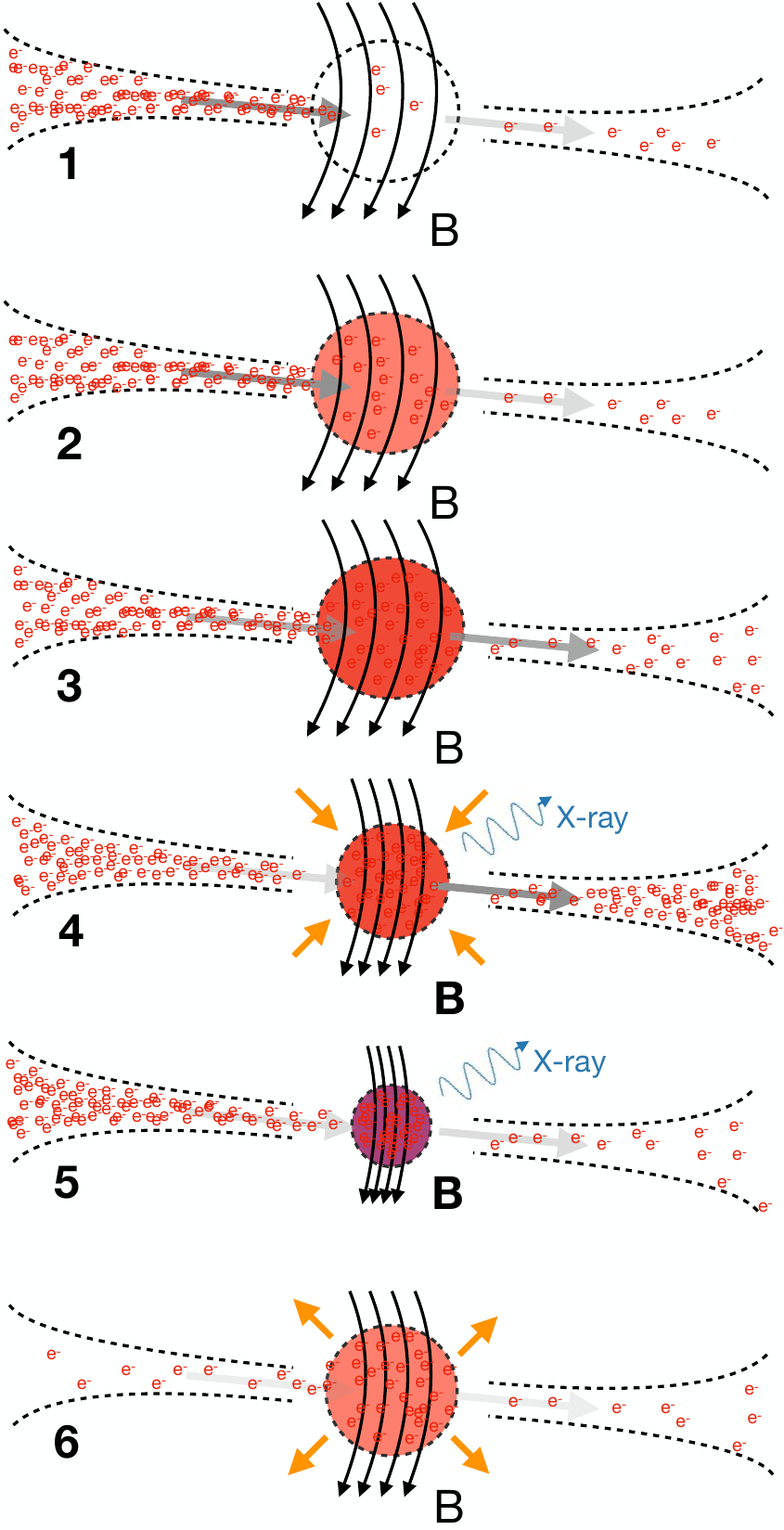}
\includegraphics[scale=0.35, angle=0]{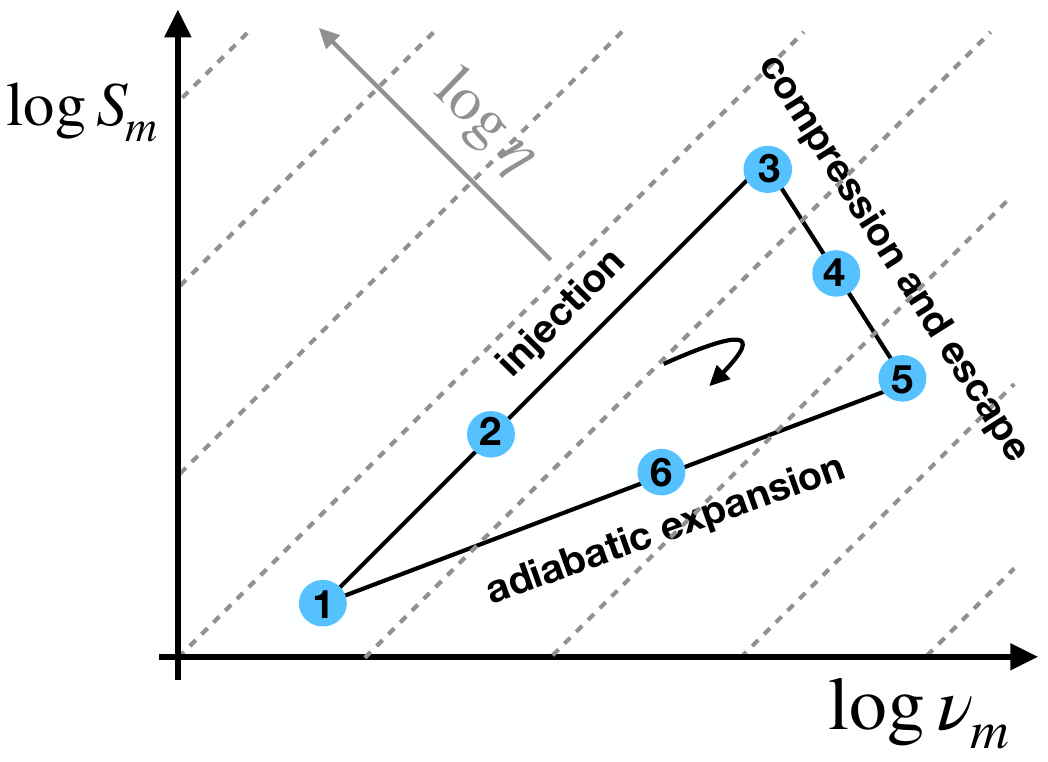}
\end{center}
\setlength{\abovecaptionskip}{-5pt}
\caption{Illustration of the evolution of our simple model for the
  compact component in Sgr A*, denoted by the dashed circle. Upper
  panels 1--6 are
  a cartoon of a possible sequence of injection, compression, and
  expansion. The steps are: 1) and 2)
  injection of electrons with a non-thermal energy distribution into
  a spherical region with a uniform magnetic flux density giving
  detectable submm and NIR emission. 3) and 4) further injection,
  compression, and increase
  of magnetic flux under electron escape giving increased self-Compton
  scattering efficiency and detectable X-ray emission. 5) minimum
  size with highest magnetic flux density under ongoing injection
  followed shortly after by maximum in electron density and in NIR
  and X-ray emission. 6) adiabatic expansion, no injection giving maximum
  in submm emission.  The
  funnels on the two sides of the synchrotron sphere are idealized
  and not implying actual symmetry of the inflow or
  outflow of particles. In particular, particle escape under
  compression might well be isotropic.  The bottom panel
  shows the corresponding cycle of the 
  self-absorption turnover position in the $S_m$--$\nu_m$
  plane.  Steps are labeled as above, and the diagonal dashed lines are
  lines of constant~$\eta$. }\label{cartoon12}
\end{figure}

Our three-step source evolution is illustrated in
Figure~\ref{cartoon12},  including
transitions between injection and compression and compression and
expansion. The model qualitatively predicts submm light curves 
correlated with the NIR and shows a range of possible delays between
the two bands. It is, indeed, the simplest source evolution model that
can reproduce the observed phenomenology without generating
artificial symmetries in the light curves or too-strict
correlations.
 
\begin{table}[htb]
\begin{center}
\caption{Constants} \label{const}
\begin{tabular}{lll}
\hline
\hline
Constant & Value & Unit \\
\hline

$C_1$ & $16.08$  & MHz~$\mu$G$^{-1}$~GeV$^{-2}$ \\
$C_2$ & $1.16540 \times 10^{-26}$ & GeV~$\mu$G$^{-1}$~sr$^{-1}$ \\
$C_3$ & $1.19732 \times 10^{-10}$ & yr$^{-1}$~$\mu$G$^{-2}$~GeV$^{-1}$\\
$C_4$ & $1.358688 \times 10^{-10}$ & mas$^{-1}$ \\
$C_5$ & $6.241506 \times 10^{-17}$ & GeV~Jy$^{-1}$ \\
$C_6$ & $0.0098087$ & mas~$(1.2\times 10^{10}~{\rm m})^{-1}$\\
$C_7$ & $525949$ & minutes/yr\\
${R}_{S}$ & $1.2\times 10^{10}$ & m\\
$D_s$ & $2.523\times10^{20}$& m\\
$k_X$ & 3.39 & (factor) \\
$A_K$ & 2.46 & magnitudes\\
$A_M$ & 1.00 & magnitudes\\
\hline
\end{tabular}
\end{center}
\end{table}

\subsection{Model Implementation}\label{impl}

We implemented the Section~\ref{simple} synchrotron--SSC
mechanism  in a semi-empirical model. The model is
semi-empirical because
\begin{itemize}
\item
the opacity evolution model is derived from submm--NIR phenomenology;
\item it is based on two generic, log-normal red-noise processes drawn
  from the same random numbers according to Section~\ref{xncorr}
  for which we can provide only empirical reasoning.
\end{itemize}
One of the red-noise processes is the slow process drawn from a PSD
with the parameters $\gamma_{\rm{slow}}$ and $f_{b, \rm{slow}}$:
\begin{equation}\label{slowprodef}
S_{\rm{thin}}(t) = \exp[\sigma_{\rm{slow}} y_{\rm{slow}}(t) + \mu_{\rm{slow}}]\;\rm{Jy} \;\;,
\end{equation}
with $y_{\rm{slow}}(t)$ the slow Gaussian process and
$\mu_{\rm{slow}}$ and $\sigma_{\rm{slow}}$ the log-normal
parameters. This process represents the variability of the optically thin
part of the synchrotron spectrum and the timescales set by the
injection process and expansion cooling. We chose a log-normal PDF
because the synchrotron equations are power laws, and a log-normal
process remains log-normal under multiplication and
exponentiation. The other red-noise process is the fast process drawn from a PSD with
the parameters $\gamma_{\rm{fast}}$ and $f_{b, \rm{fast}}$:
\begin{equation}\label{fastprodef}
\nu_2(t) = \nu_{\rm{NIR}} \cdot \exp[\sigma_{\rm{fast}} y_{\rm{fast}}(t) +
\mu_{\rm{fast}}] + \max(\nu_m, \nu_{\rm{min}}) \;\;, 
\end{equation}
with $y_{\rm{fast}}(t)$ the fast Gaussian process and
$\mu_{\rm{fast}}$ and $\sigma_{\rm{fast}}$ its log-normal
parameters. This process represents the variable location of the
synchrotron cooling cutoff that is the result of the two competing
processes at the high frequency tail of the synchrotron spectrum: the
tail of the injection spectrum and the cooling through synchrotron
emission. In this case, we used a log-normal PDF as well to ensure
$\nu_2(t) > 0$. However, it is a three-parameter log-normal PDF to
account for the fact that the cooling cutoff frequency should be
larger than $\max(\nu_m, \nu_{\rm{min}})$ at all times, i.e., larger
than both the self-absorption turnover frequency and the
transition frequency from synchrotron cooling to expansion
cooling. The latter can be derived from the equation for the critical
frequency for a given electron energy
\begin{equation}
\nu_c = C_1 B E^2
\end{equation}
and the equation for the time after which an electron of initial
energy $E_0$ has cooled to $1/e$ of its initial energy 
\begin{equation}
t_{1/e} = (C_3 B^2 E_0)^{-1} (e-1) \cdot C_7 \;\;.
\end{equation}
We then define $\nu_{\rm{min}}$ as the frequency in GHz where the $1/e$
cooling time is equal to the correlation time scale of the slow process: 
\begin{equation} 
\nu_{\rm{min}} =  10^{-3}C_1 C_7^{2}\cdot \frac{f_{b, \rm{slow}}^2\;(e-1)^2}{(C_3)^2 B^3}  \;\;. 
\end{equation}
Figure~2 of \cite{1975gaun.book..211M} shows the spectrum of an
individual electron. The definition above guarantees
that the synchrotron cooling break $\nu_2$ can occur only at
frequencies that have enough time to cool through synchrotron
emission during an episode of injection, compression, and expansion
(with a typical duration of $1/f_{b, \rm{slow}}$).
We define
\begin{equation}
  \beta[\eta(t)] \equiv c^2 \cdot (C_1)^{1/2} \cdot [\eta(t) 
  \cdot C_4]^{-2} \cdot F' \times 10^{-12} \cdot C_5 \;\;,
\end{equation}
with
\begin{equation}\label{eta}
\eta(t) = \frac{\theta(t)}{B(t)^{1/4}} \;\; 
\end{equation}
and $F'$ from Equation~\ref{Fprime}.
We can now derive expressions for the combined quantities
\begin{equation}\label{kappaL}
  \kappa_{\nu} L = \nu^{-(\gamma_{e}+4)/2} \cdot \beta[\eta(t)] \cdot S_{\rm{thin}}(t) \cdot \nu_{\rm{thin}}^{(\gamma_{e}-1)/2}
\end{equation}
and
\begin{equation}\label{epsoverkappa}
  \frac{\epsilon_{\nu}}{\kappa_{\nu}} \cdot \Delta\Omega = \frac{\nu^{5/2}}{\beta[\eta(t)]} \;\;.
\end{equation}
With these and $\nu_2(t)$,
Equations~\ref{radtrans} and \ref{fluxdensity} give the
time-variable synchrotron flux density at each frequency.
Similarly, $\eta(t)$ and $S_{\rm{thin}}(t)$ give
\begin{align}
\nu_m(t)& = \,0.001\cdot\label{nu_m}\\
  &\left[C_5 \cdot \beta[\eta(t)] \cdot S_{\rm{thin}}(t)  
  \left(\nu_{\rm{NIR}} \cdot 1000\right)^{\frac{\gamma_{e}-1}{2}}\right]^{{2}/({\gamma_{e}+4})} \nonumber 
\end{align}
and 
\begin{equation}\label{S_m}
S_m(t) = S_{\rm{thin}}(t) \cdot 
  \left[\nu_m(t)/\nu_{\rm{NIR}}\right]^{-\frac{\gamma_{e}-1}{2}} \;\;,
\end{equation}
and $\theta(t)$ and Equation~\ref{SSCEq} give the
power-law section of the SSC SED.

\begin{figure}[h!]
\begin{center}
\includegraphics[scale=0.60, angle=0]{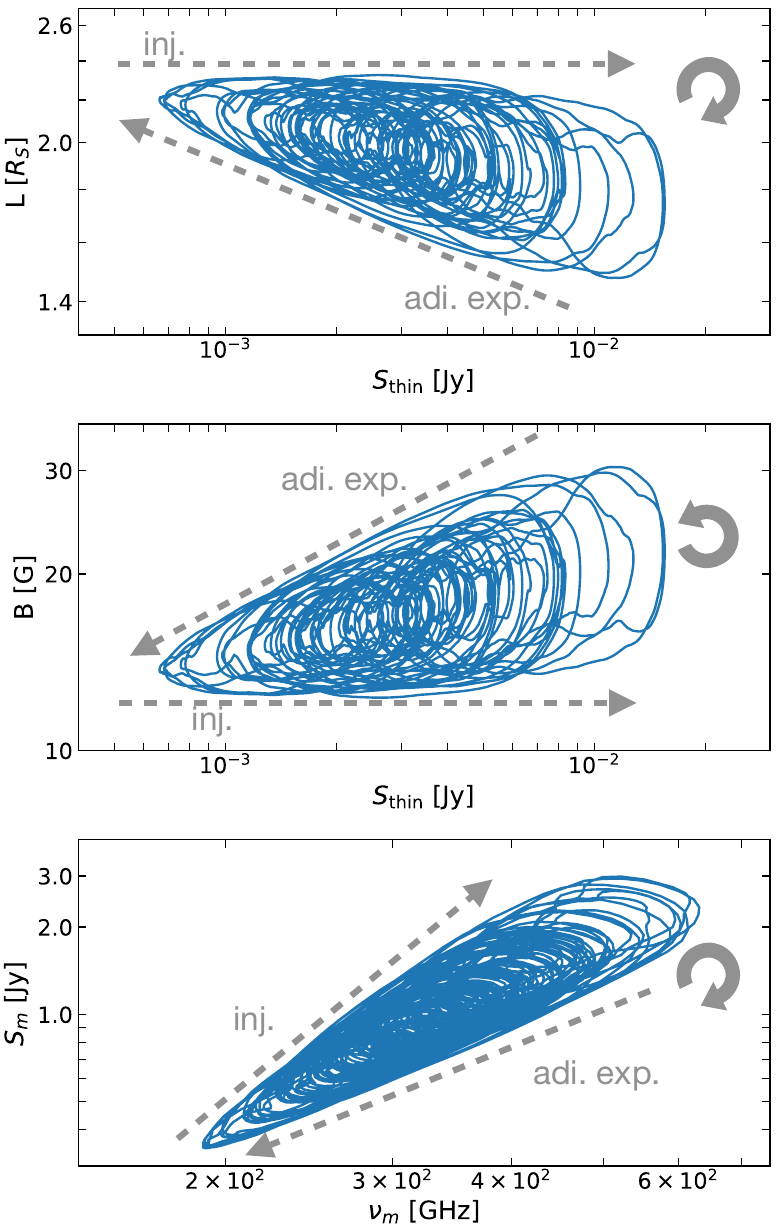}
\end{center}
\setlength{\abovecaptionskip}{-5pt}
\caption{Cyclic evolution of model parameters. Top panel shows source
  size $L$ versus optically thin synchrotron flux density (neglecting
  interstellar extinction), middle
  panel shows magnetic flux density versus the same, and bottom panel shows
  self-absorption turnover in the $(\nu_m$,$S_m)$ plane. All
  parameter time sequences (blue loops) were
  derived according to Equations~\ref{loopeqs} to
  \ref{varindtheta}. The grey
  dashed lines indicate the trajectories for pure injection
  ($\Delta\eta = 0$) and pure adiabatic expansion as labeled. The circular
arrows indicate the direction of time evolution in each panel.}\label{loops}
\end{figure}

We parameterized the cyclic source evolution model by linear functions
with variable slopes in the logarithmic $(S_{\rm{thin}},\theta)$ and
$(S_{\rm{thin}},B)$ planes (i.e., power-laws with variable indices):
\begin{align}
B(t) & = B_0 
\left[\frac{S_{\rm{thin}}(t)}{S_{\rm{thin},0}}\right]^{u(t)} \label{loopeqs}
\\ 
u(t) & = \frac{1}{2}\left(\frac{2}{2\gamma_{\rm{e}}+1}
  \right) \\
  & \cdot  \left\{\tanh\left[\frac{\dot{S}_{\rm{thin}}(t)}{2S_{\rm{thin}}(t)} \cdot \rm{min} \right]+1\right\} 
  \label{varindB}\\ 
L(t) & = L_0 \cdot
\left[\frac{S_{\rm{thin}}(t)}{S_{\rm{thin},0}}\right]^{-v(t)} \label{thetaeq}\\
v(t) & =
  \frac{1}{2}\left(\frac{1}{2\gamma_{\rm{e}}+1}
  \right) \\
  & \cdot \left\{\tanh\left[\frac{\dot{S}_{\rm{thin}}(t)}{2S_{\rm{thin}}(t)} \cdot \rm{min} \right]+1\right\} \label{varindtheta}
\;\;. 
\end{align}
(The super-dot notation means time derivative in units of
minutes$^{-1}$.)
In Equation~\ref{loopeqs} for $B$,
the power-law index $u$ was constrained to the interval of
$[0,\frac{2}{2\gamma_{\rm{e}}+1}]$. On average it is
$\frac{1}{2\gamma_{\rm{e}}+1}$ with deviations to either a
larger or smaller index depending on the fractional derivative
$\frac{\dot{S}_{\rm{thin}}(t)}{S_{\rm{thin}}(t)}$ in units of $\rm{min}^{-1}$. Similarly, the
index $v$ in Equation~\ref{thetaeq} for $L$ was constrained to
$[0,\frac{1}{2\gamma_{\rm{e}}+1}]$,  averaging
$\frac{1}{4\gamma_{\rm{e}}+2}$. $B_0$ and $\theta_0$ are scaling
parameters with units G and $R_S$, respectively. This cyclic model is
a simple empirical scenario based on the cases in
Section~\ref{opac}. The  extreme Cases~1 and~3 are
asymptotically approached for very rapid changes. Depending on
$\dot{S}_{\rm{thin}}(t)$, our
implementation allows for mixing of the different cases (e.g., slight
compression during injection, etc.) and smooth transitions between
the three steps of the source evolution. Figure~\ref{loops} shows trajectories in the
$(S_{\rm{thin}},\theta_0)$ plane, the $(S_{\rm{thin}},B_0)$ plane,
and the $(\nu_m$,$S_m)$ plane calculated according to
Equations~\ref{loopeqs} to~\ref{varindtheta}.

The dependence
of $S(E_{\rm{keV}})$ on $S_{\rm{thin}}$ for the peak flux densities
(and the flux-density minima) comes from setting  $\dot{S}_{\rm{thin}}(t) = 0$. Then
\begin{align}
\nu_m & \propto (S_{\rm{thin}})^{\frac{2}{5 - 2\alpha}} \cdot \eta^{\frac{-4}{5-2\alpha}} \;\;,\\
B &\propto (S_{\rm{thin}})^{\frac{1}{3-4\alpha}} \;\;,\\
\theta & \propto (S_{\rm{thin}})^{-\frac{1}{(6-8\alpha)}} \;\;,
\end{align}
and for $\alpha=-1$,
\begin{align}
S(E_{\rm{keV}}) &\propto (S_{\rm{thin}})^{1.86} \;\;.
\end{align}

With this model implementation, we are able to predict light curves
at submm and NIR light curves according to
Equation~\ref{fluxdensity} and X-ray light curves according to
Equation~\ref{SSCEq} and compare them via a suited distance function
to the observed light curves in all three wavelength regimes. A 
high-level block diagram for generating light curves according to this
model is shown in Figure~\ref{cartoon34}b. For the X-rays, to
convert predicted flux densities to count rates, we used the
conversion from the pileup-free count rate to the absorbed energy
flux in the 2--8~keV band  $c_{\rm{conv}} = 2.81 \times
10^{-11}$~ergs~cm$^{-2}$~s$^{-1}$~cps$^{-1}$
\citep{Yuan2015}.  To
account for interstellar extinction, we multiplied  by 3.39, the effective
correction 
derived from the luminosity calculated from $c_{\rm{conv}}$ and the
unabsorbed luminosity  \citep{Yuan2015} and used the
relation
\begin{equation}
  \Lambda(t) = \frac{S(E_{\rm{keV}})}{\rm{Jy}} 
  \frac{\nu_{8\rm{keV}}^{\alpha}-\nu_{2\rm{keV}}^{\alpha}}
  {\nu_{5\rm{keV}}^{\alpha} (1+\alpha) 
  \cdot 2.81 \cdot 3.39 \times 10^{12}}~\rm{cps}\;\;.
\end{equation}
Observed count rates were then  calculated according to
Equations~\ref{cpseq} to \ref{cpsmeaseq}, but we assumed that the
steady X-ray background is equal for all three detector modes of
\Ch\ (i.e., $\chi_{I} =  \chi_{G} = \chi_{S}$).
We used $e_I$ as the reference effective area for \Ch.

The free parameters in our analysis are listed in
Table~\ref{results}. In order to 
compare the model predictions to observed NIR flux density, we applied extinction
magnitudes $A_{{K}}$ and $A_{M}$ given in Table~\ref{const}. Because these
values have significant uncertainties, we let the fitting
modify the extinction magnitudes via
parameters $\Delta A_{{K}}$, and $\Delta A_{{M}}$.
We set parameter 
$S_{\rm{thin},0} = 0.1 \cdot \exp(\mu_{\rm{slow}} - \sigma_{\rm{slow}}^2 )$~Jy,
the mode of the flux-density distribution of $S_{\rm{thin}}$.  This lets the fitter
set $S_{\rm{thin}}(t)$ wherever $\mu_{\rm{slow}}$ and
$\sigma_{\rm{slow}}$ say it should be while not having $B(t)$
(Equation~\ref{loopeqs}) or $\theta(t)$ (Equation~\ref{thetaeq})
take on improbable values.
The exponent of $0.5$ in the
exponential term of Equation~\ref{fluxdensity} and the factor 0.5 in
the argument of the $\tanh$ in Equations~\ref{varindB} and
\ref{varindtheta} are fiducial parameters which we did not fit. A
mismatch in the exponent can be absorbed by
the parameters $\mu_{\rm{fast}}$ and $\sigma_{\rm{fast}}$, at least
where $\nu_2(t) \gg \max(\nu_m, \nu_{\rm{min}})$. Where $\nu_2(t)$
comes close to its minimum, we estimate the error to be not greater
than $10\%$. However, close to the minimum of $\nu_2(t)$, the NIR flux
density is far below the detection limit, and the variability is
entirely dominated by the measurement noise. The factor of 0.5 in the
argument of the $\tanh$ gives reasonable results, and testing showed
that the fitting results of the other parameters are not strongly
influenced by its exact value. To fit for this parameter we would
need detailed information on the covariance at several mm to submm
wavelengths and a suited distance function.
Our data are not sufficient to constrain this factor.

Finally, Equations~\ref{varindB} and \ref{varindtheta} make use of
the derivative $\dot{S}_{\rm{thin}}(t)$. We
take advantage of the fact that we are generating the mock light
curves via FFT from Fourier coefficients. To compute the derivative
from the same Fourier coefficients, we can use the fundamental
relation from Fourier theory:
\begin{equation}
F\left[\frac{{d}}{{d}x}g(x)\right] = 2\pi i f_x G(F_x) \;\;
\end{equation}
with $F$ the Fourier transform and $G(f_x) = F[g(x)]$. However, this
equation cannot simply be applied to the discrete case, and for that we
followed the method of wavenumber modification 
\citep{Sunaina_2018}. \added{This determines $\dot{y}_{\rm{slow}}(t)$ and with Equation~\ref{slowprodef} $\dot{S}_{\rm{thin}}(t)$.}

\added{In summary, the steps to create the cyclic synchrotron--SSC model are as follows:

\begin{itemize}
\item
$y_{\rm{slow}}(t)$ is used to generate $S_{\rm{thin}}(t)$ according to Equation~\ref{slowprodef};
\item
$S_{\rm{thin}}(t)$, $\nu_2(t)$, and $\eta(t)$ are used to derived the spectrum, according to Equations~\ref{kappaL} and \ref{epsoverkappa} together with \ref{fluxdensity} and \ref{radtrans} (and for the SSC Equations~\ref{nu_m} and \ref{S_m} together with \ref{SSCEq}).
\item
$B(t)$ and $L(t)$, and thus $\eta(t)$, are cyclic by construction and depend on $S_{\rm{thin}}(t)$ and its derivative according to Equations~\ref{eta} and \ref{loopeqs}--\ref{varindtheta}.
\end{itemize}
}

\subsection{The Distance Function }\label{dist}

As in Section~\ref{XPSD}, to use ABC we need a distance
function. Our distance function used several components to
guide the ABC algorithm to a valid set of posteriors:
\begin{itemize}
\item The distance between the predicted and the observed structure
  functions as defined in Equation~\ref{distdef}. We calculated the
  structure functions and distances for the three NIR, two submm (lower panel of Figure~\ref{Xsf}), and
  three X-ray datasets separately. We added these eight distances
  multiplied by normalized weights, which were the quadratic
  difference between the logarithm of the maximum value and the
  logarithm of the minimum value of each of the eight observed
  structure functions $V_j$:
\begin{equation}
D_{\rm{sf}} = 
  \frac{\sum_{j} \left[\max{\log(V_j)} - \min{\log(V_j)}\right]^2
    \phi_j}
   {\sum_{j} \left[\max{\log(V_j)} - \min{\log(V_j)}\right]^2}\;\;
\end{equation}
with $D_{\rm{sf}}$ the total distance calculated from structure
functions and $\phi_j$ the structure function distance for each
dataset. For each structure function, we used uniform weights $w_i=1$ for
all bins except for the last bins of the NIR and X-ray data (i.e.,
the bin with $\tau > 20$~minutes for the Keck data and the one
with $\tau > 128$~minutes for all other NIR and X-ray data),
which had $w_i=3$. This approach led to uniformly converging
fits at all time lags and in all bands.
(See also the discussion by \citealt{2018ApJ...863...15W},
Appendix~B.2.)
\item To help to constrain the log-normal PDF of $\nu_2(t)$, we
  determined two other quantities to describe the empirical
  distribution of observed $K$-band flux densities: the skewness of the
  distribution and the fraction of positive flux densities. Both
  values are sensitive to the white noise characteristics, and we
  determined them only for the VLT dataset, which is the larger of the
  two $K$-band datasets.  The fraction of positive flux
  densities is 0.9. This value was derived from the most recent
  analysis of the intrinsic flux density distribution from the {VLTI/GRAVITY} interferometer (\citealt{2020A&A...638A...2G}). VLTI/GRAVITY
  data show significantly less noise than the data from single-dish
  telescopes, and its 1~mas angular resolution makes it virtually
  free of source confusion. The VLTI/GRAVITY data show a median flux density of
  $1.1 \pm 0.3$~mJy dereddened, and the peak of the distribution is
  ${\sim}0.4$~mJy dereddened. While the empirical distribution of
  VLTI/GRAVITY flux densities is always positive, convolving it with a
  Gaussian noise of $\sigma_{\rm{NACO}} \approx 0.3$~mJy dereddened
  will result in a tail of $\sim$10\% negative flux
  densities.  We counted the
  fraction of  $n_{\rm{>0}}/n$ points in each VLT mock light curve
  above zero and the quadratic difference:
\begin{equation}
D_{>0} = (n_{\rm{>0}}/n-0.9)^2 \;\;.
\end{equation}
For the skewness,
\begin{align}
\rm{Sk}_{\rm{VLT, obs}} & = 
  \frac{\frac{1}{n} 
  \sum_{i=1}^{n}(S_i-\overline{S})^3}{\left[{\frac{1}{n-1}
  \sum_{i=1}^{n}(S_i-\overline{S})^2}\right]^{3/2}} \nonumber \\
&= 3.44 \;\;
\end{align}
with $S_i$ the observed flux densities and $\overline{S}$ the mean
flux density.  For each VLT mock light
curve, we calculated the skewness and the quadratic difference 
\begin{equation}
D_{\rm{sk}} = (\rm{Sk}_{\rm{VLT, sim}}-\rm{Sk}_{\rm{VLT, obs}})^2 \;\;
\end{equation}
and added the result to the distance function.
\item Following \citet[][their discussion of Case~3 in
  Section~4.4]{2018ApJ...863...15W}, we additionally used spectral information from
  simultaneous $K$- and $M$-band data. At an average (observed) flux
  density $S_{{K}} = 0.15$~mJy, \citeauthor{2018ApJ...863...15W}\
  found the ratio of (observed) NIR flux densities at $M$ and $K$ bands to
  be $\Re_{\rm{obs}} = 12 \pm 0.5$. We calculated this ratio for each
  parameter set from our mock data.  Because here we are not using
  simple log-normal PDFs for modeling the flux densities (as
  \citeauthor{2018ApJ...863...15W}\ did), we cannot derive this ratio
  analytically from model parameters. Instead, we simulated for each
  parameter set equally sampled, measurement-noise-free $K$- and $M$-band
  light curves of 10\,000 minutes duration each and determined
  $\Re(M/K, S_{{K}})$ as the average ratio over a
  suited flux-density range. The flux-density range of the
   \citeauthor{2018ApJ...863...15W}\ simultaneous datasets is
  $\sim$0.07--0.27~mJy. However, it is difficult to determine
  background flux-density levels for single-dish data. After
  comparing the median of the NACO flux-density distribution with the
  newest VLTI/GRAVITY study (\citealt{2020A&A...638A...2G}), we adopted
  a background correction factor based on a flux-density range of
  [0.0--0.17]~mJy. We then defined the distance
\begin{equation}
  D_{\rm{sp}} = 
  \left[\Re(M/K, S_{\rm{K}}) - \Re_{\rm{obs}} + \mathcal{N}(0,0.5^2)\right]^2 \;\;,
\end{equation}
with $\mathcal{N}(0,0.5^2)$ a normal random number to account for the
uncertainty in $\Re_{\rm{obs}}$.
\end{itemize}

The final distance function was
\begin{equation}
D_{\rm{tot}} = D_{\rm{sf}} + 10 \cdot D_{>0} + D_{\rm{sk}} + D_{\rm{sp}}/50 \;\;.
\label{Dtot}
\end{equation}
The factors of 10 and 50 are empirical and ensured that none of the
additional distance terms starts to dominate $D_{\rm{sf}}$. This
distance function does not include metrics to quantify the degree of
correlation between wavelengths, and we restricted the $230$~GHz
structure function to time lags $<$50~minutes. We will come back to these
two points in the discussion.

\begin{figure}[h!]
\begin{center}
\includegraphics[scale=0.5, angle=0]{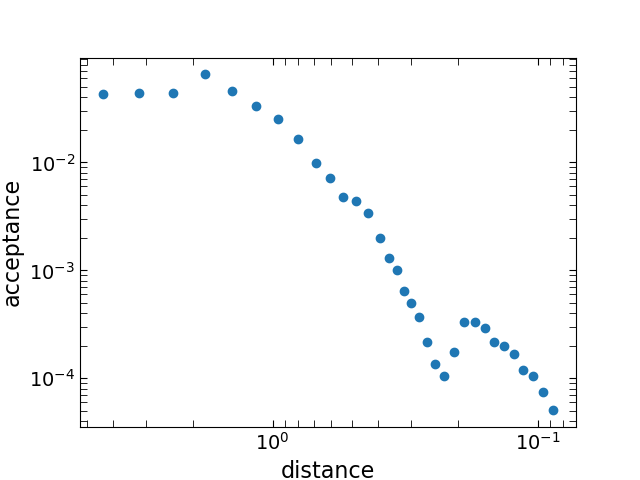}
\end{center}
\setlength{\abovecaptionskip}{-5pt}
\caption{Performance of the ABC sampler. Ordinate is the
  acceptance rate, and abscissa 
  is the 45\% quantile of all distances  ($D_{\rm tot}$, Eq.~\ref{Dtot}) of the
  particular iteration. Towards the smallest distances the sample
  found $<$1 in 10\,000 light curves that satisfied the distance
  threshold.}\label{acceptance}
\end{figure}

\begin{figure*}[h!]
\begin{center}
\includegraphics[scale=0.15, angle=0]{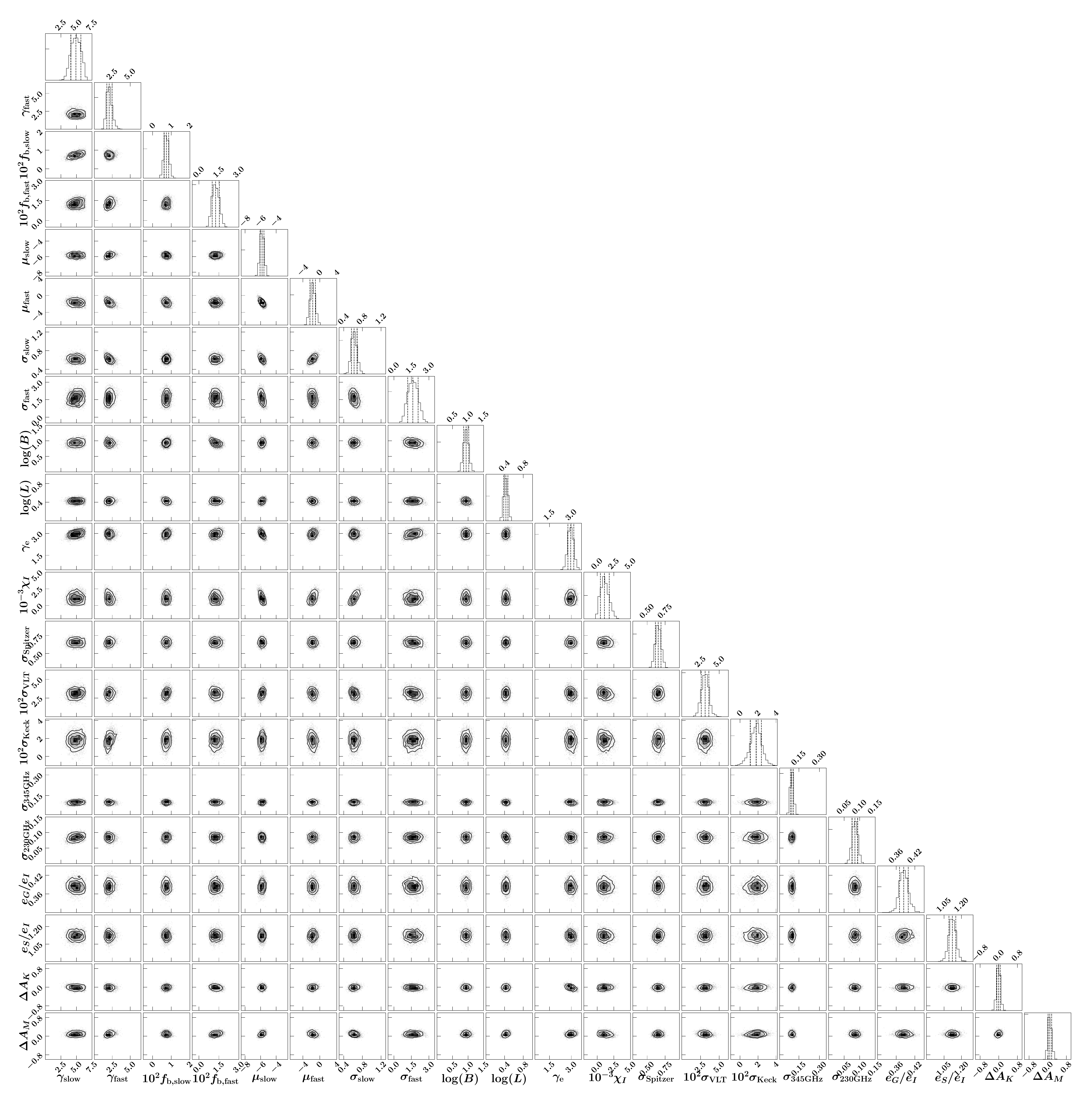}
\end{center}
\setlength{\abovecaptionskip}{-5pt}
\caption{Results for the Bayesian structure function fit for the
  synchrotron--SSC model. Contours show the joint posterior
  probability density for each parameter pair, and panels along the
  right upper edge show histograms of the marginalized posterior of
  each parameter defined in Table~\ref{results}. For each histogram,
  the dashed lines mark the 16\%, 50\%, and 84\%
  quantiles. Instrumental $\sigma$ parameters are as observed, not
  extinction-corrected.}\label{corner_plot_rm}
\end{figure*}

\section{Results}\label{res}

We have implemented the described model and distance function in our
C++ code. The analytic nature of the model allows us to calculate
many realizations as large as the observed dataset in reasonable
time. However, because of the iterative nature of the ABC and the
need for a sufficiently large particle system, we had to run our code
on the VLBI compute cluster of the Max-Planck-Institut f\"ur Radioastronomie.
The model and the distance function are the result of more
than fifty test runs that all had to be executed over several days on
200--400 cores (10--20 nodes).  Over the course of two years, we estimate
the total CPU time (including all runs necessary for testing,
developing, and implementing the model) to be ${\sim}10^6$~hours. To
analyze the results, we also implemented a Python version of
our model. This version allows us to quickly visualize and diagnose
the ABC results, including a tool for generating animations of the
timing of the SED, the light curves, and the derived physical
parameters.

\begin{figure}[h!]
\begin{center}
\includegraphics[width=0.4\textwidth, angle=0]{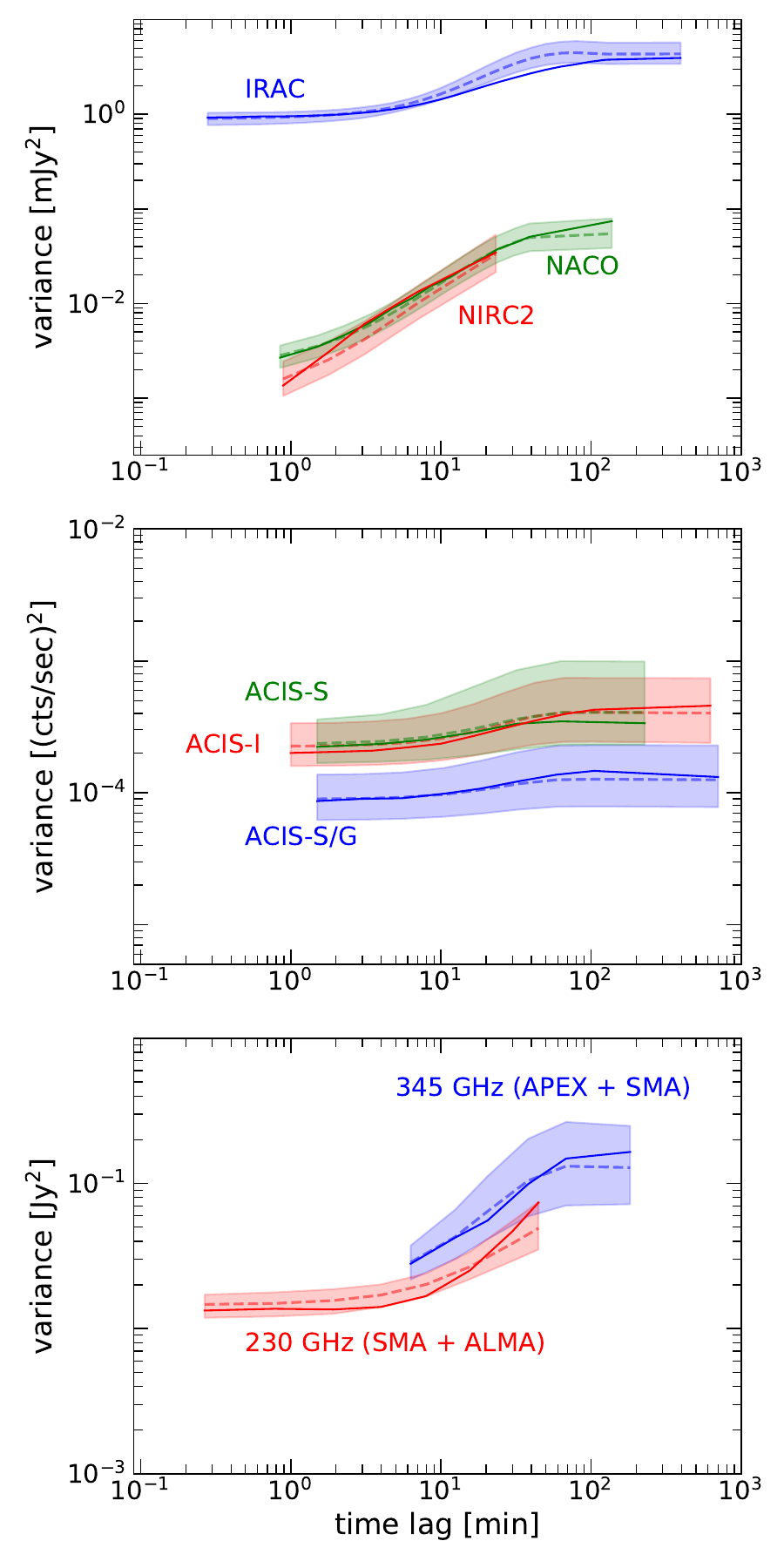}
\end{center}
\setlength{\abovecaptionskip}{-12pt}
\caption{Structure functions (Eq.~\ref{eq:strfundefa}) for the eight
  datasets. Panels from top to bottom show NIR, X-ray, and submm, all
  as observed, not corrected for extinction or pileup.
  Solid lines show the observed SFs (as presented in
  Figure~\ref{Xsf}), and corresponding dashed curves show the median
  of 10\,000 SFs for the respective mock
  datasets. The shaded envelopes denote the model $68\%$ credible
  intervals for each time lag. 
  The details of generating
  the SFs, including the choice of time-lag ranges,
  are described in Section~\ref{dist}. The slope of each SF
  relates to the slope of the underlying PSD but also
  depends on the underlying white noise level.}\label{sf_pred}
\end{figure}

\begin{figure}[b]
\begin{center}
\includegraphics[scale=0.55, angle=0]{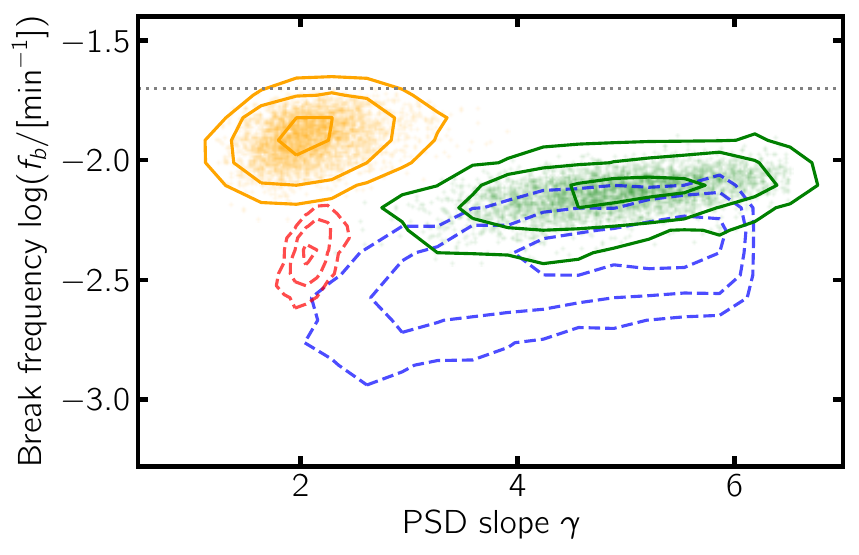}
\end{center}
\setlength{\abovecaptionskip}{-5pt}
\caption{Credible contours (68\%, 95\%, 99\%) for the parameters
  $\gamma_{1}$ and $f_{b}$. Green and orange show the PSD parameters
  of our synchrotron--SSC model (Section~\ref{radmodel} and
  Table~\ref{results}), green showing the slow process (identified
  with X-rays) and orange the fast process (identified with NIR).
  Blue contours are for our X-ray-only analysis
  (Section~\ref{xncorr}), and red  contours are for the NIR analysis
  of \cite[][their Case~3]{2018ApJ...863...15W}. All posteriors have
  been marginalized over all other
  parameters. The horizontal grey dotted line marks the frequency that corresponds to a periodic timescale of 50 minutes, which is the orbital timescale of the astrometric motion measured by \cite{2018A&A...618L..10G}.}\label{X_NIR_contours_overv}
\end{figure}

Table~\ref{results}  lists the priors of all parameters. Priors for all
physical parameters  except $B_0$ and $L_0$ were flat and wide.
For  $B_0$ and $L_0$, we used log-flat
priors.
Parameters for the log-normal red noise processes were also flat and
wide.  We labeled these parameters 
``slow'' and ``fast,'' but we used the exact same priors for
both PSDs. In other words, we did not force the PSDs to take on the filter
relation discussed in Section~\ref{xncorr}.
For parameters specifying instrumental characteristics, we used
narrow Gaussian priors.

The final ABC run presented here is the result of 33 iterations, 32
with a particle size $n=500$ and the last with
$n=5000$. (The model computations had to use fewer particles than the
X-ray-only model because submm and NIR light curves had to be
computed in addition to X-rays.) Figure~\ref{acceptance} shows the evolution
of the distance and the acceptance rate. We reached
$D_{\rm{tot}}<0.09$ at an acceptance rate of
$5\times10^{-5}$.  (This means the final iteration generated $10^8$~light
curves in order find 5\,000 it could accept.)
The local maximum in the acceptance near
$D_{\rm{tot}}<0.2$ is a hint that at this stage,
the algorithm still ``learned'' relevant information.

The posterior estimates of all parameters converged to
well-constrained, peaked distributions. Figure~\ref{corner_plot_rm}
shows the posteriors and pairwise correlations of all parameters.
The medians and 1$\sigma$ credible
intervals are listed in Table~\ref{results}. Figure~\ref{sf_pred}
shows the structure functions with the corresponding 1$\sigma$
envelopes drawn from the posterior. The fit
describes the observed structure functions well with only the structure
function of the \Sp/IRAC data lying partly outside the 1$\sigma$
envelope.  This discrepancy is not statistically significant, and
part of it
could be caused by our exact choice of synchrotron cooling
cutoff (Equation~\ref{fluxdensity}). The
true shape of this cutoff depends on fine details of the process, e.g., the injection spectrum
and the exact cooling mechanism. These depend on the geometry
of the magnetic field lines and other specifics. The shape we chose
predicts a slightly higher variability in the middle range of time lags
at 4.5~$\micron$. In the absence of more specific information on the
coevolution of the spectral indices in the IR bands, it is
not possible to determine a more realistic scenario.

The physical parameters of the model are tightly constrained:
$\gamma_{e} = 2.95 \pm 0.2$, $B_0 =
8.6^{+1.8}_{-1.4}$~G, and $L_0 = 2.7 \pm 0.3~R_S$.
Figure~\ref{X_NIR_contours_overv} shows the posteriors of
the PSD parameters.
The PSD slope of the slow process comes out a little steeper in the combined model than
when analyzing the X-ray data alone but not significantly so.  A
bigger difference is between the fast and slow process timescales and the timescale from the earlier NIR analysis.  
However, both timescales ($135^{+26}_{-20}$~minutes and $82^{+28}_{-16}$~minutes, respectively) are consistent 
with the earliest report of the NIR break timescale of $154^{+124}_{-87}$ by \cite{2009ApJ...694L..87M}.

\begin{figure}[h]
\begin{center}
\includegraphics[scale=0.5, angle=0]{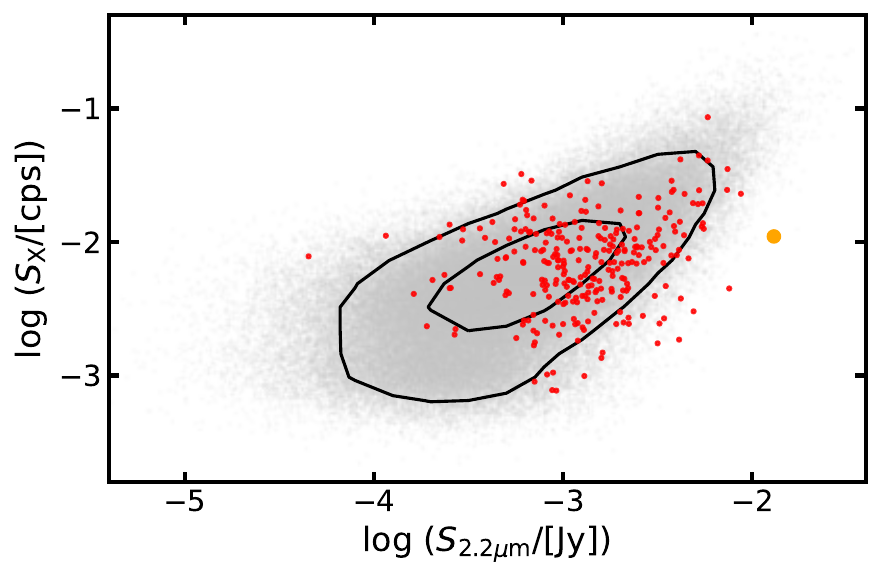}
\includegraphics[scale=0.5, angle=0]{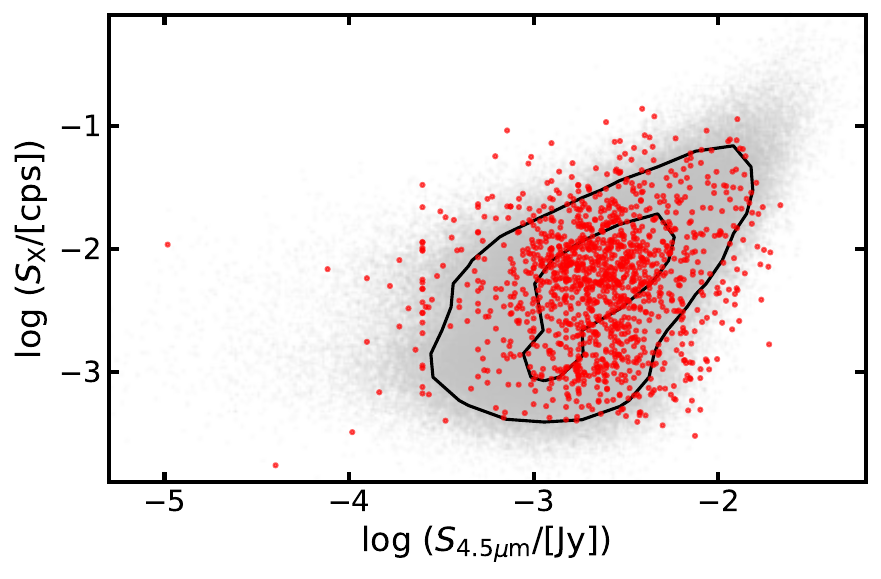}
\end{center}
\setlength{\abovecaptionskip}{-5pt}
\caption{Probability distributions of simultaneous NIR and X-ray
  measurements. Grey points show distributions from 1000 mock
  multi-wavelength light curves (Section~\ref{res}), and lines show
  their 68\% and 95\% credible contours. Red points show real
  measurements from simultaneous observations. All values, X-ray and
  NIR, mock and real, are intrinsic with the real data corrected for
  extinction by the canonical factors given in
  Table~\ref{const}. Upper panel shows $[S_{{K}}, S_X]$
  with data from \citet{2018ApJ...864...58F}. The large orange data
  point shows the peak value of the NIR flare and its X-ray
  counterpart observed by those authors (their Table~3). The even
  more extreme event observed by \citealt{2019ApJ...882L..27D} had no
  simultaneous X-ray observations.  At a corrected
  $S_{2.2~\micron}=60$~mJy, that observation would be just to the
  right of the plot area. The bottom panel shows $[S_{{M}},
  S_X]$ with observed data from
  \cite{2019ApJ...871..161B}.  To be consistent with the presentation
  of the X-ray data in the respective publications, the X-ray count
  rates are shown in 600~s bins for the upper panel and 300~s bins
  for the lower.  Even in those bin widths, X-ray counts can be
  zero. In order to show a continuous posterior in log space, we
  added $\chi_{S} = 0.001$~cps to the count rate and a Gaussian random variable
  of $\sigma = 0.18$ to the values of the logarithm to dither both the mock and observed X-ray
  data.}\label{x_ir_corr}
\end{figure}

\begin{figure}[h]
\begin{center}
\includegraphics[scale=0.5, angle=0]{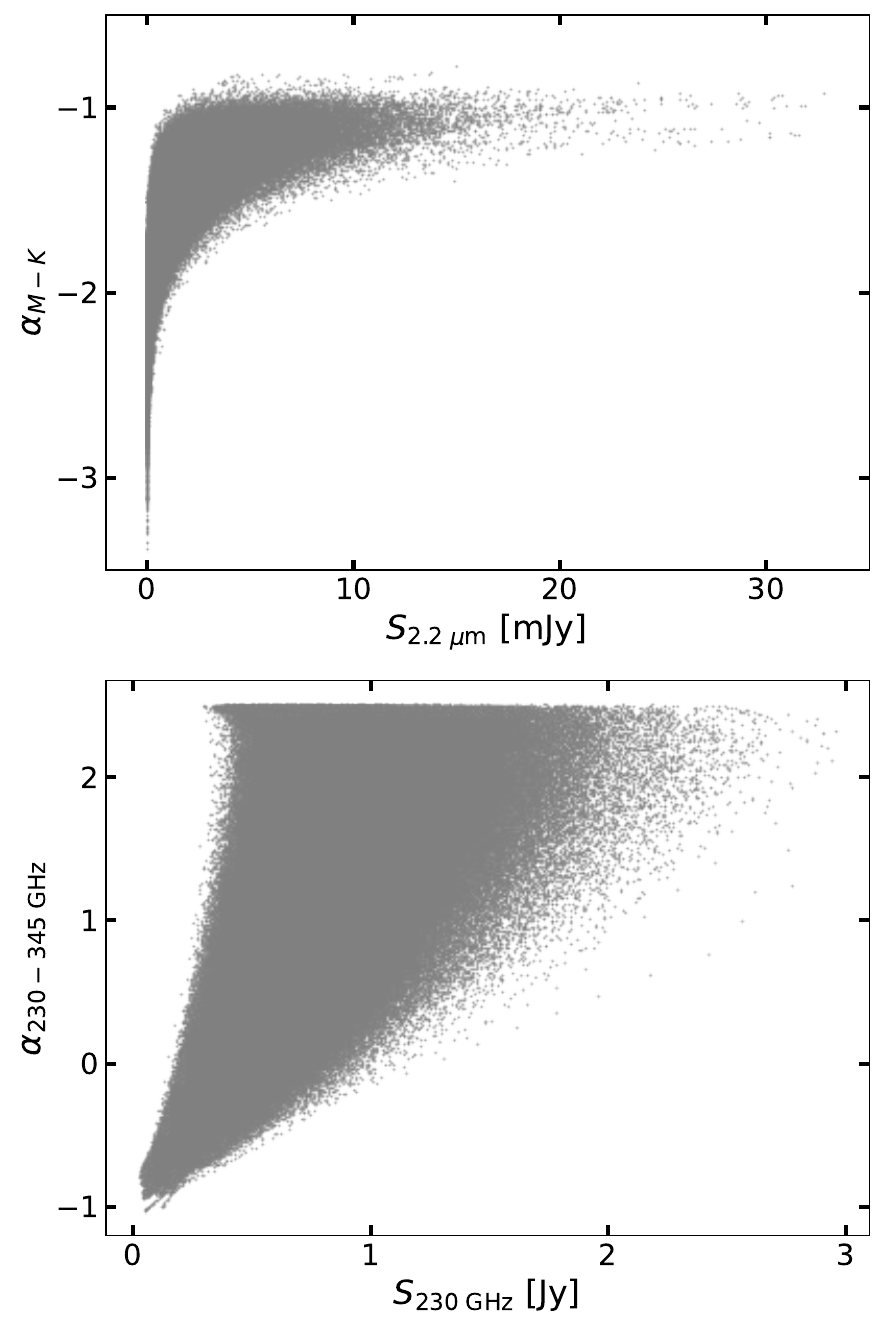}
\end{center}
\setlength{\abovecaptionskip}{-5pt}
\caption{Posterior distributions calculated from mock light curves
  (Section~\ref{res}). Points in the upper panel show NIR spectral
  index versus flux density $S_{\rm{2.2~\micron}}$, and those in the lower panel
  shows the submm spectral index versus
  $S_{\rm{230~GHz}}$. Quantities shown are intrinsic, and observations have to be corrected for
    extinction before comparison.}\label{specinds}
\end{figure}

In order to determine posteriors of observable and physical
quantities, we used the model to 
construct 1000 mock light curves of 700~minutes duration in all
bands.  The results are given in Table~\ref{results}.  The mock light
curves were then ``observed'' at uniform intervals, unlike the real data.
\replaced{This may give the observed data higher variance than predicted by
the mock data.}{This may give the mock data lower variance than the real data.}
Figure~\ref{x_ir_corr} compares the posterior of
$(\log{S_{\rm{NIR}}}, \log{S_{{X}}})$ to
observed flux-density pairs.
Figure~\ref{specinds} shows  calculated NIR and submm spectra
indices. The model NIR spectral index becomes 
flatter towards brighter flux densities with a maximum around
$\alpha_{\rm{NIR}} = -1$, while the submm spectral index is a wide
scatter cloud terminated sharply at $\alpha_{\rm{submm}} = +2.5$.
These are as expected for a synchrotron--SSC source.

\begin{figure}[h]
\begin{center}
\includegraphics[width=.45\textwidth]{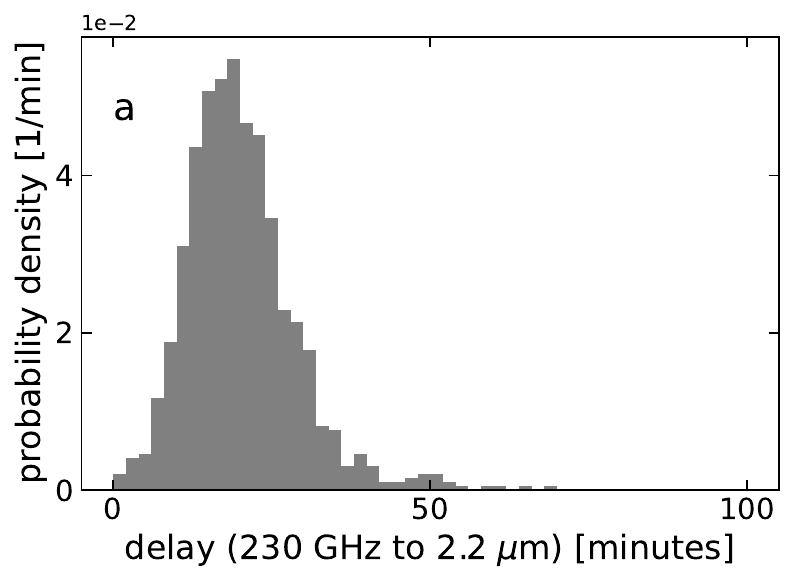}
\end{center}
\setlength{\abovecaptionskip}{-5pt}
\caption{Posterior distribution of time lags from NIR to submm from mock data, positive means NIR
  leads.}\label{post_delay}
\end{figure}

\begin{figure*}[h]
\begin{center}
\includegraphics[width=1.0\textwidth]{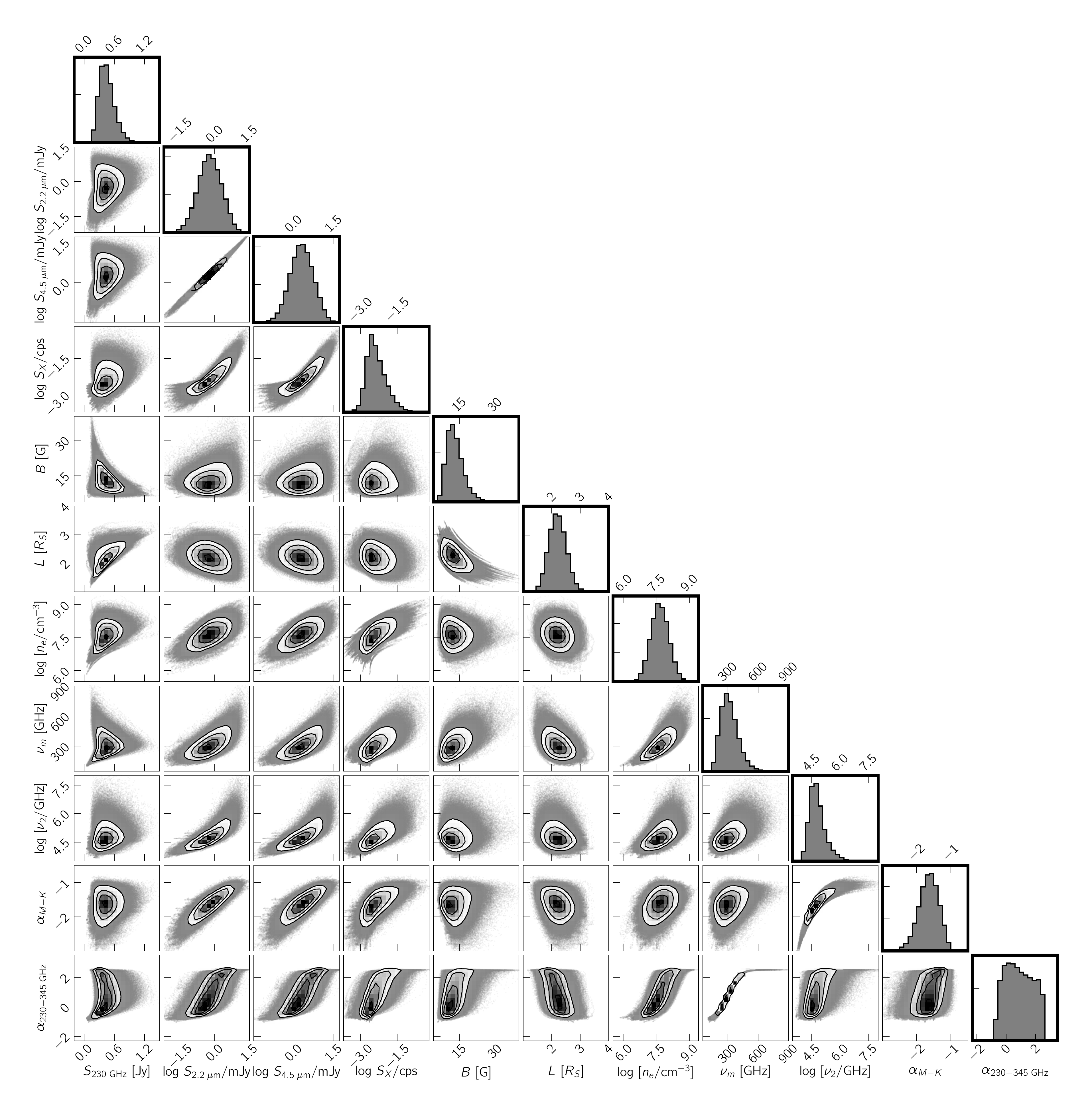}
\end{center}
\setlength{\abovecaptionskip}{-5pt}
\caption{Posterior distributions of flux densities and derived parameters for the
  synchrotron--SSC model (Section~\ref{res}) calculated from the mock
  light curves. These posteriors represent both the variance
  propagated from the posterior as well as the variance from the
  variability. The large skewness of the NIR flux-density
  distribution biases the parameters towards small NIR fluxes.
  Panels show histograms and pairwise correlations of the 230~GHz flux density,  the 2.2~$\micron$ and 4.5~$\micron$ flux densities, the X-ray flux density,  the magnetic flux density, the source size in units of $R_S$,  the density of high-energy electrons in the source volume, the frequency of maximum synchrotron flux density,  the frequency of
  cooling cutoff of synchrotron spectrum, the intrinsic 4.5 to 2.2~\micron\ 
  spectral index, and the 230 to 345~GHz spectral
  index.}\label{posteriors}
\end{figure*}

Figures~\ref{post_delay} and \ref{posteriors} show the posterior distributions of physical
and observable parameters based on the models and mock light
curves. The delay (Figure~\ref{post_delay}) was
computed from the maximum of the cross-correlation function of the
230~GHz flux densities with $S_{\rm{thin}}$ for each 700~minute light
curve. We accepted only delays with peaks at least a factor of 5 higher
than the standard deviation of the cross-correlation function.  The
high-energy electron density of Figure~\ref{posteriors} was calculated according to:
\begin{align}
n_0(t) &= \frac{12 \cdot S_{\rm{thin}}(t)}{\pi \theta^3 \cdot d}
\frac{(\nu_{\rm{NIR}}/C_1)^{\frac{\gamma_{e}-1}{2}}}{C_2 \cdot
  B^{\frac{\gamma+1}{2}} G'(\infty,0,\gamma_{e}) }
\;\;,  \\ 
n_{e}(t) &= \frac{n_0(t)}{(1-\gamma_{e})} \cdot \left\{\left[\frac{\nu_2(t)}{C_1
      \cdot B}\right]^{(1-\gamma_{e})} - E_{\rm{min}}^{(1-\gamma_{e})}\right\} \nonumber \\
& \mathrm{for}\;\; \gamma_{e} > 1 \;\;. 
\end{align}
All other quantities were calculated according to the equations in Section~\ref{impl}.

The synchrotron--SSC model gives reasonable values for physical and
observable parameters.
The submm to NIR time delay (Figure~\ref{post_delay} ) is a positively skewed distribution in the interval of
[0,100]~minutes and peaks at ${\sim} 20$~minutes.  Figure~\ref{posteriors} shows other parameters. The magnetic flux density $B$
varies between 6 and 30~G and peaks around 13~G. The source diameter
$L(t)$ is a symmetric distribution around $2.2~R_{S}$
with values between 1.3 and 3.4~$R_{S}$. The synchrotron electron density $n_{e}$ is
distributed around $4\times10^{7}$~$\rm{cm}^{-3}$ and varies by somewhat
more than an order of magnitude in both directions. The spectrum
shows a self-absorption turnover $\nu_m$  most often near 280~GHz but
ranging between 100~GHz and 750~GHz. The cooling cutoff $\nu_2 \sim
3 \times 10^4$~GHz and rarely goes below $5 \times 10^3$~GHz or
exceeds $10^6$~GHz. The distributions of $\alpha_{\rm{NIR}}$ and
$\alpha_{\rm{submm}}$ are complementary to Figure~\ref{specinds}.
Table~\ref{results} 
lists medians and 1-$\sigma$ credible
levels for  each of these quantities.

Even two uncorrelated light curves will sometimes show peaks near the
same time.
The ``false alarm probability''
gives the probability of finding such spurious correlations
at any time lag due to the white noise and the autocorrelation (the
red noise) in each of the two bands.  Figure~\ref{cross_corr} shows  the 95\% false alarm probability
for the submm--NIR light curves.
This was derived by calculating 
the DCC of the 230~GHz and 4.5~\micron\ mock data in the
same way as for
the observed data but using pairs of 700~minute light curves that do
not belong to each other. By modeling five datasets with
the cadence of the observed data, we  generated 200 DCC
functions from our 1000 light curves. At each time lag we then
determined the 95\% quantile.

\begin{figure}[h]
\begin{center}
\includegraphics[scale=0.42, angle=0]{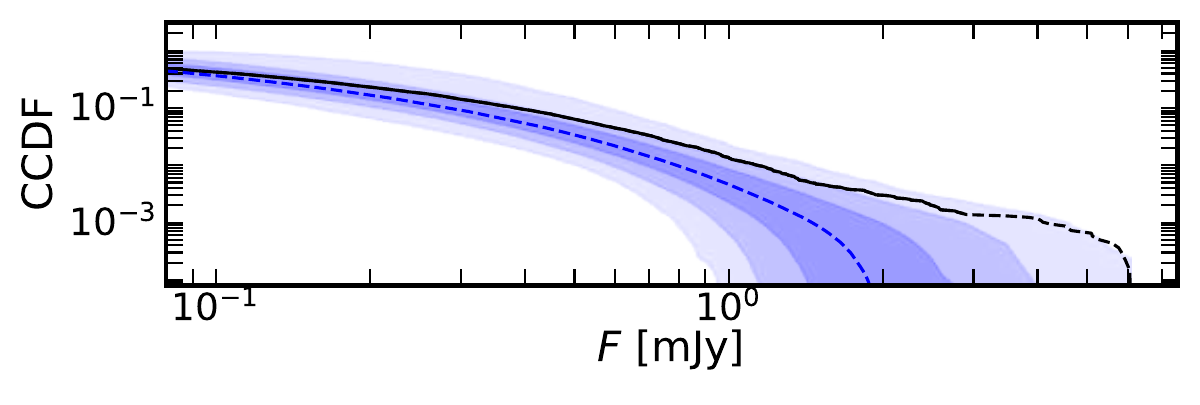}
\end{center}
\setlength{\abovecaptionskip}{-5pt}
\caption{Cumulative distribution functions (CCDFs) of Sgr~A* 2.12~\micron\
  flux densities (as observed, not corrected for extinction). The black upper line shows the CDF observed by the VLT
  and Keck with the dashed section  indicating the CDF needed to match
  the single largest flux-density
  ever observed (\citealt{2019ApJ...882L..27D}). The lower
  dashed blue line shows the median CDFs from the mock light curves. Shaded
  areas show $68\%$, $95\%$, and 99\% credible intervals
  (Table~\ref{results}).}\label{NIR_CDF}
\end{figure}

Figure~\ref{NIR_CDF} shows the posterior of complementary cumulative distribution functions (CCDF)
of $K$-band flux densities in comparison to the empirical CCDF from
NIRC2 and NACO data. For all fluxes, the model is
3$\sigma$ consistent with the observed data, and up to about 20~mJy
(derredened) it is 2$\sigma$ consistent. However, 
the model does not predict the brightest flux densities to be as
frequent as we observe them.

\begin{figure*}[h]
\begin{center}
\includegraphics[scale=0.42, angle=0]{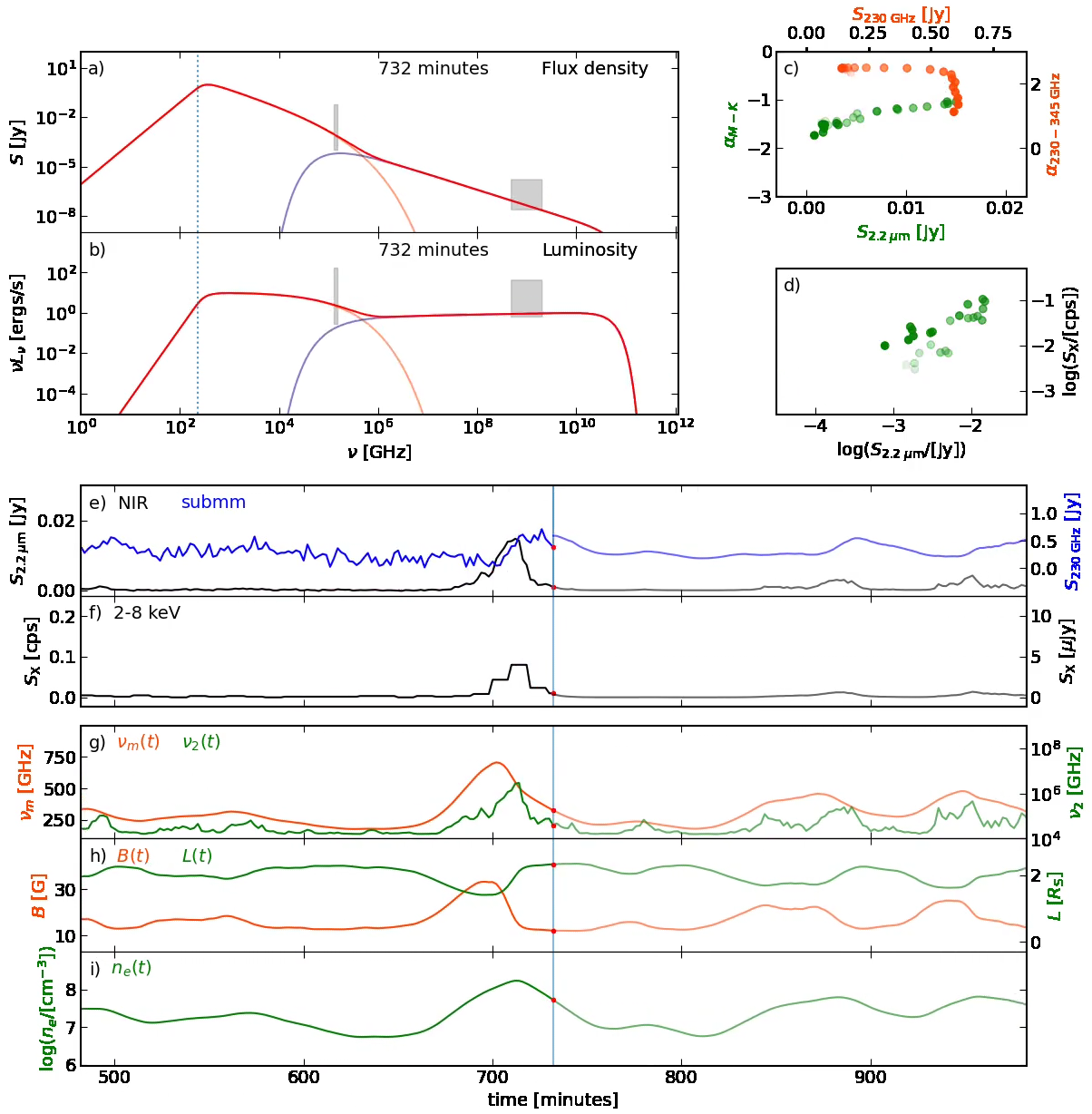}
\end{center}
\setlength{\abovecaptionskip}{-5pt}
\caption{Snapshot of the time evolution of the SED (generated from a mock data realization from posterior
  particle 1655) generated with our Python package
  (Appendix~\ref{A1}).  Panels a) and b) show the observed SED at
  minute 732 of the simulated evolution, a) in flux density and b) in luminosity units. Orange
  lines show the synchrotron emission, blue the SSC emission, and red
  their sum. The dashed line marks $\nu = 230$~GHz, the short grey
  solid line the typical range of NIR variability when the source
  flux is above the detection limit, and the grey rectangle the
  typical range of peak 2--8~keV flux densities.
  Panel c) shows spectral index versus flux density for minutes 683
  to 732 of the mock light curve with the earlier points being more
  transparent. Green shows NIR, and orange shows submm.  Panel d)
  shows X-ray flux ``as measured'' (cps) versus $S_{2.2~\micron}$,
  again for minutes 683 to 732 with the earlier points being more
  transparent.
  Panels e)--i) show time series of various parameters as
  labeled. Left and right ordinate labels are color-coded to match
  the quantities plotted. The vertical line marks minute 732.
  Points to the right of that are shown as intrinsic values, while in
  panels e)--g), points to the left are shown ``as measured,'' i.e.,
  with measurement noise added.  (This is not applicable to panels h
  and i.)  Red points represent a ``measurement'' at minute
  732. The X-ray count rate is averaged over 600 s bins.
\added{An animation of this figure is available online. The animation is 5:33 minutes long and shows 2250 minutes of source evolution.  Panels a) and b) show the up and down of the synchrotron and SSC components of the spectrum as well as changes in the self-absorption turnover frequency and the fluctuation of the cutoff frequency with $\gamma_{\rm{max}}$.  Panels c) and d) show the evolution of 50-minute-long traces of the spectral index and X-ray flux density vs.  NIR flux density.  Panels e) to i) show graphs of time series of the various quantities that run with time from right to left.}
 }\label{anim}
\end{figure*}

Finally Figure~\ref{anim} shows a snapshot of our animation.
The animation is the most concise way to
illustrate the properties of our model. However, presenting model
light curves this way  is based on picking a particular
parameter set from the posterior. It is the nature of the Bayesian
approach that this posterior includes particles that might not
generate one or another aspect of the observed data because, after
all, the observed data could be an unlikely realization of the
underlying process. In order to allow the reader to become more
familiar with the phenomenology of the model and the presented
posterior, we are publishing the particle system (the ``chain'' of parameters),
the observed data, and our Python implementation of the model and of the
animation code. The details of the code
and the data repository are described in  Appendix~\ref{A1}. 

\section{Discussion}\label{discussion}

\subsection{Fitting results}

The ABC algorithm results in constrained distributions of all
parameters and describes the structure functions of nearly all datasets
within the 1$\sigma$ envelopes. The only exception
is \Sp/IRAC, where parts of the structure
functions show slightly
less variability than the model at  timescales $\sim$40~minutes. This
deviation is not significant. The values of all parameters
are reasonable and consistent with previous analyses (e.g.,
\citealt{2012A&A...537A..52E}) and prior knowledge:  magnetic flux
density of $5<B<30$~G, source sizes
${<}4~R_{S}$, the upper limit derived from 1.3~mm VLBI observations (\citealt{2008Natur.455...78D,2018ApJ...859...60L}),
and  $n_{e}=4\times 10^7$~{cm}$^{-3}$, comparable
to \replaced{the solar corona}{the densities derived from the radio to submm luminosities (e.g., \citealt{2019ApJ...881L...2B})}.  The effective collection area of each of
the \Ch\ modes and the white noise levels in the NIR and submm are
consistent with independently determined values. The quiescent count
rate  $\chi = 1.1^{+0.8}_{-0.5} \times 10^{-3}$~cps is
identical with the background count rate of comparison apertures
close to Sgr~A*  \citep{Yuan2015}. However, \citeauthor{Yuan2015}\ found
the sum of background count rate,
truly quiescent level, and undetected variability $\chi_{\rm{I}}
\approx 4.8 \times 10^{-3}$~cps, $10\%$ of which they attributed to
undetected, weak flares. In contrast, in our analysis all X-ray
photons other than the expected background rate can be attributed to
the SSC process. At times of low activity, this  appears
quiescent.

\subsection{Timing and cross-correlation properties}

The initial idea of two correlated processes related by a low-pass
filter is consistent with our final values for $(\gamma_{\rm{slow}},
f_{b,\rm{slow}})$ and $(\gamma_{\rm{fast}}, f_{b,\rm{fast}})$ as
shown by
Figure~\ref{X_NIR_contours_overv}. \replaced{While the contours of our
pre-analysis with generic log-normal flux-density distributions
overlap, the corresponding contours of the processes in the
synchrotron--SSC model do not.}{
In section~\ref{XPSDsub} and \ref{xncorr}, the NIR and X-ray contours for the PSD parameters were estimated entirely independently, based on the assumption of a log-normal distribution of flux densities in both cases. No relation between both bands was assumed, and no characterisation of the co-variance informed the fit. Section~\ref{radmodel} re-analysed the NIR and X-ray data with additional constraints from the submm and a simultaneous fit to the autocorrelation of each band. Even then, there was no direct characterization of the co-variance (i.e., simultaneous data) between the NIR and the X-rays. The key difference was linking the probability distribution of flux densities and the PSDs of all bands by the physical synchrotron--SSC model described in section~\ref{impl}. In this second step, the slow and fast processes show clearly separated contours in the break frequency vs. slope plane, as shown in Figure~\ref{X_NIR_contours_overv}.
}
This is not the
result of the priors, which are equal for both PSDs. The break
frequencies of both processes are systematically higher than in the
pre-analysis because of their dependence on the underlying flux-density
model, which  is given by the radiative model. In particular, the
fast process is marginally consistent with a timescale of 50 minutes,
which is the orbital timescale of the astrometric motion measured by
VLTI/GRAVITY (\citealt{2018A&A...618L..10G}).

The consequence of the low-pass filter is the decorrelation of the two
processes towards higher frequencies. We can provide some qualitative
reasoning for this behavior. The variability is caused by injection
of non-thermal electrons whose energy distribution
is truncated at some maximum energy $E_{\rm{max}}$.  This cuts off
the synchrotron spectrum at a frequency close to the NIR. The
injected electrons are subject to both expansion cooling and synchrotron
cooling. The former affects the entire spectrum, while the latter (in the
time frame of injection and expansion cooling) affects only the
highest frequencies.  The
timescale of expansion cooling depends on the rather low expansion
velocity, but the synchrotron cooling timescale at frequencies
above the NIR is mere seconds to minutes. Therefore rapid fluctuations of the
injection process will be tracked by NIR emission, but the submm emission
will track only the long-term average. It should be possible to
derive a time-dependent model of the injection process from our
semi-empirical variability model, but this is beyond the scope of our
analysis.

The distance function used for our ABC implementation does not
use any measure of correlation between the bands. Instead, the
correlation is naturally built into the model. The exact
phenomenology of the correlation, however, depends on the parameters
that the ABC algorithm finds in the attempt to describe the structure
functions of all bands. Figure~\ref{x_ir_corr} shows a scatter plot
of the posterior NIR and X-ray flux-density pairs. The  observations are
in good agreement with the posterior with the exception of a rare, very high
$K$-band flux density at a low X-ray level 
\citep[][their Figure~3]{2018ApJ...864...58F}. Estimating the
probability of this event is difficult, given the different cadences
of the mock and real light curves, but the event is a hint that the real
source is more complicated than our purely cyclic model.

\cite{2015ApJ...799..199N} pointed out that the relation between NIR
and X-ray fluxes is expected to be mildly non-linear. Assuming power-law
distributions for the flux-density distributions in the NIR and the
X-ray, they derived the exponent of the power-law dependence of X-ray
flux densities on NIR flux densities (their Equations~8--11). In the
case of a power-law index of ${\sim} 2$ for the X-ray distribution and
${\sim} 4$ for the NIR
(\citealt{2012ApJS..203...18W,2015ApJ...799..199N}), the dependence
should be $S(E_{\rm{keV}}) \propto S_{\rm{NIR}}^3$. Our simple model predicts
an exponent between 1.7 and~2.0 (Section~\ref{opac}). The higher
non-linearity in the relation of observed NIR flux densities to
simultaneously measured X-ray flux densities can, however,
be explained by the exponential cooling cutoff
(Equation~\ref{fluxdensity}).

In synchrotron--SSC
models, the X-rays have the same spectral index as the optically thin
part of the synchrotron spectrum. In their careful analysis of the
X-ray spectral slope, \cite{2017MNRAS.468.2447P} found a photon index
$\Gamma = 2.27\pm 0.12$, which corresponds to a spectral index of
$\alpha_{X} = -1.27\pm 0.12$,  about 1$\sigma$
consistent with our $\alpha = -0.98\pm0.1$. \cite{2019ApJ...886...96H} analyzed the two brightest X-ray
flares detected so far and found $\Gamma = 2.06\pm 0.14$ and $\Gamma = 2.03\pm 0.27$, respectively, 
again consistent with our results.

\cite{2019ApJ...871..161B} offered another way to investigate the NIR
to X-ray correlation. Their 
cross-correlation analysis of the simultaneous \Ch\ and \Sp\
data  found  a tendency for the X-ray peaks to
precede the NIR. However, \citeauthor{2019ApJ...871..161B}\ could not
claim a statistically significant delay given the large
uncertainties. \citeauthor{2019ApJ...871..161B}\ also presented an
overview of earlier results, many of which showed
the X-ray to follow the NIR but again without high significance. Our 
700~minute mock light curves show
no significant delay between the X-ray and NIR  bands.

Our model predicts correlation between the NIR and submm light curves
as well. We  quantify this in terms of
the cross-correlation  between 
$S_{\rm{thin}}$ and the corresponding 230~GHz mock light curves
as shown in 
Figure~\ref{posteriors}a. Observed delays range from
almost synchronous variability (as observed by
\citealt{2018ApJ...864...58F} and in this paper in the case of the
345~GHz SMA data from 2014 Jun~17) to delays up to 90~minutes (as found
by \citealt{2008A&A...492..337E}). The model distribution peaks at
${\sim}22$~minutes with a FWHM of ${\sim}10$~min, which corresponds nicely to
the result we obtained from our ${\sim}48$~hours of synchronous NIR and
230~GHz data. The cross-correlation in Figure~\ref{cross_corr} shows
a pronounced peak at ${\sim}27$~min.  The observed light curves have
a maximum correlation  $R =
0.21$. However, the observed NIR flux densities, if our model is
correct, are not identical with the slow process $S_{\rm{thin}}(t)$
but include the faster varying effects of the cooling cutoff. In
order to approximate the quantity $S_{\rm{thin}}(t)$, we also present
the cross-correlation of the 230~GHz data with the logarithm of the
low-pass-filtered NIR flux densities. This results in a
significant increase of the correlation to $\rm{R}=0.43$. The 95\% false
alarm probability levels calculated from our model and the posterior
in Section~\ref{results} indicate that the data  are consistent
with our model assumptions\footnote{However, we have not tested the
significance of the correlation independently (with, e.g., a Granger
causality test). The cross-correlation of Figure~\ref{cross_corr}
cannot be used as independent evidence for the existence of this type
of correlation. All we can conclude here is that the \Sp--SMA
synchronous dataset is consistent with a delay that is predicted by
our model.}.

\subsection{SED and Adiabatic expansion}

\begin{figure*}[h]
\begin{center}
\includegraphics[scale=0.8, angle=0]{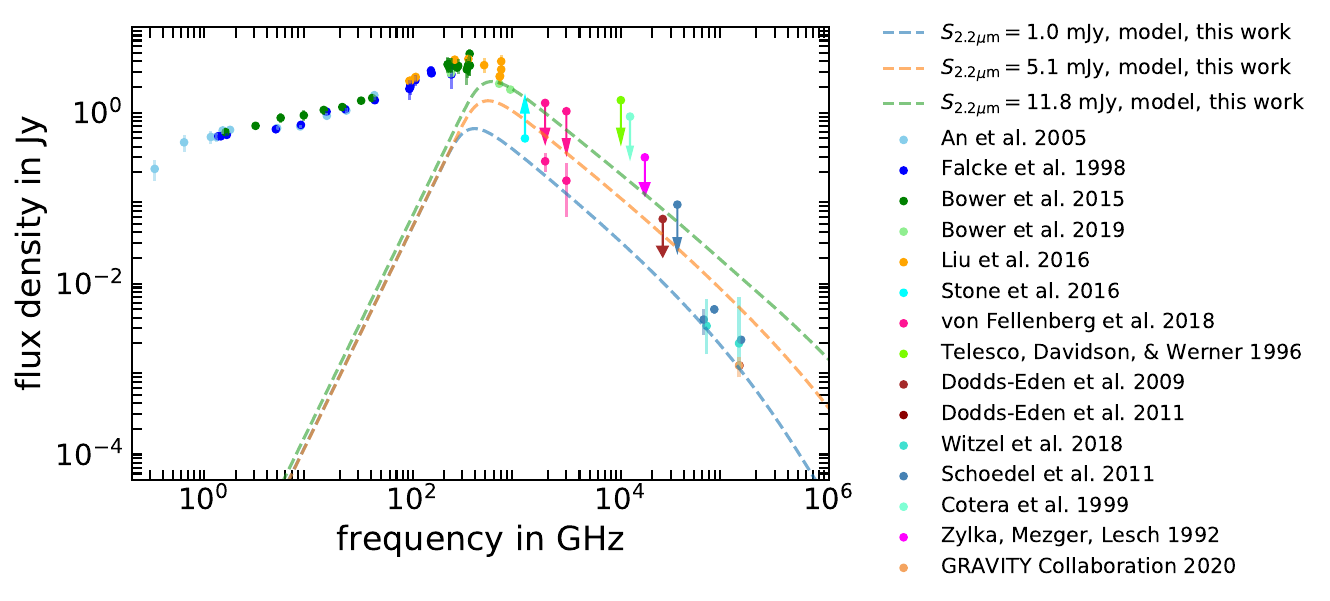}
\end{center}
\setlength{\abovecaptionskip}{-5pt}
\caption{Variable SED in comparison to overall SED of Sgr A*. The dashed lines show snapshots of the
SED from our model for three different (dereddened) NIR flux-density levels. Points show SED
values taken from
\cite{1992A&A...261..119Z,1996ApJ...456..541T,1998ApJ...499..731F,1999ASPC..186..240C,2005ApJ...634L..49A,2009ApJ...698..676D,2011A&A...532A..83S,2011ApJ...728...37D,2015ApJ...802...69B,2016A&A...593A.107L,2016ApJ...825...32S,2018ApJ...862..129V,2018ApJ...863...15W,2019ApJ...881L...2B,2020A&A...638A...2G}.
 The (dereddened) NIR points represent different
attempts to quantify the mode (peak) of the NIR flux-density
distribution (compare discussion in
\citealt{2018ApJ...863...15W}). At the Herschel bands (100, 160, and
250~$\micron$) two numbers are presented: the amplitude of the
detected variable signal (blue lower limit, grey points) that serve
as an lower limit of the steady flux density, and for 100 and
160~$\micron$ upper limits of the steady flux density derived from
the assumption of 15\% variability
(\citealt{2018ApJ...862..129V}).}\label{SED_obs}
\end{figure*}

Figure~\ref{SED_obs}  compares the SED of the 
compact component modeled here with the so-called steady SED of (selected)
literature values. A minimum
requirement for our model \replaced{should be that it does not}{is that it cannot} violate the
steady SED. \deleted{in the submm or radio.  
For a source that is variable in
all parts of the spectrum, it is a matter of definition what can be
considered the steady flux. While in the radio the variability
amplitude is a small fraction of the mean flux, this problem is 
prominent NIR, where the source is
entirely dominated by variable flux, and no truly steady point source
has been identified. Additionally, observations of the mean flux
(especially in the presence of variability) are limited by their
duration, their resolution, extended flux surrounding the Galactic
center, and the atmosphere. Despite the difficulties, we see a certain degree of
scatter at radio to submm flux densities which most likely are caused
by intrinsic variability. }
At  frequencies above submm,
the only detections of \Sg\ are far-infrared measurements with \Her\
(\citealt{2016ApJ...825...32S,2018ApJ...862..129V}) and NIR and X-ray
measurements as presented here.  Our
model \replaced{can describe}{is consistent with} the variability amplitudes \deleted{at NIR and FIR
frequencies well} and \deleted{even follows} the spectral index \deleted{suggested by the
observed data in the} \added{in the} FIR while not violating any upper limits in the
radio or submm part of the SED. 
In the submm, our model component significantly
contributes to the overall variability but cannot explain the
entire flux density. At lower radio frequencies ({$<$300}~GHz),
the variable contribution decreases quickly, and the SED is dominated by one or
more other source components, \added{presumably the quasi-steady emission from the accretion flow \deleted{, which might contribute to the excess
submm variability at longer timescales that we cannot account for}.
This implies that the compact source region we have modeled is immersed in submm photons from the neighboring plasma. We have modeled this scenario with a thermal and non-thermal synchrotron model with SSC and IC scattering of the ambient submm photons from the thermal component by the non-thermal electrons of the compact component. For this model we did not use the analytical approximations described in this paper but instead used numerical integration to properly calculate the flanks of the SSC and IC spectra. We assumed that the non-thermal electrons are exposed  to a constant submm photon density as if located at the center of a sphere of a larger volume of thermal electrons. The thermal component of the spectrum was fitted to the cm to mm observations. For a rather bright NIR state, the resulting IC component is six to seven orders of magnitude dimmer than the SSC component. If the two electron populations are separated in space by some distance, the photon density and consequently the IC emission would be even lower. Therefore, the IC component from the ambient photons can be safely neglected here.}

Several papers have discussed expanding plasmon flare evolution models for
Sgr~A*. The earliest, by \cite{2006ApJ...650..189Y},  described
the temporal evolution of flares in the cm wavelength regime.
\cite{2006A&A...450..535E, 2008A&A...492..337E, 2012A&A...537A..52E}
explained a possible correlation and delay between NIR and a submm
flares. \cite{2008ApJ...682..373M} presented an example of simultaneous IR
and submm data with  a delay (IR leading, submm
following) of $20\pm5$~minutes, consistent with the data presented
here and with our model and also discussed adiabatic expansion. All
these expanding plasmon models result in typical expansion velocities
of ${\sim} 0.01c$, with $c$ the speed of light. Peak expansion
velocities derived from our time dependent model reach
${\sim} 0.01c$ and are consistent with the earlier estimates.

In contrast to \cite{2006ApJ...650..189Y} and \cite{2009ApJ...706..348Y}, \cite{2009A&A...496...77F}
interpreted 20--40~minute delays in the 20 to 40~GHz regime in the context
of the frequency dependence of VLBI sizes and saw evidence for a
relativistic outflow. \cite{2015A&A...576A..41B} even followed the 
realtime progression of variability maxima from 100 to 19~GHz and
similarly derived relativistic outflow velocities of up to ${\sim}
0.77c$, i.e.,  a jet. 

Whether the similar delays found
at radio frequencies and in the submm to NIR regime indicate a direct
relation between the variability of those regimes or are merely
coincidental cannot be decided here.  \added{As Figure~\ref{SED_obs} demonstrates, for frequencies {$<$100}~GHz and $\alpha_{\rm{Radio}} = +2.5$, the variable component seldom contributes significant flux. Our posterior of the self-absorption peak position $\nu_m$ (Figure~\ref{posteriors}) predicts values as low as 90~GHz with measurable contributions down to 40--50~GHz. However, predictions for the radio part crucially depend on the actual $\alpha_{\rm{Radio}}$ of the optically thick branch, which for electron density profiles other than constant with radius can differ from 2.5 and is very uncertain. }
Also, we cannot determine whether
the compact component
responsible for the fast, high-frequency variability is located in a
jet or in the accretion disk.

The model predicts the self-absorption turnover to vary between \replaced{150~GHz
and 700~GHz}{90~GHz
and 1~THz}, yielding a wide range of spectral indices at these
wavelengths. On the other hand, in the NIR the spectal index is
typically steeper than the optically thin spectral index $\alpha =
(1- \gamma_{e})/2$ due to the cooling cutoff. Both regimes, one used
by mm VLBI, the other by VLTI/GRAVITY, are limited in the sense that
they are subject to rapid changes in the state of the radiative
transfer: the submm by its regular changes between the optically
thick and thin branch of the spectrum which potentially causes rapid
changes in the source structure and the NIR by fast synchrotron
cooling, which makes it hard to follow the compact component of
Sgr~A* in its entire evolution.

The median brightness temperature at the self-absorption turnover $T_b =
1.8\times10^{10}$~K. It varies between 80\% and 90\%
of the equipartition brightness temperature $T_{\rm{eq}}$.
Assuming a pure electron plasma and a Doppler factor of~1,
this value suggests that the source is energetically dominated by
the magnetic flux.

\subsection{Inconsistencies}

The model spectral index in the NIR has an upper limit
$\alpha_{\rm{NIR}} \lesssim -0.8$ with many bright flares showing
spectral indices as steep as $\alpha_{\rm{NIR}} \approx -1.2$ (Figure~\ref{NIR_CDF}). The
upper limit of $\alpha_{\rm{NIR}}$ is defined by the spectral index
$\alpha = (1-\gamma_{e})/2$ of the optically thin spectrum without
cooling cutoff. The NIR  spectral index reaches the
optically thin limit when the cooling cutoff 
is at high energies.
In our model fit, $\gamma_{e}$  is determined by the variance of the submm
variability relative to the variance of the NIR variability. $\alpha
\approx -1.0$ is steeper than the canonical NIR spectral index for
bright flares of $\alpha_{\rm{NIR}} \approx -0.65$
(\citealt{2005ApJ...628..246E,2005ApJ...635.1087G,2006ApJ...640L.163G,2006ApJ...642L.145K,2007ApJ...667..900H,2011AA...532A..26B,2014IAUS..303..274W}).
However, spectral index analyses in the NIR have
two main sources of systematic uncertainty: the extinction correction
and background flux levels. While the extinction correction is
precise enough and cannot explain this discrepancy, unaccounted
background flux makes a big difference. However,
\cite{2019ApJ...882L..27D}  reported 
$\alpha_{\rm{NIR}} \approx -0.4$ (with canonical
extinction correction) 
 at such high flux-density levels that
background contamination does not play a role.
Such a flat spectral index could be the
result of the SSC scattering contributing to the NIR.
The lower limit of the power-law section of the SSC spectrum
is given in Equation~\ref{SSC_limits}.  As seen in
Figure~\ref{SSC_SED}, for source parameters discussed here, the SSC
spectrum sets in around NIR frequencies and might become
dominant once in awhile.

Another, potentially related, problem is the submm variability
variance: at 230~GHz the variance at longest timescales cannot be
reproduced by the model. This is the reason we did not include
the last bin of the 230~GHz structure function in our distance
function. Furthermore, Figure~\ref{submmobs} shows significant
offsets in average flux density between the individual SMA and ALMA
epochs. If real, these offsets would point to a long term variability
component that is not accounted for in our model. Several
explanations are possible: 
\begin{itemize}
\item
We modeled the synchrotron source as a sphere homogeneously filled with
electrons, resulting in an optically thick spectral index 
$\alpha_{\rm{thick}} = +2.5$. In the case of a different electron
density profile, this slope can be significantly different with
consequences for the position of the self-absorption turnover and the
relative flux densities between the 230~GHz, 345~GHz, and the NIR. 
\item
A separate mechanism, perhaps a second electron population, is
required to produce the non-varying emission, which accounts for 
most of the submm and longer-wavelength flux most of the time. This second (or
rather primary) component is not needed to account for most of the
observed submm variability, but it might contribute, especially on
long timescales.  This would
reduce the variability associated with
the compact source, resulting in a
flatter optically thin spectral index. 
\item
Instead of one zone undergoing cycles, there might be distinct zones
coming and going with different initial conditions, explaining the
distinct levels of the submm epochs. 
\end{itemize}

Finally, our model gives the brightest observed NIR flare
so far \citep{2019ApJ...882L..27D} only ${\sim}10^{-3}$
probability. In other words, the brightest observed NIR states of Sgr~A* are
somewhat underrepresented in our mock light curves. \cite{2011ApJ...728...37D} and
\cite{2020A&A...638A...2G} interpreted the NIR flux-density
distribution as the result of two states with variability from two
distinct physical processes. Whether this is the case or instead very
bright events are only extreme cases of the process described here
cannot be determined. In order to generate extreme events more
regularly, it would suffice to replace log-normality of the
underlying fast and slow processes with a suited
distribution. However, proper inclusion of extreme values in the
statistics is difficult because the probability of rare events is very uncertain.

\section{Summary}\label{summary}

This paper has presented a comprehensive analysis of submm,
NIR, and X-ray light curves of Sgr~A* including two new epochs of
ALMA data and four new epochs of SMA data.  These include about 2 days of
simultaneously observed NIR and submm light curves, which show definite
correlation between submm and NIR variability and are
consistent with a median delay of $\sim$20~minutes of the submm 
with respect to the NIR. 

This paper is also the first analysis of the X-ray PSD. The X-ray and
NIR variations are correlated but with correlation decreasing at
short variability timescales.

A simple but physically consistent model explains most features of \Sg's variability.  The model was constructed to match the variability amplitudes and timing properties (structure functions) of the NIR and X-ray emission considered separately.  Given those, the model predicts the submm variability properties, the time lag between NIR and submm variations, the submm and NIR spectral indices, and the observed correlation between NIR and X-ray variability.  The model consists of a single zone in an external magnetic field with random injections of high energy electrons and cyclic expansion and contraction.  The radio emission is optically thick synchrotron, the NIR is optically thin synchrotron, and X-rays arise from synchrotron self-Compton emission.  The variability comes from varying density of high-energy electrons and source size, which affect the amplitude of synchrotron emission, the optical depth, the synchrotron cutoff frequency, and the self-Compton amplitude.  The minimum required magnetic field is $\sim$8.5~G, the maximum source size is $\sim$2.7~$R_S$, and the high-energy electrons have a power-law index $\sim$2.95.  The variability processes have timescales of roughly 82 and 135 minutes and are ruled by the
tradeoff between electron injection and expansion cooling in the submm and X-rays and by the tradeoff between injection and synchrotron cooling in the NIR.  General relativity plays no role in the model.

The predicted changes in source parameters are moderate,
typically about a factor two in source size and magnetic flux density
and an order of magnitude in electron density. The submm radiation of
this compact component of Sgr~A* changes from optically thick to thin
and back at the variability timescales.  This likely creates
complex changes in the intrinsic source structure during VLBI observations.

Postulates of the model include the PSD of the underlying trigger
processes and that rises and falls are symmetric in time. The model
allows for but does not require different correlation times at
different frequencies and time lags between frequencies.  Where the
model may fall short is that the most extreme
observations are improbable (but not impossible), and the observed NIR
spectral index at high flux densities is flatter than the model
predicts. That last could probably be remedied by allowing SSC emission
at NIR frequencies.

All in all, the model's shortcomings are related to rare observations
such as the brightest flux-density states and therefore are documented insufficiently
in our sample to warrant a more complex model. For the vast majority
of observations, the model presented here is a sufficient
representation.  At the least, it provides a baseline to
help identify extraordinary variability, e.g., states dominated by
effects of general relativity in the future.
Relativistic effects alone cannot be the origin of the variability
because they cannot explain the rich phenomenology of correlations
between the wavelengths. Our model in its current form does not
require any relativistic boosting or other effects to describe the
variable flux densities. Whether and how this is commensurable with the
findings of apparent circular motion close to the last stable orbit
by \cite{2018A&A...618L..10G} has to be investigated.

Physical conclusions from this work include:
\begin{itemize}
\item There is a compact component with size 2--$3R_S$ that dominates the high
frequency regime. 
\item There is more evidence now for SSC generating the X-rays (and maybe contributing to the NIR).
\item The NIR is linked to the submm variability, and expansion is  a
  strong candidate process to explain the correlation.
\end{itemize}

Some things the model does not explain include:
\begin{itemize}
\item Long-timescale variability in the submm.
\item Perhaps the most extreme NIR flux densities.
\end{itemize}

This paper has presented
a quantitative assessment of what a simple model can achieve.  It
can serve as a starting point for further work to test and improve the
model.
Theoretical work should investigate acceleration mechanisms for electrons and whether suitable populations of high-energy electrons can be generated and on what timescales.

The model presented here was enabled and its parameters determined by
the vast library of light-curve monitoring data accumulated over the
past two decades.  Simultaneous light curves at different wavelengths
were especially valuable for testing (though not deriving) the
model. Further work should include better simultaneous measurements
of the X-ray and NIR spectral indices, which should be identical if
our model is correct. Monitoring \Sg's variability simultaneously
from $\sim$2 to $\sim$30~\micron\ with the {\it James Webb Space
  Telescope} should test whether the model's predicted NIR spectral
indices are accurate, clarify the relevance of synchrotron cooling
for this part of the spectrum, and show whether the self-Compton
emission can contribute to the NIR.
\clearpage

\acknowledgments 
\added{We thank the anonymous referee, the statistical editor, and the data editor for their valuable and constructive comments.}
The authors are grateful to Eduardo Ros for valuable comments on the
manuscript. We thank Ziqian Hua, Axel Weiss, Mikhail
Lisakov, Nicholas MacDonald, Stefan Gillessen, Sebastiano von
Fellenberg, Yigit Dallilar, Gabriele Ponti, Arno Witzel
and Silke Britzen for fruitful discussions. We thank Helge Rottmann
and the MPIfR Correlator team for their help and support in running
our code on the VLBI computer cluster.
This work is based on observations made with the
\SST, which is operated by the Jet Propulsion
Laboratory, California Institute of Technology under a contract with
NASA. Support for this work was provided by NASA through an award
issued by JPL/Caltech. We thank the staff of the Spitzer Science Center
for their help in planning and executing these demanding
observations.  This publication is based on data of the Submillimeter
Array. The SMA is a joint project between the Smithsonian
Astrophysical Observatory and the Academia Sinica Institute of
Astronomy and Astrophysics, and is funded by the Smithsonian
Institution and the Academia Sinica.  This paper makes use of the following ALMA data: ADS/JAO.ALMA\#2011.0.00887.S and ADS/JAO.ALMA\#2017.1.00503.S. ALMA is a partnership of ESO (representing its member states), NSF (USA) and NINS (Japan), together with NRC (Canada), MOST and ASIAA (Taiwan), and KASI (Republic of Korea), in cooperation with the Republic of Chile. The Joint ALMA Observatory is operated by ESO, AUI/NRAO and NAOJ. \added{Support for this work was provided by the National Aeronautics and Space Administration through Chandra Award Numbers GO6-17136B and GO7-18135B, issued by the Chandra X-ray Center, which is operated by the Smithsonian Astrophysical Observatory for and on behalf of the National Aeronautics Space Administration under contract NAS8-03060.  P.W., G.G.F.,  M.A.G.,  J.L.H., and H.A.S.  acknowledge grant 80NSSC18K0416 of the National Aeronautics Space Administration.  G.M.,  E.E.B., T.D.,  A.G.,  and M.R.M. acknowledge grant AST-1909554 of the National Science Foundation.}

\vfill\null
\phantom{xx}\facilities{Spitzer(IRAC), ALMA, SMA, APEX, Chandra(ACIS), VLT:Yepun(NACO), Keck:II(NIRC2)}
\software{SED Animation v.1.0 \citep{gw_2021_10.17617/1.kctx3s25}, FFmpeg (\citealt{tomar2006converting}),
Jupyter Notebook (\citealt{Kluyver:2016aa}), Python 3
(\citealt{10.5555/1593511}) with the packages math, decimal, mpmath,
and time, Numerical Python (numpy,
\citealt{oliphant2006guide,van2011numpy}), Matplotlib
(\citealt{Hunter:2007}), and Scientific Python (scipy,
\citealt{2020SciPy-NMeth}), C++, MPI, WebPlotDgitizer (\citealt{Rohatgi2020}).}

\clearpage

\bibliographystyle{apj}
\bibliography{mybib_gc}{}

\appendix

\section{Supplemental Code Package for Generating SED Animations}\label{A1}

As a supplement to this article, we have created a repository
containing the python code with which we generated Figure~\ref{anim}
and the linked animation (SED Animation v1.0, Max Planck Digital Library, doi:10.17617/1.kctx3s25, developed on https://gitlab.mpcdf.mpg.de). The repository can be found at \url{https://doi.org/10.17617/1.kctx3s25}.

The repository contains:
\begin{itemize}
\item
a python library of classes and functions for 
\subitem
--generating synchrotron and SSC spectra from model and
empirical data, 
\subitem
--generating  time series and their auto- and cross-correlations,
\subitem
--making simple caculations and plotting auxiliary data and
calculation results,  
\subitem
--handling the posterior particle system  from  our final ABC run, and
\subitem
--generating movies of the evolution of the data over time;

\item
a Jupyter notebook with simple examples showing how to generate an SED
and an animation from the posterior; and

\item
the posterior particle system of 5\,000 parameter combinations with
appropriate weights. 
\end{itemize}
This package is tailored to the case of Sgr~A* and
the model described above. However, we hope it might be of use for
different applications, and we publish the code under a 3-Clause BSD
License that will permit free use. 
The code uses the FFmpeg library (\citealt{tomar2006converting}),
Jupyter Notebook (\citealt{Kluyver:2016aa}), Python 3
(\citealt{10.5555/1593511}) with the packages math, decimal, mpmath,
and time, Numerical Python (numpy,
\citealt{oliphant2006guide,van2011numpy}), Matplotlib
(\citealt{Hunter:2007}), and Scientific Python (scipy,
\citealt{2020SciPy-NMeth}). 
\end{CJK*}
\end{document}